\documentclass[times]{nmeauth}

\usepackage{moreverb}

\usepackage[colorlinks,bookmarksopen,bookmarksnumbered,citecolor=red,urlcolor=red]{hyperref}

\newcommand\BibTeX{{\rmfamily B\kern-.05em \textsc{i\kern-.025em b}\kern-.08em
T\kern-.1667em\lower.7ex\hbox{E}\kern-.125emX}}

\usepackage[font=footnotesize]{subfig}
\usepackage{stmaryrd} 
\usepackage{xcolor,graphicx}

\usepackage{algorithm}
\usepackage{algorithmic}

\usepackage{amssymb,amsmath,amstext, amsfonts, mathrsfs, mathtools}
\usepackage{arydshln}

\usepackage{cite}

\usepackage{anyfontsize}

\usepackage{varwidth}

\algsetup{linenosize=\scriptsize}
\definecolor{algorithmcolor}{gray}{0.55} 

\newcommand{\algorithmicbreak}{\color{algorithmcolor}\textbf{break}\color{black}}

\renewcommand{\algorithmicwhile}{\color{algorithmcolor}\textbf{while}\color{black}}

\newcommand{\algorithmicbreakwhile}{\algorithmicbreak\ \algorithmicwhile}

\renewcommand{\div}{\nabla \cdot}     
\newcommand{\Div}{\div}               
\newcommand{\grad}{\nabla}            
\newcommand{\Grad}{\grad}             

        % ...

\usepackage{letltxmacro}
\renewcommand*{\mathellipsis}{%
  \mathinner{{\ldotp}{\ldotp}{\ldotp}}%
}
\makeatletter
\@ifdefinable{\org@ldots}{%
  \LetLtxMacro\org@ldots\ldots
  \DeclareRobustCommand*{\ldots}{%
    \ifmmode
      \expandafter\my@ldots
    \else
      \expandafter\textellipsis
    \fi
  }%
}
\newcommand*{\neghalfmskip}{%
  \nonscript\mskip-.5\muexpr\thinmuskip\relax%
}
\newcommand*{\my@ldots}{%
  \mathellipsis
  \@ifnextchar,\neghalfmskip{%
  \@ifnextchar:\neghalfmskip{%
  \@ifnextchar;\neghalfmskip{%
  \@ifnextchar.\neghalfmskip{%
  \@ifnextchar!\neghalfmskip{%
  \@ifnextchar?\neghalfmskip{%
    \rightdelim@
    \ifgtest@
      \mskip-.5\muexpr\thinmuskip\relax
    \fi
  }}}}}}%
}
\makeatother

\newcommand \deriv[2]  { \tfrac{\mathrm d{#1}}{\mathrm d{#2}}    }   
\newcommand \derivsecond[2]  { \frac{\mathrm d^2{#1}}{\mathrm d{#2}^2}    }   
\newcommand \partderiv[2]{\tfrac{\partial #1}{\partial #2}      }   

\newcommand{\DSecondTime}[1]{\derivsecond{#1}{t}}                   
\newcommand{\DTime}[1]{\deriv{#1}{t}}                   
\newcommand{\dTime}[1]{\partderiv{#1}{t}}               
\newcommand{\dtimedot}[1]{\dot{#1}}                             
\newcommand{\dsecondtimedot}[1]{\ddot{#1}}                             

\newcommand{\restr}[2]{ \left. #1 \right|_{#2}}

\DeclareMathOperator{\RE}{Re}		

\newcommand{\jump}[1]{\ensuremath{[\![#1]\!]} }

\newcommand{\meanavg}[1]{\ensuremath{\left\{#1\right\}}_{\mathrm{m}}}

\DeclareMathOperator{\traceop}{tr}		
\newcommand{\trace}[1]{\traceop{(#1)}}

\newcommand{\vel}{\boldsymbol{u}}				

\newcommand{\x}{\boldsymbol{x}}
\newcommand{\n}{\boldsymbol{n}}
\newcommand{\bodyf}{\bff}

\newcommand{\stress}{\boldsymbol{\sigma}}
\newcommand{\visc}[1]{\boldsymbol{\epsilon}(#1)}
\newcommand{\I}{\boldsymbol{I}}

\newcommand{\scalL}[2]{(#1,#2 )}                       
\newcommand{\scalbound}[3]{\langle #1,#2 \rangle_{#3}} 

\newcommand{\euclidian}[1]{\|#1\|}

\newcommand{\onehalf}{\frac{1}{2}}

\newcommand{\eqdef}{:=}      

\newcommand{\sd}{\mathrm{s}} 
\newcommand{\fd}{\mathrm{f}} 
\newcommand{\fs}{\mathrm{fs}} 

\newcommand{\Dom}{\Omega}      
\newcommand{\Domh}{\Dom_h}     

\newcommand{\Int}{\Gamma}           

\newcommand{\Domf}{\Dom^\fd}          
\newcommand{\Doms}{\Dom^\sd}          
\newcommand{\DomsRef}{\Dom_0^\sd}          

\newcommand{\IntN}{\Int_{\mathrm{N}}}          
\newcommand{\IntD}{\Int_{\mathrm{D}}}          

\newcommand{\Intfs}{\Int^{\fs}}      

\newcommand{\IntDRef}{\Int_{\mathrm{D},0}}     
\newcommand{\IntNRef}{\Int_{\mathrm{N},0}}     

\newcommand{\bfhN}{\bfh_{\mathrm{N}}} 
\newcommand{\bfgD}{\bfg_{\mathrm{D}}} 

\newcommand{\bfsigma}{{\pmb\sigma}}

\newcommand{\bfepsilon}{{\pmb\epsilon}}

\newcommand{\bfc}{\boldsymbol{c}}
\newcommand{\bfu}{\boldsymbol{u}}
\newcommand{\bff}{\boldsymbol{f}}
\newcommand{\bfg}{\boldsymbol{g}}
\newcommand{\bfh}{\boldsymbol{h}}
\newcommand{\bfv}{\boldsymbol{v}}

\newcommand{\bfw}{\boldsymbol{w}}

\newcommand{\bfd}{\boldsymbol{d}}
\newcommand{\bfa}{\boldsymbol{a}}
\newcommand{\bfn}{\boldsymbol{n}}

\newcommand{\bfr}{\boldsymbol{r}}

\newcommand{\bfP}{\boldsymbol{P}} 
\newcommand{\bfS}{\boldsymbol{S}} 
\newcommand{\bfF}{\boldsymbol{F}} 
\newcommand{\bfR}{\boldsymbol{R}} 
\newcommand{\bfU}{\boldsymbol{U}} 
\newcommand{\bfC}{\boldsymbol{C}} 
\newcommand{\bfE}{\boldsymbol{E}} 
\newcommand{\bfN}{\boldsymbol{N}} 
\newcommand{\bfG}{\boldsymbol{G}} 
\newcommand{\bfH}{\boldsymbol{H}} 

\newcommand{\bfM}{\boldsymbol{M}} 
\newcommand{\bfD}{\boldsymbol{D}} 
\newcommand{\bfA}{\boldsymbol{A}} 

\newcommand{\bfL}{\boldsymbol{L}} 

\newcommand{\bfx}{\boldsymbol{x}} 
\newcommand{\bfX}{\boldsymbol{X}} 
\newcommand{\bfChi}{\boldsymbol{\chi}} 
\newcommand{\bfPhi}{\boldsymbol{\Phi}} 
\newcommand{\bfvarphi}{\boldsymbol{\varphi}} 

\newcommand{\bfzero}{\boldsymbol{0}}

\newcommand{\R}{\mathbb{R}}

\newcommand{\foralls}{\forall\,}

\newcommand{\mcG}{\mathcal{G}}
\newcommand{\mcF}{\mathcal{F}}
\newcommand{\mcT}{\mathcal{T}}
\newcommand{\mcV}{\mathcal{V}}
\newcommand{\mcW}{\mathcal{W}}
\newcommand{\mcX}{\mathcal{X}}
\newcommand{\mcQ}{\mathcal{Q}}

\newcommand{\mcC}{\mathcal{C}}

\newcommand{\mcA}{\mathcal{A}}

\newcommand{\mcD}{\mathcal{D}}
\newcommand{\mcB}{\mathcal{B}}

\newcommand{\mcL}{\mathcal{L}}

\newcommand{\nablan}{\partial_{\bfn}}

\newcommand{\breakeq   }[3]{\ifthenelse{\equal{#1}{break}}{\nonumber\\&#2}{#3}}
\newcommand{\addifbreak}[2]{\ifthenelse{\equal{#1}{break}}{}{#2}}

\newcommand{\lat}[1]{#1} 
\newcommand{\ie}{\lat{i.e.\@}} 
 
\newcommand{\eg}{\lat{e.g.\@}}

\newcommand{\name}[1]{{\textsc{#1}}} 

\newcommand{\Secref}[1]{Section\,\ref{#1}}   
\newcommand{\Figref}[1]{Figure\,\ref{#1}}    
\newcommand{\figref}[1]{\Figref{#1}}         
\newcommand{\Eqref}[1]{\eqref{#1}}           
\newcommand{\Algoref}[1]{Algorithm\,\ref{#1}}

\newcounter{defcount}

\newcounter{remcount}

\newtheorem{definition}[defcount]{Definition}  
\newtheorem{remark}[remcount]{Remark}
\definecolor{darkblue}{rgb}{0.27 0.52 0.60}

\begin{document}

\runningheads{B.~Schott et al.}{A hybrid Eulerian-ALE approach to fluid-structure interaction}

\title{A monolithic approach to fluid-structure interaction based on\\ a hybrid Eulerian-ALE fluid domain decomposition\\ involving cut elements}

\author{B.~Schott\corrauth, C.~Ager and W.A.~Wall}

\address{Institute for Computational Mechanics, Technical University of Munich,\linebreak Boltzmannstra{\ss}e~15, 85747~Garching,~Germany}

\corraddr{B.~Schott, Institute for Computational Mechanics, Technical University of Munich, Boltzmannstra{\ss}e 15, D-85747 Garching, Germany. E-mail:~schott@lnm.mw.tum.de}

%
%

\begin{abstract}
A novel method for complex fluid-structure interaction (\name{FSI}) involving large structural deformation and motion is proposed.
The new approach is based on a hybrid fluid formulation that combines the advantages of purely Eulerian (fixed-grid) and
Arbitrary-Lagrangian-Eulerian (\name{ALE} moving mesh) formulations in the context of \name{FSI}.
The structure - as commonly given in Lagrangian description - is surrounded by a fine resolved layer of fluid elements based on an \name{ALE}-framework.
This \name{ALE}-fluid patch, which is embedded in an Eulerian background fluid domain, follows the deformation and motion of the structural interface.
This approximation technique is not limited to Finite Element Methods, but can can also be realized within other frameworks like Finite Volume or Discontinuous Galerkin Methods.
In this work, the surface coupling between the two disjoint fluid subdomains is imposed weakly using a stabilized Nitsche's technique in a Cut Finite Element Method (\name{CutFEM})
framework. At the fluid-solid interface, standard weak coupling of node-matching or non-matching finite element approximations can be utilized.
As the fluid subdomains can be meshed independently, a sufficient mesh quality in the vicinity of the common fluid-structure interface can be assured.
To our knowledge the proposed method is the only method (despite some overlapping domain decomposition approaches
that suffer from other issues) that allows for capturing boundary layers and flow detachment via appropriate grids around largely moving and deforming bodies and is able to do this~\eg~without the necessity of costly remeshing procedures.
In addition it might also help to safe computational costs as now background grids can be much coarser.
Various \name{FSI}-cases of rising complexity conclude the work.
For validation purpose, results have been compared to simulations using a classical \name{ALE}-fluid description or purely fixed-grid \name{CutFEM} based schemes.
\end{abstract}

%
%


\maketitle

\section{Introduction}
\label{sec:Introduction}
Fluid-structure interaction (\name{FSI}) problems involving large structural movements and deformations are of significant interest in various fields of engineering and applied sciences. 
However, an important prerequisite for achieving reliable results, especially for flows at higher Reynolds numbers, is an appropriate mesh resolution in the boundary layer.
The latter is mandatory in order to capture the wall normal gradients around the wet structure surface accurately.
An insufficient mesh quality at the fluid-structure interface likely results in an overall corrupted solution of the coupled problem.
\newpage
The essential advantage of the established Arbitrary-Lagrangian-Eulerian (ALE)-based \name{FSI}-approach,
which goes back to \cite{Hirt1974,Belytschko1978,Belytschko1980,Hughes1981,Donea1977,DoneaGiulianiHalleux1982},
is that the mesh knows about the position of the structure within the fluid domain,
such that for example the mesh can be refined towards the interface area.
However, the pre-processing of appropriate high quality meshes that satisfy often extreme requirements in the boundary layer is difficult and time-consuming,
and large structural motions can heavily distort the fluid mesh.
While this might not be a billing argument along with some meshes, it becomes crucial for example along with
boundary layer meshes that not only have extreme aspect ratios but are also placed in the region with the
highest deformations and hence are very vunerable.
Hence, costly remeshing and mesh-updating procedures have to be considered
that again are particularly challenging, \eg~in connection with boundary layer meshes.
In summary, also for such approaches optimality of a fluid mesh around the structure can often not be preserved.

The shortcoming of \name{ALE} based \name{FSI} schemes to deal with large and complex motions
was the motivation for the development of an alternative class of \name{FSI} approaches, known as fixed-grid methods.
Such methods sparked quite some interest in recent years.
For an overview of some approaches, the reader shall be referred to \cite{WallGerstenbergerGamnitzerEtAl2006}. 
Following a pure fixed-grid approach, the entire fluid domain is described in an Eulerian framework.
Since structural mesh and fluid mesh are not required being fitted at the common interface, they seem particularly suitable for large deformation \name{FSI}
\cite{Wall2008, BurmanFernandez2014, FernandezLandajuela2016}.
But unlike in the classical \name{ALE} based \name{FSI} approach, an a priori mesh refined around the wet surface can hardly be achieved.
A rather straightforward solution would be a local, adaptive mesh refinement and coarsening combined with
error estimator-based and/or heuristics-based refinement indicators, as described in, \eg, \cite{Gerstenberger2008a, Verfuerth1994}. 
Though, such an adaptive approach becomes rather inefficient for 3D problems involving large motion of the structural surface,
since large-sized regions have to be refined for several levels and the mesh updates may have to be accomplished frequently throughout the simulation.
Furthermore, common refinement algorithms operate in all spatial directions,
which would destroy the inherent grading of boundary layer meshes towards the solid body.
Other attempts to relax strong restrictions with regards to interface rotations
have been suggested, \eg, in \cite{Kloppel2011} based on sliding mesh techniques.
Another interesting variant of sliding interface-fitted meshes is the so called shear-slip mesh update method introduced in \cite{Behr1999} that reconnects nodes in the element layer next to the interface.
An interesting method for such \name{FSI} problems utilizing an \name{ALE} formulation of embedded boundary methods was proposed
by \cite{Farhat2014}, where non-interface-fitted embedded meshes
are rigidly translated and/or rotated to track the rigid component of the dynamic body motion.

A highly advantageous approach which drastically simplifies meshing around structures and perfectly suites for the creation of refinements in
FSI-interface normal direction consists in utilizing domain decomposition for the fluid field.
The idea of utilizing two independent overlapping fluid meshes allows to combine \name{ALE} and Eulerian based fluid techniques in the vicinity of the solid
and the far-field, respectively.
Chimera schemes are an example for an iterative coupling method based on an overlapping fluid decomposition,
which were introduced originally for mesh generation and for the simulation of flows around rigid bodies
(see for example \cite{Wang2000,Houzeaux2003,Steger1983}).
An extension to problems including flexible structures has been presented in \cite{Wall2008}. 
However, Chimera-like couplings have some drawbacks.
In order to obtain a converged solution after iterating between the fluid domains, an overlapping zone of two subsequent
fluid domains has to be present. This introduces an additional iteration set over the overlapping fluid
grids in order to obtain the final fluid solution.
Beside this additional cost, the overlapping domain has to be large enough to achieve a converged solution between the subdomains and this is again particularly cumbersome when highly refined boundary layer meshes should be coupled
coarse background grids.

To overcome such shortcomings, a powerful technique consists in utilizing a composite of overlapping grids,
where the solution in the background mesh is cut-off at the artificial fluid-fluid interface.
The latter is defined as the trace of an embedded grid.
This discretization technique is not limited to finite element based schemes, but can be realized in finite volume frameworks as well,
even though FEM is chosen in the present work.
The application of such a fluid discretization concept for \name{FSI} has been considered in a series of works
\cite{Gerstenberger2008a, WallGerstenbergerGamnitzerEtAl2006, Shahmiri2011, Massing2015}.
In an \name{FSI} setting, the structure is surrounded by a moving layer of fine \name{ALE}-fluid elements,
which is then embedded into the fixed-grid Eulerian background fluid grid - motivating the designation \emph{hybrid Eulerian-\name{ALE} approach}.
While the structure moves and deforms, the boundary layer mesh follows the deformation of the structural surface -
the near surface flow is captured appropriately.
However, in order to apply such fluid patches in complex \name{FSI} problems, it is crucial to satisfy high demands on the 
coupling of the separate background and embedded fluid subdomains along the shared fluid-fluid interface.
While classical Lagrange-multiplier based couplings show severe restrictions with regards to a reasonable choice of discrete function spaces
and in particular requires a careful choice of the multiplier space, stabilized schemes are often more powerful.
A stabilized stress-based Lagrange-multiplier method for coupling the fluid phases involving cut elements has been presented first in \cite{Shahmiri2011}.
To overcome restricting limitations with regards to the location of the embedded fluid patch within the background mesh,
a stable and optimal convergent Nitsche-based coupling method has been presented by \cite{SchottShahmiriKruseWall2015}.
The latter method is based on the Cut Finite Element Method (CutFEM) \cite{BurmanClausHansboEtAl2014},
which dates back to the eXtended Finite Element Method
(see \cite{SchottRasthoferGravemeierWall2015, Rasthofer2011, Krank2016, Massing2015, ChessaBelytschko2003} for various flow applications).
The fluid-fluid coupling is enforced weakly employing Nitsche's formulation \cite{Nitsche1971}
supported by additional penalty-like stabilization techniques for cut elements - the face-/edge-oriented ghost penalty stabilizations
\cite{BurmanHansbo2012,Burman2014b}.
Advancements of these stabilization techniques, acting on the inter-element jumps of velocity and pressure normal derivatives
of cut elements have been made by \cite{SchottWall2014, MassingSchottWall2016_CMAME_Arxiv_submit, Winter2018} for the incompressible Navier-Stokes equations.
The stabilized embedded fluid formulation introduced in \cite{SchottShahmiriKruseWall2015} is one prerequisite of our \name{CutFEM} based hybrid Eulerian-ALE \name{FSI} approach.

For the fluid-structure coupling, different monolithic coupling schemes are available
and the coupling between the moving \name{ALE}-fluid domain and the structure can be handled in the same way as in traditional \name{ALE} based \name{FSI} schemes,~\ie
node match of fluid and solid mesh at the common interface.
In the simplest case, common interface velocity degrees of freedom can be shared and continuity conditions can be incorporated strongly (see \eg~\cite{Kuttler2010,Gee2011}).
A more flexible scheme has been proposed in \cite{Kloppel2011}, which allows for non-conforming non-overlapping meshes,
where the interface conditions are enforced weakly utilizing a dual-mortar Lagrange multiplier method~\cite{Wohlmuth2001}.
Over the past years, also Nitsche's technique (see, \eg, \cite{BurmanFernandez2007a}) has been discussed for \name{FSI} couplings with under-resolved boundary layer regions.
While strong enforcements and therefore exact fulfillment of coupling conditions often result in oscillatory approximations of the boundary-layer solution (see discussions already for pure flow problems in \cite{Bazilevs2007, BurmanFernandezHansbo2006}),
an automatic relaxation of these constraints is preferable, which, however, still converges to the exact fulfillment with mesh refinement in a consistent sense.
Additionally introduced penalty parameters of Nitsche's method have to be scaled properly in order to be independent of the flow regime and therefore the considered problem setup.
As a further advantage of Nitsche's method over Lagrange-multiplier methods, no additional new multiplier variables are introduced to the system of equations,
which from an implementation point of view allows for an easier setup of the monolithic system and simplifies the design of efficient preconditioners.

Due to these reasons, also in this work, a Nitsche-based coupling at the fluid-solid interface is preferred,
which can be setup similar to the fluid-fluid coupling.
Such coupling techniques have been reviewed in detail in \cite{Schott2017b} in the context of unfitted \name{CutFEM} based \name{FSI} approaches
and are the second prerequisite of our hybrid Eulerian-ALE \name{FSI} scheme.

Central focus of this paper is to highlight the flexibility of this hybrid \name{FSI} scheme
for vast challenging \name{FSI} settings.
Even though the fundamental idea of utilizing fluid domain decomposition for \name{FSI} has been presented already in previous works \cite{Shahmiri2011, Massing2015, SchottShahmiriKruseWall2015},
to the best of the authors knowledge, its application to fully coupled time-dependent \name{FSI} problems
has not been presented so far and just indicated in our previous work \cite{Schott2017b} as an outlook.
In the latter publication and references therein,
important theoretical and algorithmic ingredients have been already presented, and therefore will be reviewed just briefly for clarity in the present work.
Moreover, some algorithmic peculiarities of the hybrid \name{FSI} approach will be elucidated.
In addition, since a detailed presentation and investigation of more challenging numerical simulations
was still missing so far, this is another focus of this publication.

The present paper is organized as follows:
In \Secref{sec:hybrid_eulerian_ale_fsi_approach}, we briefly discuss the limitations of discretization concepts for \name{FSI} existing so far
and propose our hybrid domain decomposition idea including the governing equations for the coupled \name{FSI} problem in its strong form.
In \Secref{sec:cutfem_based_hybrid_FSI_formulation}, we propose one potential spatial discretization technique.
It is based on a \name{CutFEM} fluid domain decomposition method and utilizes a Nitsche-type coupling of the fields at the fluid-solid and the fluid-fluid interface, respectively.
A semi-discrete stabilized form for the hybrid Eulerian-ALE \name{FSI} problem is presented
and algorithmic steps for the monolithic solution of the coupled hybrid \name{FSI} system are discussed. 
We demonstrate several numerical examples of increasing complexity in order to verify our method and highlight the capability and the potential of our approach
in \Secref{sec:numerical_examples}.
Finally, conclusions are drawn in \Secref{sec:conclusions}.

\section{A hybrid Eulerian-ALE fluid-structure interaction approach}
\label{sec:hybrid_eulerian_ale_fsi_approach}

\subsection{The hybrid domain decomposition idea for fluid-structure interaction}
\label{ssec:hybrid_domain_decomposition_idea}

\begin{figure}[h!]
  \centering
	\includegraphics[width=0.55\textwidth]{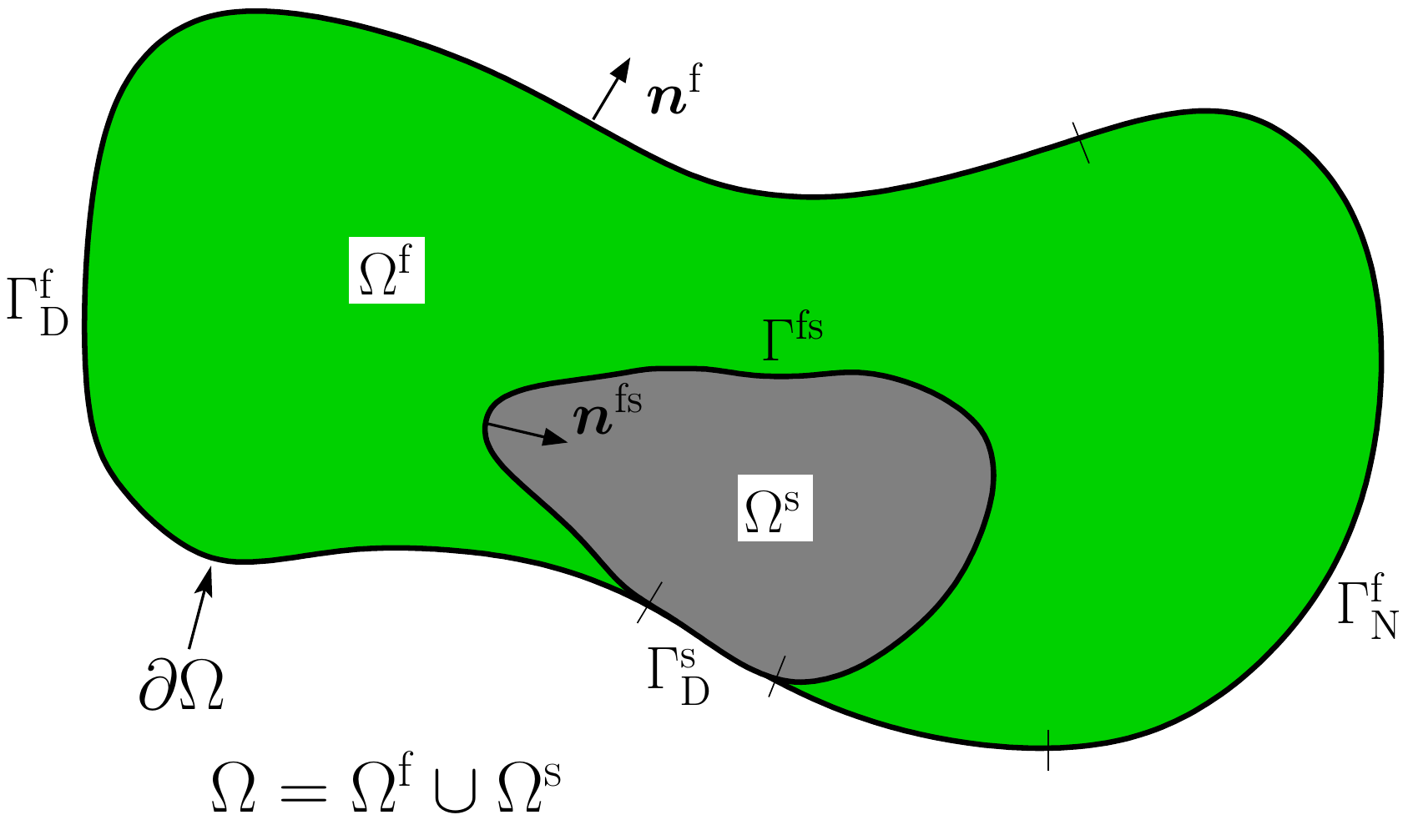}
  \caption{\name{FSI} problem settings: domains, interface and boundaries.}
  \label{fig:fsi_setting}
  \vspace{-12pt}
\end{figure}

Fluid-structure interaction belongs to the large class of surface-coupled problems.
A classical \name{FSI} problem consists of two disjoint bulk subdomains, one for the flow~$\Domf$ and one for the structure~$\Doms$
such that $\Domf\cap\Doms=\emptyset$.
The different phases interact at the common fluid-structure interface $\Intfs = \overline{\Domf}\cap\overline{\Doms}$, at which the respective fields are constraint by coupling conditions.
In addition, Dirichlet and Neumann boundary conditions for the involved fields need to be imposed at outer boundaries $\IntD^{\fd},\IntN^{\fd},\IntD^{\sd},\IntN^{\sd}$,
respectively, to complete the \name{FSI} problem, see \Figref{fig:fsi_setting}.

Common discrete approximations of the structural field use boundary-fitted meshes $\mcT_h^{\sd}$, whose boundaries fit to the domain $\partial\Dom^{\sd}(t)$
at all times~$t$. The structural kinematics are then described in a Lagrangian formalism.
Discrete approximations concepts of the coupled \name{FSI} problem usually differ in the approximation of the flow domain and the respective fields.
Most common techniques will be briefly reviewed and discussed in the following. Afterwards, as the last concept, we introduce the hybrid \name{FSI} approach.

\paragraph*{Classical Arbitrary-Lagrangian-Eulerian (ALE) flow description.}

Following an \name{ALE} based \name{FSI} approach, the fluid subdomain is approximated with a single \name{ALE} fluid mesh $\mcT_h^{\fd}$.
The latter is interface-fitted to the wet structural surface. When the structural body moves, the \name{ALE} mesh also deforms and as its boundary follows the fluid-solid interface
over time. An introduction to the \name{ALE} concept can be found, for instance, in the textbook \cite{Donea2003}.

The classical \name{ALE} based approach for \name{FSI} captivates through its simplicity
and thus is the state-of-the-art in the approximation of \name{FSI} settings.
It allows to easily obtain higher-order geometric approximations using isoparametric concepts
and the resulting schemes gain from well-established stability and best-approximation properties for the involved partial differential equations modeling continuum mechanics.

\begin{figure}
  \centering
  \subfloat[Classical \name{ALE} based \name{FSI}]{\label{fig:computational-mesh-ale-fsi}
    \begin{varwidth}{\linewidth}
	\includegraphics[width=0.40\textwidth]{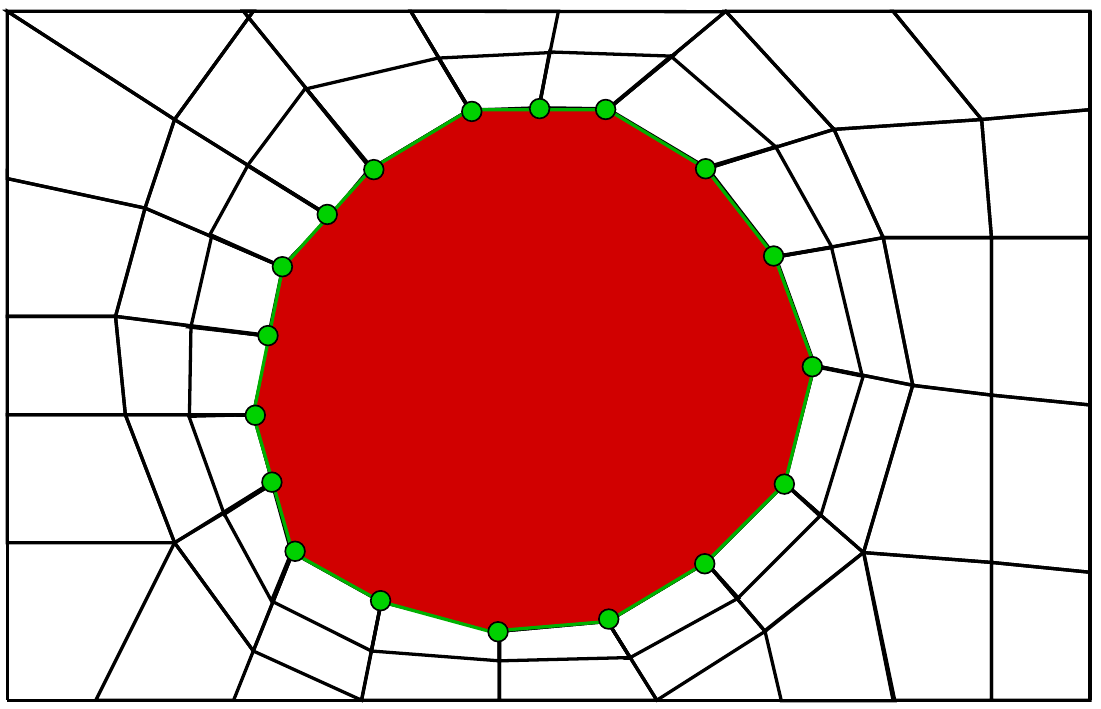}
	\quad
	\includegraphics[width=0.40\textwidth]{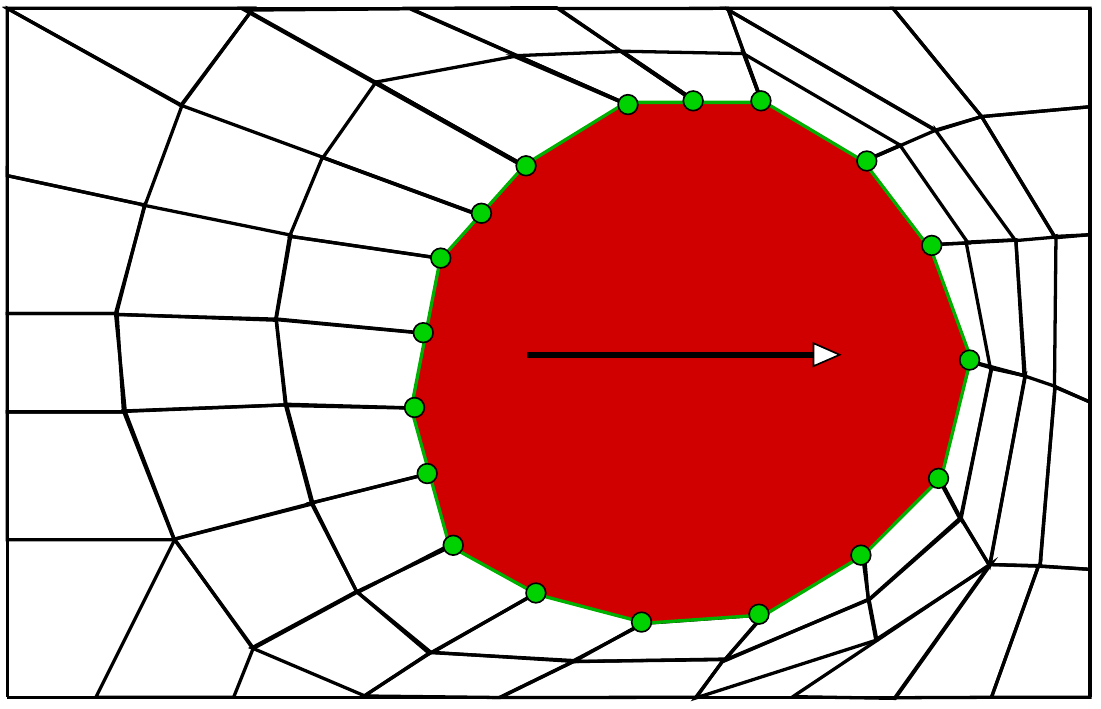}
    \end{varwidth}
}
\\
  \subfloat[Fixed-grid \name{FSI}]{\label{fig:computational-mesh-xfsi}
    \begin{varwidth}{\linewidth}
	\includegraphics[width=0.40\textwidth]{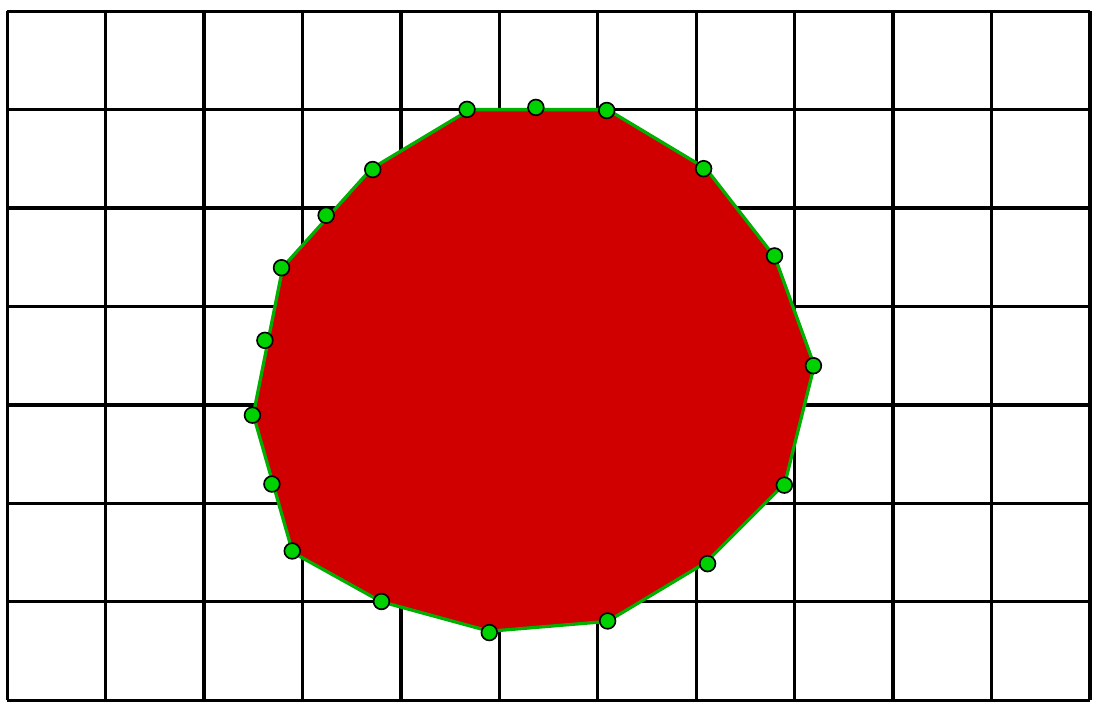}
	\quad
	\includegraphics[width=0.40\textwidth]{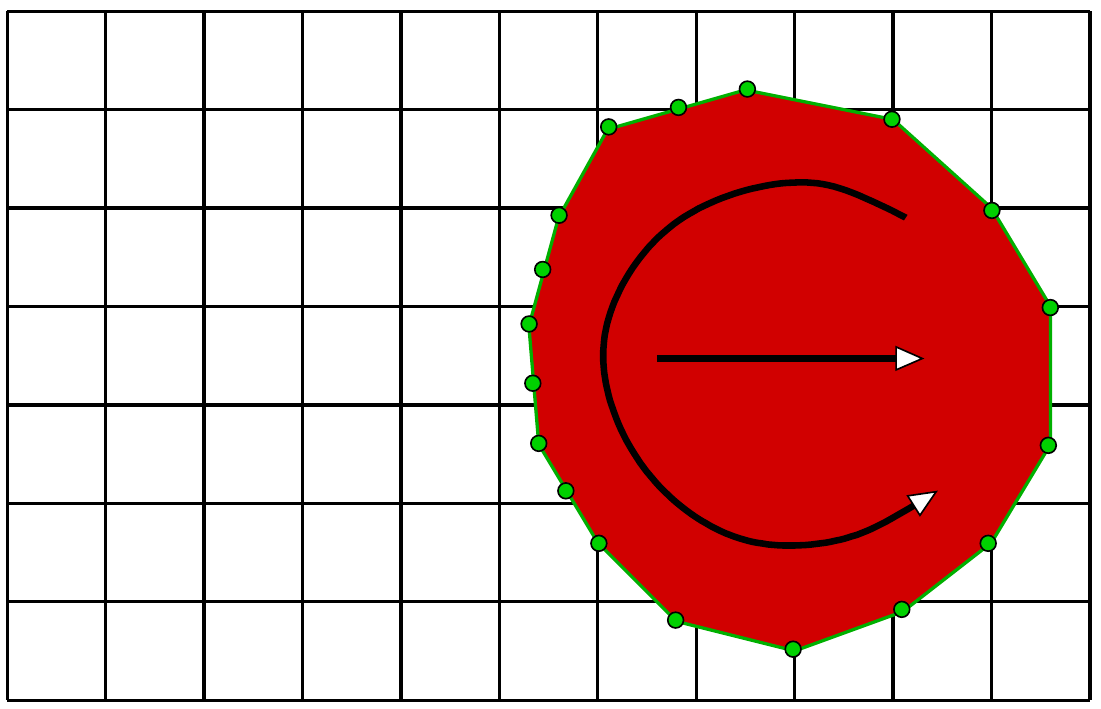}
    \end{varwidth}
}
  \caption{Different moving domain approximation techniques for \name{FSI}:
\protect\subref{fig:computational-mesh-ale-fsi}~\emph{Classical \name{ALE} based moving mesh methods}
are subjected to strict limitations regarding interface motion and deformation, otherwise the fluid mesh will distort.
\protect\subref{fig:computational-mesh-xfsi}~\emph{Fixed-grid schemes} allow for arbitrary motions of the structural body, however, lack a sufficient resolution
of the boundary layer in the vicinity of the \name{FSI} interface.}
  \label{fig:computational-mesh}
  \vspace{-12pt}
\end{figure}

Nevertheless, for complex three-dimensional domains, generating high quality computational fluid grids 
that conform to the domain boundary and are suitable for capturing boundary layers arising for high Reynolds-number flows can be often time-consuming and difficult.
In particular, if large structural motions and deformations are present, the quality of moving meshes cannot be guaranteed in general.
As the finite elements need to follow the interface in its evolution, the meshes can rapidly distort.
Then time consuming remeshing and projection steps have to be performed regularly.
A sketch of \name{ALE} based approximations of the \name{FSI}-problem is given in \Figref{fig:computational-mesh}\subref{fig:computational-mesh-ale-fsi}.

\paragraph*{Fixed-grid Eulerian flow description.}

In contrast to matching-mesh \name{ALE} based methods, pure Eulerian-based fixed-grid flow formulations are more flexible.
For non-interface-fitted approximations of the flow domain, the solution to
the problem is computed only on the active part~$\mcT_h$.
As fluid and solid meshes do not necessarily match at the interface~$\Intfs$, but may overlap, \ie~$\Domh^{\fd\ast}\cap\Domh^{\sd\ast}\neq\emptyset$,
such techniques may drastically simplify meshing of the computational domain
and can overcome the shortcomings of interface-fitted meshes with regards to large domain motions and deformations.
Therefore, such schemes are much more flexible. A visualization is given in \Figref{fig:computational-mesh}\subref{fig:computational-mesh-xfsi}.

Nevertheless, in contrast to \name{ALE} based boundary-fitted mesh techniques,
special measures are required to impose the interfacial constraints, while preserving
robustness, stability and accuracy of the resulting numerical scheme
becomes more challenging when intersecting grids.
As a major drawback of fixed-grid schemes in \name{FSI}, sufficient mesh resolution in the vicinity of the boundary layer
can be only hardly achieved at reasonable computational costs, since the location of the solid is usually unknown a priori.
This, however, is a prerequisite for the quality of the coupled \name{FSI} solution approximation.

\paragraph*{A hybrid Eulerian-ALE approach for \name{FSI}.}

In our novel hybrid \name{FSI} approach, the advantages of the classical moving mesh Arbitrary-Lagrangian-Eulerian (ALE) flow description
are combined with that of a pure fixed-grid Eulerian flow description, as will be elaborated subsequently.

\begin{figure}
  \centering
	\includegraphics[width=0.40\textwidth]{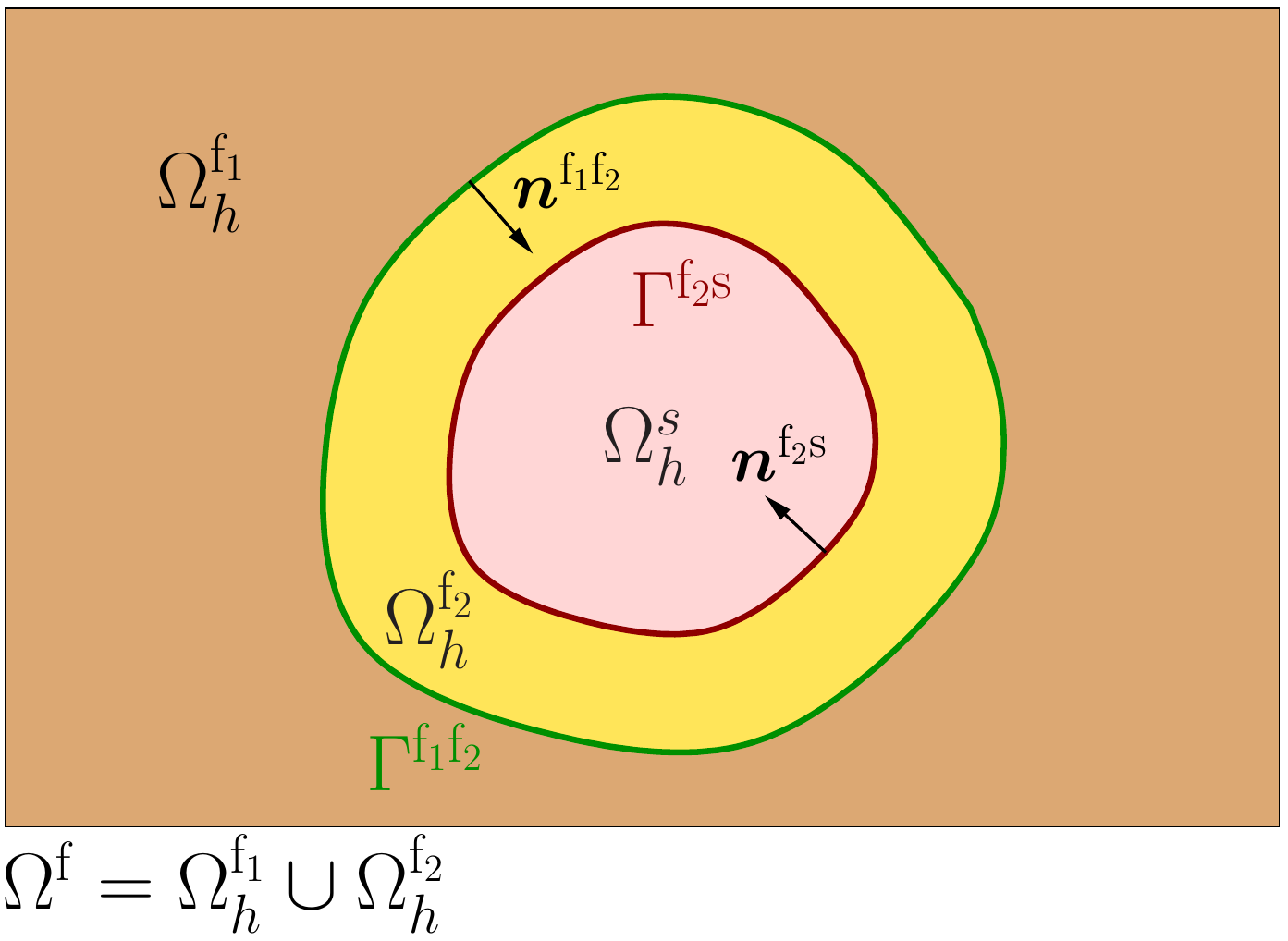}
	\quad
	\includegraphics[width=0.40\textwidth]{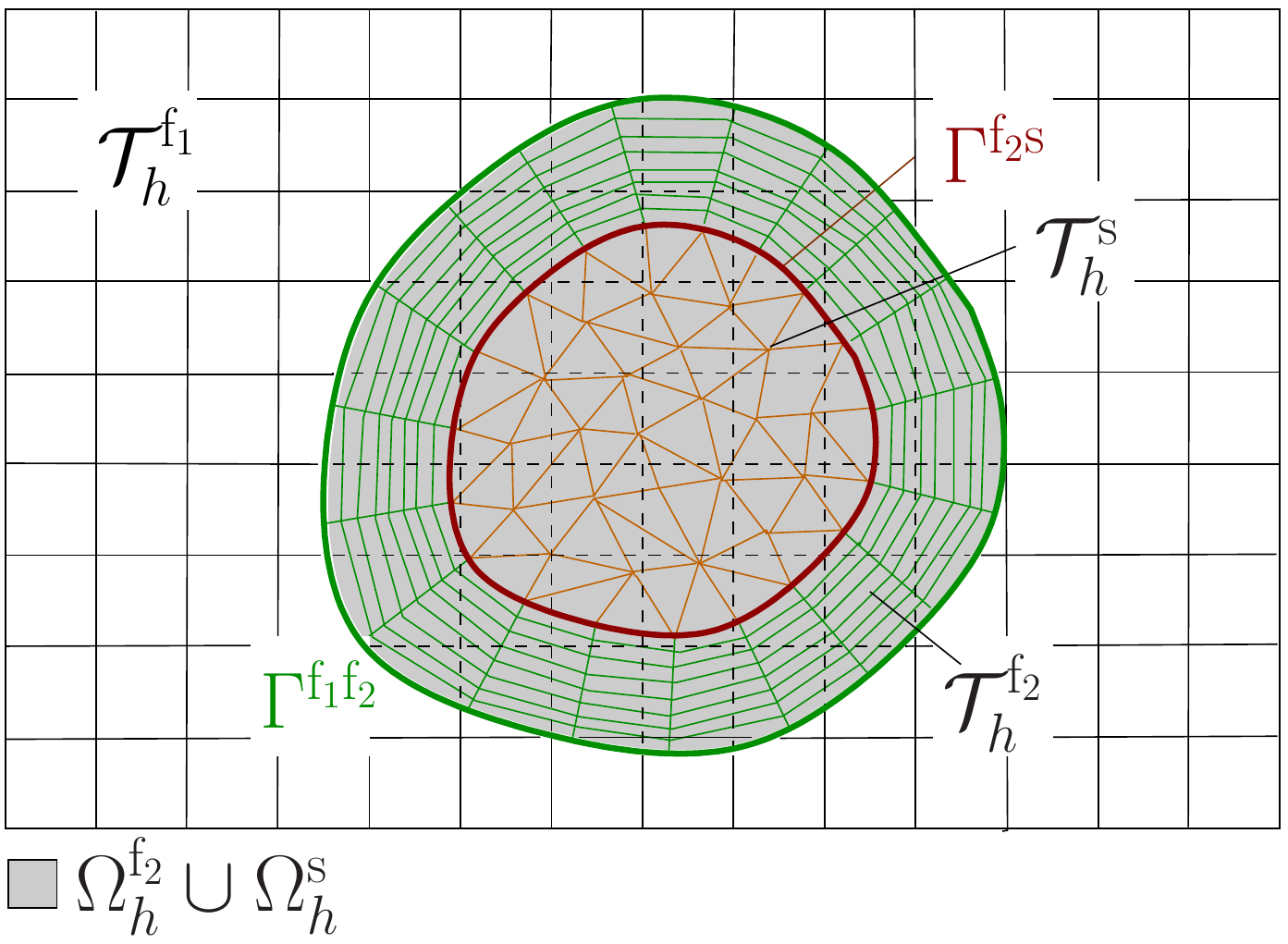}
  \caption{The hybrid Eulerian-ALE discretization concept for \name{FSI}: Domain partition (left) according to the overlapping mesh setting (right).
}
  \label{fig:computational-mesh-xffsi}
  \vspace{-12pt}
\end{figure}

In this approach, the whole physical fluid domain $\Domf$ is artificially separated into two disjoint domain parts $\Domh^{\fd_1}$ and $\Domh^{\fd_2}$,
\ie~$\Domh=\Domh^{\fd_1}\cup\Domh^{\fd_2}$,
which are approximated independently by two overlapping fluid meshes $\mcT_h^{\fd_1}$ and $\mcT_h^{\fd_2}$, as visualized in \Figref{fig:computational-mesh-xffsi}.
To benefit from the fixed-grid schemes with regards to the treatment of large structural motions, for the flow field which is far from the fluid-solid interface~$\Int^{\fd_2\sd}$,
usually a coarser fixed-grid Eulerian approximation $\mcT_h^{\fd_1}$ is utilized.
Since, coupled \name{FSI} problems require a fine-resolved approximation of wall-normal gradients in high-Reynolds-number flows to accurately capture
interfacial forces, a fluid patch $\mcT_h^{\fd_2}$ surrounding the solid body overlaps with the background fluid mesh
in a geometrically unfitted way.
At the fluid-solid interface~$\Int^{\fd_2\sd}$, structural mesh and embedded fluid mesh~$\mcT_h^{\fd_2}$ are chosen interface-fitted
and potentially even node-matching.
This fitting is preserved for all solid locations, requiring the fluid patch following the structural body in its motion and deformation.
This is realized by the use of an overlapping mesh fluid domain decomposition, where $\mcT_h^{\fd_2}$ can be embedded arbitrarily into $\mcT_h^{\fd_1}$.
Then, the solid body and its surrounding boundary layer patch can largely move and deform within the background fluid mesh.
In doing so, the fluid-fluid interface $\Int^{\fd_1 \fd_2}$ subdivides
the background mesh into an active physical part $\Domh^{\fd_1}$ and an inactive void/fictitious part $\Domh^{\fd_1,\textrm{void}}$,
where the latter is covered by the embedded fluid patch~$\Domh^{\fd_2}$ and the solid~$\Domh^{\sd}$,
such that $\Domh^{\fd_1} = \Domh^{\fd_1 \ast} \setminus \{\Domh^{\fd_2} \cup \Domh^{\sd}\} \subsetneq \Domh^{\fd_1 \ast}$
with $\Domh^{\fd_1 \ast}$ denoting the union of all background elements $ T\in\mcT_h^{\fd_1}$. 

In this work, the decisive fluid domain separation is accomplished with the help of the Cut Finite Element Method (\name{CutFEM}).
With this technique, the coupling of the two fluid approximations takes place just at the fluid-fluid interface~$\Int^{\fd_1 \fd_2}$,
and flow is not approximated twice in the overlap zone of the involved fluid meshes.
Also in opposite to overlapping domain decomposition, an iteration between the two fluid fields are needed,
but they are solved together in a single shot.
Details will be provided in \Secref{ssec:semidiscrete_Nitsche_hybrid_spatial_discretization},

Formulating the structural motion in a classical fitting-mesh Lagrangian formalism and
applying a moving mesh \name{ALE} framework to the embedded fluid patch,
a technique which is well-established in classical \name{ALE} based \name{FSI} approaches, allows the patch to follow the body in its movement.
Thus an accurate capturing of flow effects at the fluid-structure interface~$\Int^{\fd_2 \sd}$ is guaranteed.
Moreover, such fluid patches can be generated much easier than appropriate high quality meshes in classical \name{ALE} based \name{FSI} schemes.
Utilizing the \name{CutFEM} based fluid domain decomposition allows for 
independent fluid patch locations within the background mesh~$\mcT^{\fd_1}_h$.
Fluid-structure interaction involving large solid deformations in high Reynolds-number flows including boundary layer effects in the vicinity of solids
can thus be accurately simulated by such a \emph{hybrid Eulerian-ALE \name{FSI} approach}.

\subsection{Governing equations for the coupled fluid-structure interaction system}
\label{ssec:governing_equations_FSI}

In this work, we consider \name{FSI} problems governed by the transient non-linear incompressible Navier-Stokes equations for the flow field and the non-linear structural elastodynamics
equations for the solid body, complemented by appropriate boundary conditions, interfacial constraints and initial conditions.
Following the hybrid \name{FSI} approach, the flow formulation is separated into two parts utilizing Eulerian fixed-grid and \name{ALE} moving mesh descriptions, respectively.
The single-field contributions are specified below and summarized afterwards.
\medskip

\noindent\textit{The solid elastodynamics formulation.}
Solid mechanics, see \Eqref{eq:gov_eq_Solid_nonlinear-elastodynamic}--\eqref{eq:gov_eq_Solid_initial_cond_2} below, are stated in a Lagrangian formalism,
where a mapping~$\bfvarphi_t$ allows to express the motion of a material particle~$\bfX$ from reference to current configuration,
\ie~\mbox{$\Dom_0^{\sd}=\Doms(T_0)\mapsto\Doms(t)$}.
The equations are formulated in terms of the unknown displacement field~\mbox{$\bfd(\bfX,t) \eqdef \bfx_{\bfX}(\bfX,t) - \bfX$}
and its first- and second-order time derivatives, the velocity and acceleration fields
\mbox{$\bfu=\dtimedot{\bfd}(\bfX,t)$} and \mbox{$\bfa=\dsecondtimedot{\bfd}(\bfX,t)$} with \mbox{$\bfX\in\Doms(T_0)$}.

As a suitable strain-measure, \mbox{$\bfE \eqdef \tfrac{1}{2}(\bfF^T \cdot \bfF - \I)$}
denotes the {Green-Lagrange strain tensor}
and \mbox{$\bfF(\bfX,t) \eqdef \partderiv{\bfx_{\bfX}(\bfX,t)}{\bfX} = (\I + \partderiv{\bfd}{\bfX})(\bfX,t)$} the {deformation gradient tensor}.
The corresponding {second Piola--Kirchhoff stress tensor} is defined as
\mbox{$\bfS = J_{\bfX\mapsto\bfx}\bfF^{-1} \cdot \bfsigma_{\bfx} \cdot \bfF^{-T}$},
where $\bfsigma_{\bfx}$ denotes the Cauchy stresses in spatial coordinates.
In this work, we exclusively consider {Neo-Hookean} materials
with \mbox{$\bfS = 2\partderiv{\Psi}{(\bfF^T\bfF)}$}
based on a strain energy function
$\Psi(\bfF^T\bfF) \eqdef \tfrac{\mu^\sd}{2}(\trace{\bfF^T\bfF} -3) - \mu^\sd \ln (J_{\bfX\mapsto\bfx}) + \tfrac{\lambda^\sd}{2} (\ln (J_{\bfX\mapsto\bfx}))^2$,
with \mbox{$J_{\bfX\mapsto\bfx}=(\det(\bfF^T\bfF))^{1/2}$}
and Lam\'{e} parameters $\lambda^\sd$ and $\mu^\sd$, which can be expressed in terms of Young's modulus $E^\sd>0$ and Poisson's ratio $\nu^\sd\in(-1,0.5)$ as
$\lambda^\sd = E^\sd \nu^\sd ((1+\nu^\sd)(1-2\nu^\sd))^{-1} \quad\text{ and }\quad \mu^\sd= E^\sd (2(1+\nu^\sd))^{-1}$.
Further, \mbox{$\rho^{\sd} = \rho_{\bfX}^{\sd}J_{\bfX\mapsto\bfx}$} denotes the structural material density
in the initial referential configuration,
and $\nabla\cdot(\cdot)=\nabla_{\bfX}\cdot(\cdot)$ is the divergence operator with respect to material referential coordinates.

Appropriate Dirichlet and Neumann boundary data \Eqref{eq:gov_eq_Solid_bc_1}--\eqref{eq:gov_eq_Solid_bc_2}, given by $\bfgD^{\sd},\bfhN^{\sd}$,
and initial values for structural displacements and velocities \Eqref{eq:gov_eq_Solid_initial_cond_1}-\eqref{eq:gov_eq_Solid_initial_cond_2},
defined as $\bfd_0, \dot{\bfd}_0$, complement the second order initial boundary value problem.
Detailed explanations can be found, \eg, in the textbooks \cite{Wriggers2008,Zienkiewicz2000}.

\medskip
\noindent\textit{The incompressible flow formulation.}
For the description of the flow field, an Eulerian formulation for the fixed background grid~$\mcT_h^{\fd_1}$ is combined with an \name{ALE} formalism for the
embedded moving fluid patch~$\mcT_h^{\fd_2}$.
Both subdomain formulations \Eqref{eq:navier-stokes-strong-momentum}--\eqref{eq:navier-stokes-strong-mass}
are written with respect to the current domain configuration based on the more general \name{ALE} description.
Therein, \mbox{$((\bfc\cdot\Grad)\bfu)_l \eqdef  \sum_{j}{\bfc_j\cdot({\partial\bfu_l}/{\partial x_j})}$} denotes the
generalized \name{ALE} convective velocity with $\bfc\eqdef\bfu - \hat{\bfu}$, where $\hat{\bfu} \eqdef \partial_t \bfx_{\bfChi}\circ \bfPhi^{-1}$
is the velocity of the respective referential system.
The mapping therein tracks the deformation of the observed fluid domain $(\Dom^{\fd}(t),t) = \bfPhi(\Dom^{\fd}(T_0),t )$ from its initial configuration, for each subdomain independently.
While for the moving embedded patch, the grid velocity $\hat{\bfu}^{\fd_2}$ is usually non-vanishing,
for a fixed non-moving background grid it holds $\hat{\bfu}^{\fd_1}\equiv\bfzero$, which states the only difference in the embedded and background grid formulation
in the two subdomains $\Dom^{\fd_1}(t), \Domh^{\fd_2}(t)$.
An introduction to the \name{ALE} technique can be found in, \eg, \cite{DoneaGiulianiHalleux1982, Hughes1981}.

Further, $\bodyf^{\fd}$ denotes an external body force load, \mbox{$\visc{\bfu} \eqdef 1/2 \left( \Grad\vel + (\Grad\vel)^T \right)$} the symmetric strain rate tensor
and $\bfsigma(\bfu,p) = -p \I + 2\mu^{\fd} \visc{\bfu}$ the Cauchy stresses.
The dynamic viscosity is denoted with \mbox{$\mu^{\fd}=\nu^{\fd}\rho^{\fd}$}, where $\nu^{\fd}$ and $\rho^{\fd}$ are the {kinematic viscosity} and fluid density, respectively.
Appropriate Dirichlet and Neumann boundary data \Eqref{eq:gov_eq_fluid_bc_1}--\eqref{eq:gov_eq_fluid_bc_2}
are specified at all times~$t$ by functions $\bfgD^{\fd},\bfhN^{\fd}$.
The initial condition \Eqref{eq:gov_eq_fluid_initial} for the flow field is specified as $\vel_0(\bfx)$ in $\Dom^{\fd}(T_0)$.

\medskip
\noindent\textit{Coupling Conditions.}
The Cauchy stresses with respect to the current domain configuration defined on fluid and structural side of the interface are as defined above
and are denoted with \mbox{$\bfsigma(\bfu^i,p^i)$} and \mbox{$\bfsigma(\bfd^\sd\circ\bfvarphi_t^{-1})$}, respectively.
Since structural displacements are expressed with respect to its reference configuration, the temporal mapping~$\bfvarphi_t$ needs to be taken into account. 
For viscous fluids, \ie~\mbox{$\mu^{\fd}>0$}, the kinematic and dynamic interface constraints
emerge to continuity conditions \Eqref{eq:fsi_interface_condition_u_jump}--\eqref{eq:fsi_interface_condition_traction_jump} at the fluid-solid interface
and to \Eqref{eq:gov_eq_Fluid_Fluid_strong_form_5}--\eqref{eq:gov_eq_Fluid_Fluid_strong_form_6} at the fluid-fluid interface by analogy.

\begin{definition}[Strong form of the coupled hybrid \name{FSI} system]
The final hybrid coupled \name{FSI} system in its strong form reads:
Find solid displacements $\bfd:\Dom_0^{\sd}\times (T_0,T] \rightarrow \R^d$, defined in the reference configuration, satisfying
\begin{alignat}{2}
\label{eq:gov_eq_Solid_nonlinear-elastodynamic}
\rho^{\sd}\DSecondTime{\bfd} - \Div(\bfF\cdot\bfS)(\bfd) &= \rho^{\sd}\bff^{\sd} &&\qquad \foralls (\bfX,t) \in \Dom_0^{\sd}\times (T_0,T],\\
\label{eq:gov_eq_Solid_bc_1}
 \bfd &= \bfgD^{\sd}                             &&\qquad \forall (\bfX,t) \in \IntDRef^{\sd}\times (T_0,T], \\ 
\label{eq:gov_eq_Solid_bc_2}
 (\bfF\cdot\bfS)\cdot\bfN &= \bfhN^{\sd}         &&\qquad \forall (\bfX,t) \in \IntNRef^{\sd}\times (T_0,T],\\
\label{eq:gov_eq_Solid_initial_cond_1}
 \bfd(T_0)         &= \bfd_0          &&\qquad \foralls \bfX\in \Dom_0^{\sd}, \\
\label{eq:gov_eq_Solid_initial_cond_2}
 \DTime{\bfd}(T_0) &= \dot{\bfd}_0    &&\qquad \foralls \bfX\in \Dom_0^{\sd},
\end{alignat}
and flow velocity $\bfu^{i}:\Omega^{i}(t)\times t \rightarrow \R^d$ and dynamic pressure $p^i:\Omega^{i}(t)\times t \rightarrow \R$ for $i\in\{\fd_1,\fd_2\}$ such that
\begin{alignat}{2}
\label{eq:navier-stokes-strong-momentum}
\rho^{\fd}\dTime{\vel^i_{\bfChi}}\circ\bfPhi^{-1} + \rho^{\fd}(\bfc^i\cdot\Grad)\vel^i + \Grad p^i - 2\mu^{\fd}\Div\visc{\vel^i} &= \rho^{\fd}\bodyf^{\fd} \qquad && \foralls (\x,t) \in\Dom^{i}(t)\times (T_0,T], \\
\label{eq:navier-stokes-strong-mass}
\Div\vel^i &= 0 \qquad && \foralls (\x,t) \in \Dom^{i}(t)\times (T_0,T], \\
\label{eq:gov_eq_fluid_bc_1}
 \vel^i           &= \bfgD^{\fd}       \qquad &&\forall (\x,t) \in \IntD^{i}\times (T_0,T], \\ 
\label{eq:gov_eq_fluid_bc_2}
 \stress\cdot\n &= \bfhN^{\fd}       \qquad &&\forall (\x,t) \in \IntN^{i}\times (T_0,T], \\
\label{eq:gov_eq_fluid_initial}
 \vel^i(\bfx,0) &= \vel^i_0(\bfx)  \qquad &&\foralls \bfx\in \Dom^{i}(T_0),
\end{alignat}
subjected to kinematic and dynamic interface constraints at the fluid-solid interface $\Int^{\fd_2 \sd}$
\begin{alignat}{2}
\label{eq:fsi_interface_condition_u_jump}
\jump{\bfu} &= \bfu^{\fd_2} - \tfrac{\mathrm d{\bfd^\sd}}{\mathrm d{t}}\circ\bfvarphi_t^{-1} = \bfzero&&\qquad \foralls \bfx\in\Int^{\fd_2 s}(t),\\
\label{eq:fsi_interface_condition_traction_jump}
\jump{\bfsigma}\cdot\bfn^{\fd_2 s} &= (\bfsigma(\bfu^{\fd_2},p^{\fd_2}) - \bfsigma(\bfd^\sd\circ\bfvarphi_t^{-1}))\cdot\bfn^{\fd_2 s} = \bfzero  &&\qquad \foralls \bfx\in\Int^{\fd_2 s}(t),
\end{alignat}
and equivalently at the artificial fluid-fluid interface $\Int^{\fd_1 \fd_2}$
\begin{alignat}{2}
\label{eq:gov_eq_Fluid_Fluid_strong_form_5}
 \jump{\bfu} = \bfu^i -\bfu^j   &= \bfzero                                                             \qquad &&\foralls \bfx \in \Int^{\fd_1\fd_2}(t), \\
\label{eq:gov_eq_Fluid_Fluid_strong_form_6}
 \jump{\bfsigma(\bfu,p)}\cdot\bfn^{ij} = (\bfsigma(\bfu^i,p^i) - \bfsigma(\bfu^j,p^j)) \cdot\bfn^{ij} &= \bfzero   \qquad &&\foralls \bfx \in \Int^{\fd_1\fd_2}(t).
\end{alignat}
\end{definition}

\section{A Cut Finite Element Method (\name{CutFEM}) based hybrid \name{FSI} formulation}
\label{sec:cutfem_based_hybrid_FSI_formulation}

In this section, we present one potential spatial discretization technique based on the framework of Finite Element Methods (\name{FEM}s).
We would like to highlight, however, that our hybrid Eulerian-ALE discretization concept for multiphysics problems is not limited
to Finite Element based schemes, but even possible to realize with, for instance, Finite Volume or Discontinuous Galerkin based techniques.
Despite the variety of potential applicable finite-dimensional approximation frameworks,
a prerequisite for this hybrid concept is to enable finite elements, cells or volumes to get intersected by an overlapping embedded fluid patch,
and thus to enable a sharp disjoint domain decomposition of the fluid region.

The hybrid \name{FSI} approximation proposed in this work is based on a \name{CutFEM} fluid domain decomposition technique developed in our previous work \cite{SchottShahmiriKruseWall2015}
(see also \cite{Hansbo2003} for couplings in elliptic non-moving problems) and on the Nitsche-based weak coupling of the solid phase to the embedded fluid patch solution, as presented in detail in \cite{Schott2017b}.
It should be mentioned however that any fluid solid coupling scheme from fitting, \ie~classical \name{ALE} based \name{FSI}
approaches could be used as well.
Combining both interface coupling methods with suitable bulk-stabilized forms on cut background meshes for the transient incompressible Navier-Stokes equations,
as developed in \cite{SchottWall2014, MassingSchottWall2016_CMAME_Arxiv_submit}, provide a highly accurate and robust hybrid \name{FSI} approach.

Before providing our spatial discretization of the \name{FSI} problem, it is worthwhile to mention that
the number of future approximation techniques based on this hybrid concept are highly diverse and allow for various promising novel concepts.
While in the present work, just low-order continuous flow approximations based on Lagrangian finite elements are utilized,
our approach also enables to approximate embedded fluid patch and background mesh with different finite element schemes,
like for instance, different types and shapes of elements, interpolation functions, continuous and discontinuous or even enriched function spaces.
The subsequent presentation of our \name{CutFEM FSI} method is kept short with regards to numerical details, but still reviews the most important ingredients.

\subsection{Semi-discrete Nitsche-type hybrid spatial discretization}
\label{ssec:semidiscrete_Nitsche_hybrid_spatial_discretization}

Corresponding to the disjoint hybrid domain partition \mbox{$\Dom_{h} \eqdef \Dom_h^{\fd_1} \dot{\cup} \Dom_h^{\fd_2} \dot{\cup} \Dom_h^{\sd}$},
let \mbox{$\mcW_{\bfgD,h} \eqdef \mcW_{\bfgD,h}^{\fd_1} \oplus \mcW_{\bfgD,h}^{\fd_2} \oplus \mcW_{\bfgD,h}^\sd$} be the
associated space of admissible discrete \name{FSI} solutions,
consisting of the two flow approximation spaces, according to the background and the embedded mesh, and a structural approximation space.

Either of the incorporated single-mesh fluid approximations consists of a product space
\mbox{$\mcW_{\bfgD,h}^{\fd_i}\eqdef\mcV_{\bfgD,h}^{\fd_i}\times\mcQ_h^{\fd_i}$} for velocity and pressure with boundary conditions~\Eqref{eq:gov_eq_fluid_bc_1}
assumed enforced strongly.
In this work, velocity and pressure are approximated with continuous equal-order interpolations on quadrilateral or hexahedral meshes of polynomial order~$k=1$
on both families of meshes $\{\mcT_h^{\fd_1}\}_h$ and $\{\mcT_h^{\fd_2}\}_h$.

For the structural approximation in reference configuration,
let $\{\mcT_h^\sd\}_h$ be a family of boundary/interface-fitted quasi-uniform meshes,
each approximating \mbox{$\DomsRef\approx\Dom^\sd_{0,h}=\cup_{T\in\mcT^\sd_h}{T}$}.
Displacements and velocities are approximated on linearly-interpolated continuous isoparametric finite element spaces
\begin{equation}
 \mcX_{0,h} = \{ x_h \in C^0(\overline{\Dom^\sd_{0,h}}): \restr{x_h}{T}=v_{\hat{T}}\circ S_T^{-1}(t) \text{ with } v_{\hat{T}}\in \mathbb{V}^k(\hat{T})\, \foralls T \in \mcT_h^\sd \}
\end{equation}
where \mbox{$S_T(t):\hat{T}\mapsto T$} are isoparametric mappings to the element parameter space.
Taking into account the respective trace values according to the strongly imposed Dirichlet constraints~\Eqref{eq:gov_eq_Solid_bc_1},
the discrete function spaces for solid displacements~$\bfd_h$ and velocities~$\dtimedot{\bfd}_h$ result in 
\mbox{$\mcD_{\bfgD,h} \eqdef [\mcX_{0,h}]^d \cap \mcD_{\bfgD}$}, \mbox{$\mcD_h \eqdef [\mcX_{0,h}]^d \cap \mcD$}, respectively,
and the test function space to \mbox{$\mcD_{\bfzero,h}\eqdef [\mcX_{0,h}]^d \cap \mcD_{\bfzero}$}.

\begin{definition}[Semi-discrete Nitsche-type hybrid \name{FSI} formulation]
\label{def:hybrid_FSI_operators}
The Nitsche-type stabilized formulation for the hybrid \name{FSI} problem setting reads as follows:
for any time \mbox{$t\in(T_0,T]$}, find background fluid velocity and pressure \mbox{$U_h^{\fd_1}(t)=(\bfu_h^{\fd_1}(t),p_h^{\fd_1}(t))\in\mcW_{\bfgD,h}^{\fd_1}$},
embedded fluid patch velocity and pressure \mbox{$U_h^{\fd_2}(t)=(\bfu_h^{\fd_2}(t),p_h^{\fd_2}(t))\in\mcW_{\bfgD,h}^{\fd_2}$}
and solid displacement and velocity \mbox{$D_h(t)=(\bfd_h(t),\dtimedot{\bfd}_h(t))\in\mcW_{\bfgD,h}^\sd$}
such that $\foralls(V_h^{\fd_1},V_h^{\fd_2},W_h)=(\bfv_h^{\fd_1},q_h^{\fd_1},\bfv_h^{\fd_2},q_h^{\fd_2},\bfw_h)\in\mcW_{\bfzero,h}^{\fd_1} \oplus \mcW_{\bfzero,h}^{\fd_2}\oplus \mcW_{\bfzero,h}^\sd$

\begin{equation}
\label{eq:chap_5_Nitsche_CIP_GP_CUTFEM_form_multiplefluidsolid_1}
   {\mcA}_h^{\textrm{FSI}}((U_h^{\fd_1},U_h^{\fd_2},D_h),(V_h^{\fd_1},V_h^{\fd_2},W_h)) = {\mcL}_h^{\textrm{FSI}}((U_h^{\fd_1},U_h^{\fd_2}),(V_h^{\fd_1},V_h^{\fd_2},W_h)),
\end{equation}
where
\begin{align}
 {\mcA}_h^{\textrm{FSI}}((U_h^{\fd_1},U_h^{\fd_2},D_h),(V_h^{\fd_1},V_h^{\fd_2},W_h)) &\eqdef
 \mcA_h^{\fd_1}(U_h^{\fd_1},V_h^{\fd_1}) + \mcA_h^{\fd_2}(U_h^{\fd_2},V_h^{\fd_2}) + \mcA_h^{\sd}(D_h,W_h) \\
 & \quad\quad+ \mcC_h^{\fd_1\fd_2}((U_h^{\fd_1},U_h^{\fd_2}),(V_h^{\fd_1},V_h^{\fd_2}))  \\
 & \quad\quad+ \mcC_h^{\fd_2 \sd}((U_h^{\fd_2},D_h),(V_h^{\fd_2},W_h)), \\
{\mcL}_h^{\textrm{FSI}}((U_h^{\fd_1},U_h^{\fd_2}),(V_h^{\fd_1},V_h^{\fd_2},W_h)) &\eqdef
\mcL_h^{\fd_1}(U_h^{\fd_1},V_h^{\fd_1}) + \mcL_h^{\fd_2}(U_h^{\fd_2},V_h^{\fd_2}) + \mcL_h^{\sd}(W_h)
\end{align}
with single-mesh stabilized fluid operators \mbox{$\mcA_h^{i}-\mcL_h^{i}$} for $i\in\{\fd_1,\fd_2\}$ (see Definitions~\ref{def:fluid_background_CutFEM_operators}
and \ref{def:fluid_embedded_FEM_operators}),
a structural finite element approximation \mbox{$\mcA_h^{\sd}-\mcL_h^{\sd}$} (see Definition~\ref{def:structural_FEM_operators})
and Nitsche-type interface coupling terms $\mcC_h^{\fd_1\fd_2},\mcC_h^{\fd_2 \sd}$ for the fluid-fluid interface~$\Int^{\fd_1\fd_2}$
(see definition~\ref{def:Nitsche_fluid-fluid}) and the fluid-solid interface~$\Int^{\fd_2\sd}$ (see Definition~\ref{def:Nitsche_fluid-solid}), respectively.
It needs to be pointed out that due to the homogeneity of the coupling conditions,
no right-hand-side terms are present for the Nitsche couplings.
\end{definition}

\begin{definition}[Semi-discrete stabilized \name{CutFEM} based background fluid formulation]
\label{def:fluid_background_CutFEM_operators}
The \name{CutFEM} based semi-discrete approximation of the incompressible Navier-Stokes equations on a cut background mesh~$\mcT_h^{\fd_1}$ reads
\begin{equation}
\label{eq:chap_3_Nitsche_CUTFEM_form}
   \mcA_h^{\fd_1,\mathrm{GP}}(U_h^{\fd_1},V_h^{\fd_1}) - \mcL_h^{\fd_1}(U_h^{\fd_1},V_h^{\fd_1}) \quad \foralls V_h^{\fd_1}=(\bfv_h^{\fd_1},q_h^{\fd_1})\in\mcV^{\fd_1}_{\bfgD,h}\times\mcQ_h^{\fd_1}.
\end{equation}
In this work, a residual-based variational multiscale (\name{RBVM}) stabilized form
(see, \eg, \cite{HughesScovazziFranca2004}),
is utilized to account for different inherent instabilities \cite{BraackBurmanJohnEtAl2007,Roos2008}
arising for highly convective dominant flows and due to the use of equal-order interpolations for velocity and pressure.
The stabilization comprises \name{SUPG}/\name{PSPG} and \name{LSIC} terms.
For cut meshes, additional interface-zone stabilization in terms of the operator~$\mcG_h^{\mathrm{GP}}$ is required,
see elaborations in Remark~\ref{rem:stab_cut_elements}.

The \name{RBVM/GP}-stabilized form for a fluid mesh~$i$ reads in generalized form:
\begin{align}
\label{eq:chap_3_Nitsche_RBVM_GP_CUTFEM_form}
   &\mcA_h^{i,\mathrm{GP}}(U_h,V_h)
          \eqdef (\rho^{\fd}\dTime{\bfu_{\bfChi,h}}\circ \bfPhi^{-1}, \bfv_h)_{\Domh^{i}} + (\mcB_h^{i}+\mcG_h^{\mathrm{GP}})(\bfu_h-\hat{\bfu}_h;(\bfu_h,p_h), (\bfv_h,q_h)),\nonumber\\
&\quad
+ \sum_{T\in\mcT_h}{\Big( \rho^{\fd}\dTime{\bfu_{\bfChi,h}}\circ \bfPhi^{-1} + \bfr_{\mathrm{M}}(\bfu_h-\hat{\bfu}_{h};\bfu_h,p_h), \tau_\mathrm{M} ((\rho^{\fd}(\bfu_h-\hat{\bfu}_{h}) \cdot\nabla) \bfv_h + \nabla q_h)\Big)_{T\cap\Domh^{i}}}
\nonumber\\
&\quad
+ \sum_{T\in\mcT_h}{\Big(r_{\mathrm{C}}(\bfu_h),\tau_\mathrm{C} \nabla\cdot\bfv_h\Big)_{T\cap\Domh^{i}}},\\
\label{eq:chap_3_Nitsche_RBVM_GP_CUTFEM_form2}
   &\mcL_h^{i}(U_h,V_h)
	  \eqdef \mcL_h^{i}(\bfv_h,q_h)
+ \sum_{T\in\mcT_h}{\Big(\rho^{\fd}\bodyf^{\fd}, \tau_\mathrm{M} ((\rho^{\fd}(\bfu_h-\hat{\bfu}_{h}) \cdot \nabla \bfv_h) + \nabla q_h)\Big)_{T\cap\Domh^{i}}}
\end{align}
where, if unmistakable, the index $(\cdot)^{i}$ has been omitted to shorten the formulas.
Therein,
$\bfr_{\mathrm{M}}(\bfc_h;\bfu_h,p_h) = \rho^{\fd}(\bfc_h \cdot \nabla)\bfu_h + \nabla p_h - 2\mu^{\fd}\nabla\cdot\visc{\bfu_h}$
and $r_{\mathrm{C}}(\bfu_h) = \nabla\cdot\bfu_h$.
Appropriate piecewise constant stabilization scaling functions are given as
\begin{align}
\label{eq:residual_based_tau_definition}
 \tau_{\mathrm{M},T}(\bfc_h) &=(\left(\tfrac{2\rho^{\fd}}{\Delta t}\right)^2+(\rho^{\fd} \bfc_h)\cdot \bfG(\rho^{\fd} \bfc_h)+C_\mathrm{I}(\mu^{\fd})^2 \bfG:\bfG)^{-\onehalf}, \quad
 \tau_{\mathrm{C},T} = (\tau_\mathrm{M,T}\mathrm{tr}\left(\bfG\right))^{-1},
\end{align}
with the second rank metric tensor $G_{kl}(\bfx)=\sum_{i=1}^{d}(\partial \xi_i / \partial x_k\big|_{\bfx})(\partial \xi_i / \partial x_l\big|_{\bfx})$
and $C_\mathrm{I}=36.0$ for linearly interpolated finite elements.

The standard Galerkin terms $\mcB_h^{i}, \mcL_h^{i}$ and the \name{GP} stabilization operator~$\mcG_h^{\mathrm{GP}}$ are defined as follows:
\begin{flalign}
\label{eq:chap_3_Nitsche_CIP_GP_CUTFEM_form_Bh}
   \mcB_h^{i}(\bfc_h;(\bfu_h,p_h), (\bfv_h,q_h)) &=
(\rho^{\fd}(\bfc_h \cdot \nabla)\bfu_h, \bfv_h )_{\Domh^{i}}
+ (\bfepsilon(\bfu_h), 2\mu^{\fd} \bfepsilon(\bfv_h))_{\Domh^{i}} \nonumber\\
 & \quad
- (p_h, \nabla\cdot\bfv_h)_{\Domh^{i}}
+ (q_h, \nabla\cdot\bfu_h)_{\Domh^{i}} \\
\label{eq:chap_3_Nitsche_CIP_GP_CUTFEM_form_Gh}
   \mcG_h^{\mathrm{GP}} (\bfc_h;(\bfu_h,p_h), (\bfv_h,q_h)) &= (g_c + g_u + g_p )(\bfc_h;(\bfu_h,p_h), (\bfv_h,q_h)), \\
\label{eq:chap_3_Nitsche_CIP_GP_CUTFEM_form_Lh}
   \mcL_h^{i}(\bfc_h;(\bfv_h,q_h)) &=
	      (\rho^{\fd} \bodyf^\fd, \bfv_h)_{\Domh^{i}} + \langle \bfhN^\fd ,\bfv_h\rangle_{\IntN^{i}},
\end{flalign}
where in the interface-zone, facets~$F$ located next to intersected elements~$T_F^+,T_F^-$, \ie~$F\in\mcF_\Int\eqdef\{F\in\mcF_i~|~T_F^+\cap \Int\neq\emptyset \vee T_F^-\cap \Int\neq\emptyset\}$,
are stabilized by face-jump penalty terms
\begin{align}
\label{eq:chap_3_Nitsche_CIP_GP_CUTFEM_form:g_beta}
   g_c (\bfc_h;\bfu_h,\bfv_h)           &= \gamma_{c} \sum_{F\in\mcF_\Int} \sum_{1\leqslant j \leqslant k}{  \rho^{\fd}( \nu + \phi_{c,F} c_{\infty,F}^2 + \sigma h_F^2)  h_F^{2j-1}        \langle \jump{\nablan^j \bfu_h},\jump{\nablan^j \bfv_h} \rangle_F}, \\
   g_u (\bfc_h;\bfu_h,\bfv_h)               &= \gamma_{u}     \sum_{F\in\mcF_\Int} \sum_{0\leqslant j \leqslant k-1}{ \phi_{u,F}    \rho^{\fd}                 h_F^{2j+1}        \langle \jump{\nabla \cdot \nablan^j\bfu_h},\jump{\nabla\cdot \nablan^j \bfv_h} \rangle_F}, \\
   g_p (\bfc_h;p_h,q_h)                     &= \gamma_{p}     \sum_{F\in\mcF_\Int} \sum_{1\leqslant j \leqslant k}{   \phi_{p,F}    (\rho^{\fd})^{-1}            h_F^{2j-1}        \langle \jump{\nablan^j p_h},\jump{\nablan^j q_h} \rangle_F}.
\end{align}
Therein, it is set \mbox{$c_{\infty,F} \eqdef \|\bfc_h \|_{0,\infty,F}$}
and
\begin{equation}
 \phi_T(\bfc_h)= \nu + c_u (\|\bfc_h\|_{0,\infty,T} h_T) + c_{\sigma} (\sigma h_T^2), \quad\phi_{c,T}=\phi_{p,T}=h_T^2 \phi_T^{-1}, \quad\phi_{u,T}=\phi_T \text{ and }\sigma=1/(\theta\Delta t).
 \label{eq:cip_scaling}
\end{equation}
Further, $\jump{\cdot}$ denotes the jump of quantities across interior facets~$F$
and the subscript~$(\cdot)_F$ in stabilization scalings indicates to take the mean over quantities from both adjacent elements~$T$.
For details, the reader is referred to, \eg, \cite{MassingSchottWall2016_CMAME_Arxiv_submit,Schott2017b, Winter2018}.
\end{definition}

\begin{remark}
Since the background mesh~$\mcT_h^{\fd_1}$ is assumed to remain fixed over time, as exclusively considered throughout this work,
the time derivative occurring in \Eqref{eq:chap_3_Nitsche_RBVM_GP_CUTFEM_form}
simplifies to $(\dTime{\bfu_{\bfChi,h}}\circ \bfPhi^{-1}, \bfv_h)= (\dTime{\bfu_h}, \bfv_h)$, \ie~the grid velocity \mbox{$\hat{\bfu}_h=\bfzero$} vanishes
such that \mbox{$\bfc_h =\bfu_h$} in a pure Eulerian consideration.
Note, however, that it might be also an option to allow even the background mesh to move, as considered for instance in an approach shown in \cite{Schott2017b}.
It should be mentioned that all further numerical or algorithmic steps for this have been already addressed in \cite{Schott2017b}, however, are not considered
further for simplicity in this paper.
\end{remark}

\begin{remark}[Stabilization of cut clements]
\label{rem:stab_cut_elements}
Due to the intersection of elements $T\in\mcT_h^{\fd_1}$ by the fluid-fluid interface $\Int^{\fd_1 \fd_2}$,
additional stabilization measures are required. Ghost-penalty stabilizations~$\mcG_h^{\mathrm{GP}}$ \Eqref{eq:chap_3_Nitsche_CIP_GP_CUTFEM_form_Gh},
as developed in \cite{MassingSchottWall2016_CMAME_Arxiv_submit, SchottWall2014},
penalize jumps of normal derivatives of order~$j$ across interior facets~$F$ in the vicinity of the interface and thus ensure
well system conditioning, stability and optimality of the approximation independent of the mesh intersection.
For further details on this technique, the reader is referred to, \eg, \cite{BeckerBurmanHansbo2009, Burman2010, BurmanHansbo2012, Burman2014b}.
\end{remark}

\begin{definition}[Semi-discrete stabilized \name{FEM} based embedded fluid formulation]
\label{def:fluid_embedded_FEM_operators}
The semi-discrete fluid operators for the fitted-mesh approximation of the embedded fluid patch~$\mcT_h^{\fd_2}$ is comprised in
\begin{equation}
\label{eq:chap_3_Nitsche_RBVM_FEM_form}
   \mcA_h^{\fd_2}(U_h,V_h) - \mcL_h^{\fd_2}(U_h,V_h),
\end{equation}
where respective operators $\mcA_h^{\fd_2}, \mcL_h^{\fd_2}$ are as defined in 
\Eqref{eq:chap_3_Nitsche_RBVM_GP_CUTFEM_form} and \Eqref{eq:chap_3_Nitsche_RBVM_GP_CUTFEM_form2}
with $\mcG_h^{\mathrm{GP}}\equiv 0$, since embedded meshes are uncut and do not require interface zone stabilization.
\end{definition}

\begin{remark}[Grid velocity for moving fluid patch]
\label{def:fluid_embedded_grid_displacement}
Similar to the Lagrangian formalism for structures,
the motion of the fluid grid $\Dom^{\fd_2}(T_0)\mapsto\Dom^{\fd_2}(t)$ is tracked in terms
of the mapping $\bfPhi(\bfChi,t)$.
Introducing a fluid patch bulk displacement field~\mbox{$\bfd^{\fd}(\bfChi,t) \eqdef \bfx_{\bfChi}(\bfChi,t) - \bfChi$} describing the computational grid motion,
the \name{ALE} time derivative $\partial_t \bfx_{\bfChi}$ can be simply expressed in terms of the fluid domain displacements
resulting in \mbox{$\hat{\bfu}_{\bfChi} = \partial_t \bfx_{\bfChi} = \partial_t (\bfd^{\fd}(\bfChi,t) +  \bfChi) = \partial_t \bfd^{\fd}(\bfChi,t)$}
for the grid velocity. For their approximation, a classical linear isoparametric finite element concept is utilized.

Since the fluid patch and the solid mesh need to match at~$\Int^{\fd_2\sd}$ at all times,
the fluid domain displacement field~$\bfd^{\fd_2}$ is constraint being equal to the solid displacement field~$\bfd^{\sd}$,~\ie
\begin{alignat}{2}
\label{eq:ALE-constraint_FSI}
\bfd^{\fd_2}\circ \bfPhi^{-1}(\bfx,t) & = \bfd^{\sd} \circ \bfvarphi^{-1}(\bfx,t) \quad\foralls (\bfx,t) \in (\Gamma^{\fd\sd}(t),t).
\end{alignat}
For the motion of the non-constraint part of the fluid mesh,
a pseudo-structure mesh update algorithm as also used in \cite{Kloppel2011} has been used.
It is important to appreciate that mesh distortions are much smaller than in standard \name{ALE} cases as the outer boundary
of this mesh is free to move.
\end{remark}

\begin{definition}[Structural finite element approximation]
\label{def:structural_FEM_operators}
The space semi-discrete fitted-mesh finite element approximation of the structural elastodynamics form~\Eqref{eq:gov_eq_Solid_nonlinear-elastodynamic}--\eqref{eq:gov_eq_Solid_bc_2}
with solid displacements \mbox{$\bfd_h(t)\in\mcD_{\bfgD,h}$}
and velocities \mbox{$\dtimedot{\bfd}_h(t)=\mathrm{d}\bfd_h(t)/\mathrm{d} t\in\mcD_h$} reads,
find $D_h=(\bfd_h,\dtimedot{\bfd}_h)$ such that
\begin{align}
 \label{eq:gov_eq_Solid_weak_form_discrete}
\mcA_h^{\sd}(D_h,W_h) - \mcL_h^{\sd}(W_h) \quad  \foralls W_h=\bfw_h\in\mcD_{\bfzero,h},
\end{align}
where
\begin{align}
 \mcA_h^{\sd}(D_h,W_h) &\eqdef \scalL{\rho^\sd\DTime{\dtimedot{\bfd}_h}}{\bfw_h}_{\Dom_{0,h}^\sd} + ((\bfF\cdot\bfS)(\bfd_h),\Grad\bfw_h)_{\Doms_{0,h}}, \\
 \mcL_h^{\sd}(W_h)     &\eqdef (\rho^\sd\bff^\sd,\bfw_h)_{\Doms_{0,h}} + \scalbound{\bfhN^\sd}{\bfw_h}{\Int_{N,0,h}^\sd}
\end{align}
are evaluated on the discrete referential counterparts \mbox{$\Dom_{0,h}^\sd,\Int_{N,0,h}^\sd$}, respectively.
\end{definition}

\begin{definition}[Nitsche-type fluid domain decomposition coupling operators]
\label{def:Nitsche_fluid-fluid}
For the weak coupling of the two subdomain approximations on $\mcT_h^{\fd_1}$ and $\mcT_h^{\fd_2}$,
specified by the coupling constraints \Eqref{eq:gov_eq_Fluid_Fluid_strong_form_5}--\eqref{eq:gov_eq_Fluid_Fluid_strong_form_6},
the stabilized Nitsche-type formulation from \cite{SchottShahmiriKruseWall2015}
is applied, which is comprised in the following coupling operator
\begin{flalign}
   \mcC_h^{\fd_1\fd_2}((U_h^1,U_h^2), (V_h^1,V_h^2))
         &=
         - \langle {2\mu^{\fd}\bfepsilon(\bfu_h^{\fd_2})}\bfn^{\fd_1 \fd_2}, \jump{\bfv_h} \rangle_{\Int^{\fd_1\fd_2}}
	 + \langle {p_h^{\fd_2}}, \jump{\bfv_h}\cdot\bfn^{\fd_1 \fd_2} \rangle_{\Int^{\fd_1\fd_2}}
\label{eq:chap_4_Nitsche_CIP_GP_CUTFEM_form_fluidfluid_form_Chij_1}  \\
	 &\quad
         + \langle \jump{\bfu_h}, {2\mu^{\fd}\bfepsilon(\bfv_h^{\fd_2})}\bfn^{\fd_1 \fd_2} \rangle_{\Int^{\fd_1\fd_2}}
         -   \langle \jump{\bfu_h}\cdot\bfn^{\fd_1 \fd_2}, {q_h^{\fd_2}} \rangle_{\Int^{\fd_1\fd_2}}  
\label{eq:chap_4_Nitsche_CIP_GP_CUTFEM_form_fluidfluid_form_Chij_2} \\
         & \quad
         + \langle \gamma({\varphi^2}/2)\jump{\bfu_h},\jump{\bfv_h} \rangle_{\Int^{\fd_1\fd_2}}
\label{eq:chap_4_Nitsche_CIP_GP_CUTFEM_form_fluidfluid_form_Chij_3} \\
	 & \quad
         + \langle \gamma({\rho^{\fd}\phi^2/h}/2)\jump{\bfu_h}\cdot\bfn^{\fd_1 \fd_2}, \jump{\bfv_h}\cdot\bfn^{\fd_1 \fd_2} \rangle_{\Int^{\fd_1\fd_2}}
\label{eq:chap_4_Nitsche_CIP_GP_CUTFEM_form_fluidfluid_form_Chij_4}  \\
         & \quad
	 + \langle (\meanavg{\rho^{\fd}\bfu_h}\cdot\bfn^{\fd_1 \fd_2})\jump{\bfu_h},\meanavg{\bfv_h} \rangle_{\Int^{\fd_1\fd_2}}
\label{eq:chap_4_Nitsche_CIP_GP_CUTFEM_form_fluidfluid_form_Chij_5} \\
         & \quad
	 +  \langle \tfrac{1}{2} |(\meanavg{\rho^{\fd}\bfu_h}\cdot\bfn^{\fd_1 \fd_2})|\jump{\bfu_h},\jump{\bfv_h} \rangle_{\Int^{\fd_1\fd_2}},
\label{eq:chap_4_Nitsche_CIP_GP_CUTFEM_form_fluidfluid_form_Chij_6}
\end{flalign}
with the jump operator $\jump{x}=x^{\fd_1}-x^{\fd_2}$, the mean average operator $\meanavg{x} \eqdef \tfrac{1}{2}(x^{\fd_1}+x^{\fd_2})$
and $\bfn^{\fd_1 \fd_2}=\bfn^{\fd_1}=-\bfn^{\fd_2}$.
In the present work, the Nitsche penalty parameter is chosen as~$\gamma=50$, the fluid scaling $\phi$ is as defined in \Eqref{eq:cip_scaling}
and the scaling $\varphi$ is defined as \mbox{$\varphi^k\eqdef\mu^{\fd} (f^k)^2 \propto 1/h_k$}, where $f^k$ is obtained from a fitted mesh trace inequality.
Details can be found in the related publications \cite{SchottShahmiriKruseWall2015,Schott2017b, MassingSchottWall2016_CMAME_Arxiv_submit}.

\end{definition}

\begin{definition}[Nitsche-type fluid-structure coupling operators]
\label{def:Nitsche_fluid-solid}
The Nitsche-based fluid-solid coupling operator from \cite{Schott2017b} reads as
\begin{flalign}
 \mcC_h^{\fd_2\sd}((U_h^2,D_h),(V_h^2,W_h))
         &=
         - \langle 2\mu^{\fd}\bfepsilon(\bfu_h^{\fd_2})\bfn^{\fd_2\sd}, \jump{\bfv_h} \rangle_{\Int^{\fd_2\sd}}
	 + \langle p_h^{\fd_2}, \jump{\bfv_h}\cdot\bfn^{\fd_2\sd} \rangle_{\Int^{\fd_2\sd}}
\label{eq:chap_5_Nitsche_CIP_GP_CUTFEM_form_fluidsolid_form_Chfs_1}  \\
	 &\quad
         + \langle \jump{\bfu_h}, 2\mu^{\fd}\bfepsilon(\bfv_h^{\fd_2})\bfn^{\fd_2\sd} \rangle_{\Int^{\fd_2\sd}}
         -   \langle \jump{\bfu_h}\cdot\bfn^{\fd_2\sd}, q_h^{\fd_2} \rangle_{\Int^{\fd_2\sd}}  
\label{eq:chap_5_Nitsche_CIP_GP_CUTFEM_form_fluidsolid_form_Chfs_2} \\
         & \quad
         + \langle \gamma(\mu^{\fd}/h)\jump{\bfu_h},\jump{\bfv_h} \rangle_{\Int^{\fd_2\sd}}\\
\label{eq:chap_5_Nitsche_CIP_GP_CUTFEM_form_fluidsolid_form_Chfs_3}
	 & \quad
         + \langle \gamma(\rho^\fd\phi^2/h)\jump{\bfu_h}\cdot\bfn^{\fd_2\sd}, \jump{\bfv_h}\cdot\bfn^{\fd_2\sd} \rangle_{\Int^{\fd_2\sd}}
\end{flalign}
with \mbox{$\jump{\bfu_h} = \bfu_h^{\fd_2} - \bfu_h^\sd =  \bfu_h^{\fd_2} - \dtimedot{\bfd}_h^{\sd} \circ \bfvarphi_t^{-1}$},
\mbox{$\jump{\bfv_h} = \bfv_h^{\fd_2} - \bfw_h\circ \bfvarphi_t^{-1}$}
and $\bfn^{\fd_2 \sd}=\bfn^{\fd_2}=-\bfn^{\sd}$.
Involved fluid scalings and parameters are as stated in Definition~\ref{def:Nitsche_fluid-fluid}.
\end{definition}

\subsection{Temporal discretization}

In the present work, for the temporal discretization of the hybrid Eulerian-\name{ALE} \name{FSI} system
a Generalized\hbox{-}$\alpha$~method is applied to the structural elastodynamics part \mbox{$\mcA_h^{\sd}-\mcL_h^{\sd}$}
and a one-step-$\theta$ scheme for the stabilized fluid formulations \mbox{$\mcA_h^{\fd_1,\mathrm{GP}}-\mcL_h^{\fd_1}$}
and \mbox{$\mcA_h^{\fd_2}-\mcL_h^{\fd_2}$}. The time domain~\mbox{$(T_0, T]$} is approximated equidistantly
resulting in time step intervals \mbox{$J^n=(t^{n-1},t^n]$} of size $\Delta t$ and time levels \mbox{$t^n = T_0 + n\Delta t$} and \mbox{$t^N = T$}.
Note that, if unmistakable, occasionally the superscript indicating solid and fluid variables is omitted to shorten the presentation.

Following detailed derivations in \cite{Schott2017b} for fitted and unfitted Nitsche-type \name{FSI} approaches,
the nonlinear finite-dimensional residual of the \name{FSI} system \Eqref{eq:chap_5_Nitsche_CIP_GP_CUTFEM_form_multiplefluidsolid_1} emerges to:
find discrete vectors \mbox{$((\bfU, \bfP)^{\fd_1}, (\bfU, \bfP)^{\fd_2}, \bfD^\sd)^n$} such that
\begin{align}
\label{eq:chap_5_fsi_nonlinear_residual_multifluid}
\left[
 \begin{array}{c}
  \bfR_{(U,P)^{{\fd_1}}} \\
  \bfR_{(U,P)^{{\fd_2}}}\\
  \bfR_D
 \end{array}
\right]^n
&=
\left[
 \begin{array}{c}
   \sigma\bfR^{\fd_1}((\bfU,\bfP)^{\fd_1}) + \bfC^{\fd_1 \fd_2}((\bfU,\bfP)^{\fd_1},(\bfU,\bfP)^{\fd_2})\\
   \sigma\bfR^{\fd_2}((\bfU,\bfP)^{\fd_2}) + \bfC^{\fd_2 \fd_1}((\bfU,\bfP)^{\fd_1},(\bfU,\bfP)^{\fd_2}) + \bfC^{{\fd_2} \sd}((\bfU,\bfP)^{\fd_2},\bfD^\sd)\\
   \frac{1}{1-\alpha_f}\bfR^\sd(\bfD^\sd) + \bfC^{\sd 2}((\bfU,\bfP)^{\fd_2},\bfD^\sd) - \tfrac{\alpha_f}{1-\alpha_f} \bfF^{\sd,n-1}_{\Int^{\fd_2\sd}(t^{n-1})}
 \end{array}
\right]^n =\bfzero
\end{align}
where \mbox{$\bfC^{\fd_2 \sd},\bfC^{\sd \fd_2}$} denote splits of the fluid-structure Nitsche couplings
between fluid subdomain~$\Dom^{\fd_2}$ and the solid subdomain $\Dom^{\sd}$.
These can be identified by splitting contributions from \Eqref{eq:chap_5_Nitsche_CIP_GP_CUTFEM_form_fluidsolid_form_Chfs_1}--\eqref{eq:chap_5_Nitsche_CIP_GP_CUTFEM_form_fluidsolid_form_Chfs_3}
into fluid and structural residuals, \ie~with respect to $\bfv_h^{\fd_2}$ and $\bfw_h$.
Similar splits for the Nitsche coupling terms
\Eqref{eq:chap_4_Nitsche_CIP_GP_CUTFEM_form_fluidfluid_form_Chij_1}--\eqref{eq:chap_4_Nitsche_CIP_GP_CUTFEM_form_fluidfluid_form_Chij_6}
between the fluid phase $\Dom^{\fd_1}$ and $\Dom^{\fd_2}$
are denoted with $\bfC^{\fd_1\fd_2}$ and $\bfC^{\fd_2\fd_1}$.
Following elaborations in \cite{Schott2017b}, the previous time level structural interface force contribution
occurring in \Eqref{eq:chap_5_fsi_nonlinear_residual_multifluid},
which results from applying a generalized trapezoidal rule to the interface force approximation,
can be recovered as
\begin{align}
\label{eq:chap_5_fsi_structural_force_previous_time_step}
 \bfF^{\sd,n-1}_{\Int^{\fd_2\sd}(t^{n-1})} = -\bfC^{\sd\fd_2,n-1}((\bfU^{n-1},\bfP^{n-1}),\bfD^{n-1}).
\end{align}
Furthermore, $\bfR^{i}$, \mbox{$i=\fd_1,\fd_2$}, denote the two fluid subdomain residuals and
$\bfR^\sd$ the structural residual
according to \Eqref{eq:chap_3_Nitsche_CUTFEM_form}, \Eqref{eq:chap_3_Nitsche_RBVM_FEM_form} and \Eqref{eq:gov_eq_Solid_weak_form_discrete},
\begin{align}
\label{eq:chap_5_fsi_fluid_residual}
 \bfR^{i}(\bfU^n,\bfP^n) & = \bfM^{i,n} (\bfU^n,\bfP^n) + \sigma^{-1} \bfF^{i,n}(\bfU^n,\bfP^n)
- \bfH^{i,n-1}(\tilde{\bfU}^{n-1},\tilde{\bfA}^{n-1}), \\
\label{eq:chap_5_fsi_structural_residual}
 \bfR^\sd(\bfD^n)        & 
 = \bfM^{\sd,n} \frac{1-\alpha_m}{\beta \Delta t^2}  \bfD^{n} + (1-\alpha_f)(\bfF^{\sd,n}_{\textrm{int}}(\bfD^{n}) - \bfF^{\sd,n}_{\textrm{ext}})
- \bfH^{\sd,n-1}(\bfD^{n-1},\bfU^{n-1},\bfA^{n-1}).
\end{align}
The fluid residual $\bfR^{i}$ \Eqref{eq:chap_5_fsi_fluid_residual} for subdomain $\Domh^{i}$
contains the matrix~$\bfM^{i,n}$ resulting from the time derivative terms occurring in \Eqref{eq:chap_3_Nitsche_RBVM_GP_CUTFEM_form},
$\bfF^{i,n}$ comprises all operators associated to time level~$t^n$, \ie~standard Galerkin terms,
stabilization operators, external loads from \mbox{$\mcB_h^{i},\mcG_h^{\textrm{GP}},-\mcL_h^{i}$}.
Utilizing a \name{OST}-scheme, all terms belonging to the previous time levels are comprised in~$\bfH^{i,n-1}$
with $\sigma = (\theta \Delta t)^{-1}$.
Once $\bfU^n$ is computed, $\bfA^n$ can be updated, see \cite{Schott2017b}.

The structural residual $\bfR^{\sd}$ \Eqref{eq:chap_5_fsi_structural_residual} results from applying the \emph{Generalized-$\alpha$} (G\hbox{-}$\alpha$) method \cite{Chung1993}
to~\Eqref{eq:gov_eq_Solid_weak_form_discrete}.
The parameters are chosen as
$\beta=1/4(1-\alpha_m+\alpha_f)^2$ with
$\alpha_f=\rho_\infty/(\rho_\infty+1)$ and $\alpha_m=(2\rho_\infty-1)/(\rho_\infty+1)$
depending on the user-specified spectral radius $\rho_\infty\in[0,1]$ controlling the numerical high frequency dissipation.
Furthermore, $\bfM^{\sd,n}$ is the global mass matrix, $\bfF_{\textrm{int}}^{\sd}$ the vector of non-linear internal forces resulting from~$\mcA_h^\sd$ and
$\bfF_{\textrm{ext}}^{\sd}$ are external forces resulting from $\mcL_h^\sd$.
Previous time-level contributions are comprised in~$\bfH^{\sd,n-1}$.
Once $\bfD^{n}$ is computed, velocity and acceleration approximations $\bfU^{n}$ and $\bfA^{n}$ can be recovered
according to \cite{Schott2017b}.

The solution \mbox{$((\bfU,\bfP)^{\fd_1},(\bfU,\bfP)^{\fd_2},\bfD^{\sd})^{n}$} of \Eqref{eq:chap_5_fsi_nonlinear_residual_multifluid} is approximated iteratively
for \mbox{$m\geqslant 1$} by solving the following Newton-Raphson-like scheme for increments
\mbox{$\Delta((\bfU,\bfP)^{\fd_1},(\bfU,\bfP)^{\fd_2},\bfD^{\sd})_m^{n}$} satisfying
\begin{equation}
\label{eq:chap_5_fsi_linearized_system_multifluid}
\ADLdrawingmode{3}
\ADLactivate
\left[
\begin{array}{c | c | c}
\begin{array}{c;{2pt/2pt} c}
\bfL_{U^{\fd_1}U^{\fd_1}} & \bfL_{U^{\fd_1}P^{\fd_1}}\\ \hdashline[2pt/2pt]
\bfL_{P^{\fd_1}U^{\fd_1}} & \bfL_{P^{\fd_1}P^{\fd_1}}
\end{array}
&
\begin{array}{c;{2pt/2pt} c}
\bfL_{U^{\fd_1}U^{\fd_2}} & \bfL_{U^{\fd_1}P^{\fd_2}}\\ \hdashline[2pt/2pt]
\bfL_{P^{\fd_1}U^{\fd_2}} & \bfL_{P^{\fd_1}P^{\fd_2}}
\end{array}
&
\begin{array}{c}
\bfzero \\ \hdashline[2pt/2pt]
\bfzero
\end{array}
\\
\cline{1-3}
\begin{array}{c;{2pt/2pt} c}
\bfL_{U^{\fd_2}U^{\fd_1}} & \bfL_{U^{\fd_2}P^{\fd_1}}\\ \hdashline[2pt/2pt]
\bfL_{P^{\fd_2}U^{\fd_1}} & \bfL_{P^{\fd_2}P^{\fd_1}}
\end{array}
&
\begin{array}{c;{2pt/2pt} c}
\bfL_{U^{\fd_2}U^{\fd_2}} & \bfL_{U^{\fd_2}P^{\fd_2}}\\ \hdashline[2pt/2pt]
\bfL_{P^{\fd_2}U^{\fd_2}} & \bfL_{P^{\fd_2}P^{\fd_2}}
\end{array}
&
\begin{array}{c}
\bfL_{U^{\fd_2}D^{\sd}} \\ \hdashline[2pt/2pt]
\bfL_{P^{\fd_2}D^{\sd}}
\end{array}
\\
\cline{1-3}
\begin{array}{c ;{2pt/2pt}c} \bfzero & \bfzero \end{array} & \begin{array}{c ;{2pt/2pt}c} \bfL_{D^{\sd}U^{\fd_2}} & \bfL_{D^{\sd}P^{\fd_2}} \end{array} & \bfL_{D^{\sd}D^{\sd}} \\
\end{array}
\right]^{n}_{m}
\left[
\begin{array}{c}
\ADLdrawingmode{3}
\ADLactivate
  \Delta\bfU^{\fd_1} \\ \hdashline[2pt/2pt]
  \Delta\bfP^{\fd_1} \\ \cline{1-1}
  \Delta\bfU^{\fd_2} \\ \hdashline[2pt/2pt]
  \Delta\bfP^{\fd_2}  \\ \cline{1-1}
  \Delta\bfD^\sd
\end{array}
\right]^{n}_{m}
=
-
\left[
\begin{array}{c}
  \bfR_{U^{\fd_1}} \\ \hdashline[2pt/2pt]
  \bfR_{P^{\fd_1}} \\ \cline{1-1}
  \bfR_{U^{\fd_2}} \\ \hdashline[2pt/2pt]
  \bfR_{P^{\fd_2}} \\ \cline{1-1}
  \bfR_{D^\sd}
\end{array}
\right]^{n}_{m}
,
\end{equation}
followed by an incremental update step for the next iteration
\begin{align}
\label{eq:chap_5_fsi_linearized_system_update_multifluid}
\left[
\begin{array}{c}
  \bfU^{\fd_1} \\ \hdashline[2pt/2pt]
  \bfP^{\fd_1} \\ \cline{1-1}
  \bfU^{\fd_2} \\ \hdashline[2pt/2pt]
  \bfP^{\fd_2}  \\ \cline{1-1}
  \bfD^\sd
\end{array}
\right]^{n}_{m+1}
=
\left[
\begin{array}{c}
  \bfU^{\fd_1} \\ \hdashline[2pt/2pt]
  \bfP^{\fd_1} \\ \cline{1-1}
  \bfU^{\fd_2} \\ \hdashline[2pt/2pt]
  \bfP^{\fd_2}  \\ \cline{1-1}
  \bfD^\sd
\end{array}
\right]^{n}_{m}
+
\left[
\begin{array}{c}
  \Delta\bfU^{\fd_1} \\ \hdashline[2pt/2pt]
  \Delta\bfP^{\fd_1} \\ \cline{1-1}
  \Delta\bfU^{\fd_2} \\ \hdashline[2pt/2pt]
  \Delta\bfP^{\fd_2}  \\ \cline{1-1}
  \Delta\bfD^\sd
\end{array}
\right]^{n}_{m}.
\end{align}
Therein, \mbox{$\bfL_{xy} = \frac{\partial{\bfR_x}}{\partial y}$} denote (pseudo)-directional derivatives of residuals~$\bfR_x$ from \Eqref{eq:chap_5_fsi_linearized_system_multifluid}
with respect to the finite dimensional solution approximation~$y$, where \mbox{$x,y\in\{\bfU^{\fd_1},\bfU^{\fd_2},\bfP^{\fd_1},\bfP^{\fd_2},\bfD^{\sd}\}$}

\begin{remark}[Treatment of \name{ALE} displacement update]
It needs to be pointed out that the \name{ALE} displacements change within each Newton-type iteration~\Eqref{eq:chap_5_fsi_linearized_system_update_multifluid},
as they are constrained by the solid displacement field~$\bfD^{\sd}$ via \Eqref{eq:ALE-constraint_FSI}.
Since the latter changes non-linearly, the fluid patch displacements~$\bfD^{\fd_2}$ need to be updated accordingly.

Different strategies are available for the mesh update of the embedded fluid patch~$\mcT_h^{\fd_2}$ in literature.
The update step can be either incorporated directly into
the system of residuals \Eqref{eq:chap_5_fsi_nonlinear_residual_multifluid} as an additional block (see \eg \cite{Mayr2015})
or solved for within an extra step.
For the monolithic variant, an according additional linearization is added to \Eqref{eq:chap_5_fsi_linearized_system_multifluid}
and it is solved for the grid displacements~$\bfD^{\fd_2}$ as an additional field variable.
As an alternative technique, the mesh update in terms of the grid displacements can be performed as an extra step after solving for the Newton increment \Eqref{eq:chap_5_fsi_linearized_system_update_multifluid}.
Note that for node-matching discretizations $\mcT_h^{\fd_2}$ and $\mcT_h^{\sd}$ at $\Int^{\fd_2\sd}$,
displacement degrees of freedom associated with interface nodes are shared by $\bfD^{\sd}$ and $\bfD^{\fd_2}$.
Thus, just a subset of non-interface-aligned \name{ALE} nodes $\tilde{\bfD}^{\fd_2}\subsetneq \bfD^{\fd_2}$ needs to updated. 

In the present work, for the \name{ALE} mesh motion, a pseudo-structural mesh update technique is used.
It is solved for the grid displacements in an analogue fashion as done for the structural body with identical material properties.
For further discussions on promising alternative techniques for relaxing the \name{ALE} mesh/boundary see also Remark~\ref{rem:strategies_relaxing_ALE_boundary}
\end{remark}

\begin{remark}[Changing fluid function spaces]
\label{rem:changing_function_space}
Since both fluid meshes are coupled in a geometrically unfitted fashion, the number of active degrees of freedom associated
to the approximation on the background grid~$\mcT_h^{\fd_1}$ depends on the location of the interface~$\Int^{\fd_1 \fd_2}$.
As a result, the function space~$\mcW_h^{\fd_1}$ can change between two discrete time levels~$t^{n-1}$ and~$t^n$.
Moreover, since a monolithic Newton-type solution strategy is chosen, this issue might arise even between two subsequent iterations~$m$ and~$m+1$
of \Eqref{eq:chap_5_fsi_linearized_system_multifluid}--\eqref{eq:chap_5_fsi_linearized_system_update_multifluid}.
To overcome this issue, the strategy provided in \cite{Schott2017b}[Algorithm 1] has been applied.
It allows to adapt discrete previous time level solution fields $[\bfU^{\fd_1},\bfA^{\fd_1}]^{n-1}$
or iteration steps $[\bfU^{{\fd_1}},\bfP^{{\fd_1}}]_m$ to the current function space approximation.
The projected previous time step solution fields occurring in \Eqref{eq:chap_5_fsi_fluid_residual} are denoted with $[\tilde{\bfU}^{\fd_1},\tilde{\bfA}^{\fd_1}]^{n-1}$.
For further algorithmic details, the interested reader is referred to the original publication \cite{Schott2017b}.
\end{remark}

\subsection{Algorithmic solution procedure}

In the sequel, the major steps of the solution process for the \name{FSI} residuals \Eqref{eq:chap_5_fsi_nonlinear_residual_multifluid} are reviewed.
Our solution scheme is strongly based on techniques presented in \cite{Schott2017b} and, thus, is just briefly sketched in following.
The algorithmic procedure is summarized in \Algoref{algo:nonlinear_solution_algo_monolithic_FSI}.

\begin{algorithm}[t]
 \caption{Monolithic solution algorithm}
 \begin{algorithmic}[1]
  \STATE \name{Input}: fluid and solid initial conditions \mbox{$\bfd_0^\sd,\dtimedot{\bfd}_0^\sd,\bfu_0^\fd$} at $T_0$.
  \FOR[Time loop]{time steps~\mbox{$1\leqslant n \leqslant N$}, \mbox{$t^n=T_0+n\Delta t$}}
     \STATE Reset cycle counter to \mbox{$c=1$}.
     \WHILE[cycle over fluid function space changes]{(\NOT converged) }
      \IF[First cycle at new time level]{($c=1$)}
	\STATE \COMMENT{Predict solution fields}
	\STATE Predict structural solution (based on constant velocity) \mbox{$(\bfD,\bfU,\bfA)_{c=1}^n$}.
	\STATE Predict non-constraint fluid grid displacements \mbox{$(\tilde{\bfD}^{\fd_2})_{c=1}^n$} for $\mcT_h^{\fd_2}$.
	\STATE Update intersection dependent mesh quantities for $\mcT_h^{\fd_1}$ based on $(\Int^{\fd_1\fd_2})_{c=1}^{n}$.
	\STATE Project velocity solution of $\mcT_h^{\fd_1}$ onto $(\mcW_h^{\fd_1})_{c=1}^n$ (\cite{Schott2017b}[Algo~1]) and predict solutions \mbox{$((\bfU,\bfP)^{i})_{c=1}^n$}.
      \ELSE[The fluid function space has changed within the last cycle \mbox{$c-1$}]
	\STATE Transcribe fluid solution vectors of~$\mcT_h^{\fd_1}$ (\mbox{$(\mcW_h^{\fd_1})^{n-1} \rightarrow (\mcW_h^{\fd_1})_{c}^n$}, see \cite{Schott2017b}[Algo~1]).
	\STATE Use recent solution approximation from interrupted pass~\mbox{$c-1$} as initial guess for the following Newton-Raphson procedure. Note that
	       \mbox{$(\bfU,\bfP)_{c}^n \in (\mcW_h^\fd)_{c}^n$}.
      \ENDIF
      \STATE \COMMENT{Newton-Raphson iterations until convergence or function space changes detected}
      \STATE Perform Newton-Raphson scheme \Eqref{eq:chap_5_fsi_linearized_system_multifluid}--\eqref{eq:chap_5_fsi_linearized_system_update_multifluid} for \name{FSI} system~\Eqref{eq:chap_5_fsi_nonlinear_residual_multifluid}.
      \IF{(Newton-Raphson converged)}
	\STATE \algorithmicbreakwhile
      \ENDIF
      \STATE $c\leftarrow c+1$ \COMMENT{Fluid function space has changed}
     \ENDWHILE
     \STATE Update solution fields for solid and velocity $n\mapsto n+1$ based on G\hbox{-}$\alpha$ and \name{OST} scheme.
     \STATE Store final approximations: $(\bfD^{\sd},\bfU^{\sd},\bfA^{\sd})^n$ for the solid
            and $(\bfU^i,\bfP^i,\bfA^i)^n$ for the fluid~$(i)$ with \name{ALE} displacements~$(\bfD^{\fd_2})^n$.
  \ENDFOR
  \STATE \name{Output}: fluid and solid approximations \mbox{$\left\{((\bfU,\bfP)^{\fd_1},(\bfU,\bfP)^{\fd_2},\bfD^{\sd})^n\right\}_{1\leqslant n \leqslant N}$}.
 \end{algorithmic}
 \label{algo:nonlinear_solution_algo_monolithic_FSI}
\end{algorithm}

The core part of the full-implicit solution scheme consists of a Newton-Raphson like scheme \Eqref{eq:chap_5_fsi_linearized_system_multifluid}--\eqref{eq:chap_5_fsi_linearized_system_update_multifluid} approximating the solution
of the coupled \name{FSI} system \Eqref{eq:chap_5_fsi_nonlinear_residual_multifluid}.
It needs to be solved until convergence of residual and increments is obtained.
Following elaborations in Remark~\ref{rem:changing_function_space}, due to the displacement of the fluid-fluid interface~$\Int^{\fd_1 \fd_2}$ between iterations,
the fluid function space might change and new nodal degrees of freedom might get activated.
If this is the case, the classical Newton-Raphson scheme needs to be interrupted and the recent iteration before the function space
change occurred serves as initial guess for a new Newton-Raphson algorithm call in a new cycle~$c>1$.
To allow to restart the iterative scheme, solution iterations need to be projected from the last iteration function space
$(\mcW_h^{\fd_1})_{c-1}^n$ to $(\mcW_h^{\fd_1})^n_c$. An analogue projection needs to be performed for the previous time-step solution
between \mbox{$(\mcW_h^{\fd_1})^{n-1}$} and \mbox{$(\mcW_h^{\fd_1})_{c}^n$}. For this purpose, Algorithm~1 from \cite{Schott2017b} is applied,
where for further details the interested reader is referred to the latter publication.
Note that such projections need to be performed just for approximations on the mesh $\mcT_h^{\fd_1}$, as field approximations on the structural and the embedded fluid grid remain fixed over time and iterations.

For the first cycle at a new time level, \ie~$c=1$ at $t^n$, solutions fields of the monolithic system need to be predicted to provide a reasonable initial guess
for the new time level approximation and to accelerate the convergence of the iterative procedure.
For the structural field, in this work we choose a predictor for the velocities based on assuming constant velocities~$\bfU^{n}=\bfU^{n-1}$.
The acceleration and displacement predictors are afterwards constructed consistently with the Generalized\hbox{-}$\alpha$~method.
As result, the structural predictor yields a non-constant prediction of the solid displacements, which at $\Int^{\fd_2\sd}$ can
be used as boundary constraint~\Eqref{eq:ALE-constraint_FSI} for a predictive \name{ALE} relaxation solve for the $\mcT_h^{\fd_2}$-grid displacements,
such that the fluid grid moves to a balanced configuration and perfectly surrounds the structural body at its predicted location.
Due to the displacement of~$\Int^{\fd_1\fd_2}$, mesh related quantities of the background grid $\mcT_h^{\fd_1}$ need to be updated,
before solution fields can be transcribed to the new function space $(\mcW_h^{\fd_1})^n_{c=1}$. A potential fluid field predictor can be applied afterwards,
a step that is simplified to a constant field predictor in the present work.
After detecting convergence of the most inner Newton-Raphson iterations for residuals and increments, the solution fields associated to the fluid and solid approximations need to be updated according to the chosen temporal discretization schemes, see details in \cite{Schott2017b}.

\begin{remark}[Alternative strategies for updating $\Int^{\fd_1 \fd_2}$]\label{rem:strategies_relaxing_ALE_boundary}
    Since one of the major challenges for monolithically solved \name{CutFEM} based approaches is the potential change of approximation space~$\mcW_h^{\fd_1}$ during the iterative procedure, in the hybrid setting, even alternative strategies can be thought of.
    Instead of allowing the fluid-fluid interface~$\Int^{\fd_1\fd_2}$ to displace in all Newton iterations~\mbox{$m\geqslant 0$},
    it can be also kept fixed for~\mbox{$m>1$}, such that relaxation of the fluid grid~$\mcT_h^{\fd_2}$ takes place just in the prediction part of \Algoref{algo:nonlinear_solution_algo_monolithic_FSI}.
    This, however, puts high demands on the accuracy of the structural displacement predictor, since too large displacement increments
    in the subsequent Newton iterations could easily let distort the boundary layer patch. Note that strongly wall-refined fluid elements with high aspect ratios are used in this region.
    Thus, even though algorithmically much more demanding, an iterative full-implicit approximation of the grid displacements is preferred in this work.
    This guarantees well-posed, balanced configurations of the fluid patch during the entire Newton scheme.
\end{remark}

\section{Numerical Examples}
\label{sec:numerical_examples}

In this section, we validate the proposed hybrid Eulerian-\name{ALE} approach and demonstrate the ability of this discretization concept for \name{FSI}.
For validation purposes, our hybrid \name{FSI} approach is compared with a well-established monolithic
fitted-mesh \name{FSI} approach and a \name{CutFEM} based fixed-grid \name{FSI} scheme, which allows to deal with large structural motions.
For this goal, \name{FSI} problems with moderate deformations are chosen, which can be also computed with classical
ALE-based \name{FSI} approaches, without the need for any remeshing strategy.
Further numerical examples are provided to demonstrate the potential of the proposed approach to robustly and accurately deal
with complex large deformation \name{FSI} scenarios.
If not indicated otherwise, for all provided simulations it is chosen $\rho_\infty=1.0$ for the solid approximation and a \name{RBVM} stabilized form
is used for all fluid approximations.

It should be mentioned that examples have not been selected in order to stress the advantage of the proposed approach,
they rather have been selected in order to compare with standard approaches and to get some insight about the characteristics already in such rather simple examples. It is hopefully obvious that this approach offers amazing advantages - in terms of efficiency but also in terms of more computability - in many types of application examples.

\subsection{Compressing ball under moderate deformations}
\label{sec:fsi:numex:2Dcompcylinder}

\begin{figure}
\centering 
  \subfloat[]{\label{fig:fsi:numex:compsettingfsi}\includegraphics[width=0.40\textwidth]{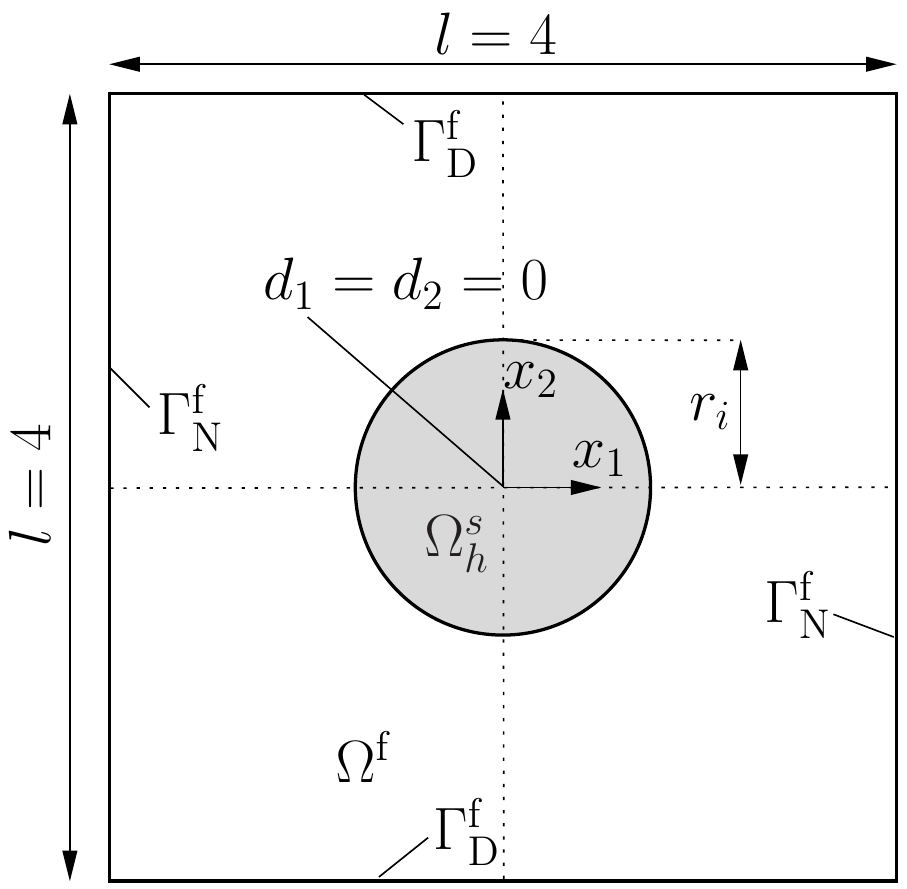}}
\qquad\qquad
  \subfloat[]{\label{fig:fsi:numex:compsettingxffsi}\includegraphics[width=0.40\textwidth]{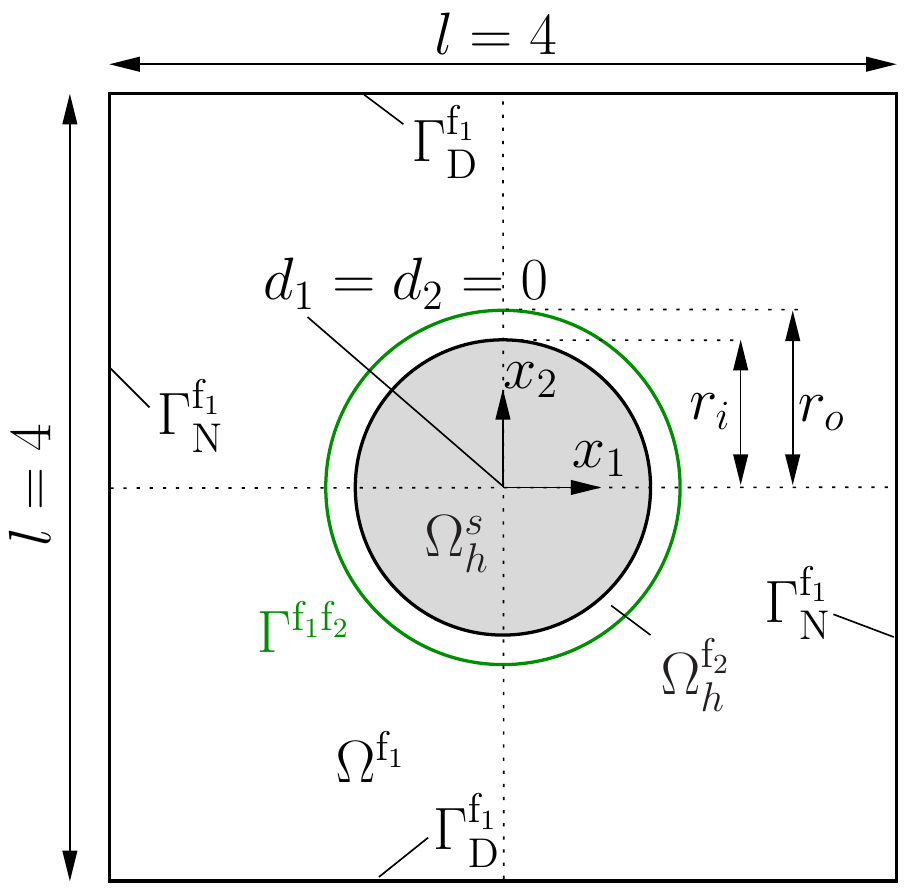}}
\caption{Compressing ball: \protect\subref{fig:fsi:numex:compsettingfsi}~classical \name{ALE} based moving mesh and \name{CutFEM} based fixed-grid \name{FSI} setup and \protect\subref{fig:fsi:numex:compsettingxffsi}~hybrid Eulerian-\name{ALE} \name{FSI} setup.
}
\label{fig:fsi:numex:2Dcompcylinder:setup}
\end{figure}
 
With the following \name{FSI} problem, we validate our monolithic hybrid Eulerian-\name{ALE} \name{FSI} approach
for moderate structural displacements by comparing the results with that of the monolithic classical \name{ALE} based \name{FSI} approach (see \cite{Schott2017b}[Approach 1]).
Since for a subsequent example provided in \Secref{sec:fsi:numex:2Dmovingcylinder} a pure fixed-grid approach is required for validation,
we consider already here the monolithic \name{CutFEM} based approach proposed in \cite{Schott2017b}[Approach 2].

\begin{figure}
\centering
\subfloat[\name{ALE} based moving mesh \name{FSI}.]{\label{fig:fsi:numex:compmeshalefsi}
      \begin{varwidth}{\linewidth} 
\includegraphics[width=0.40\textwidth, trim=0.8cm 0cm 0.8cm 0cm, clip=true]{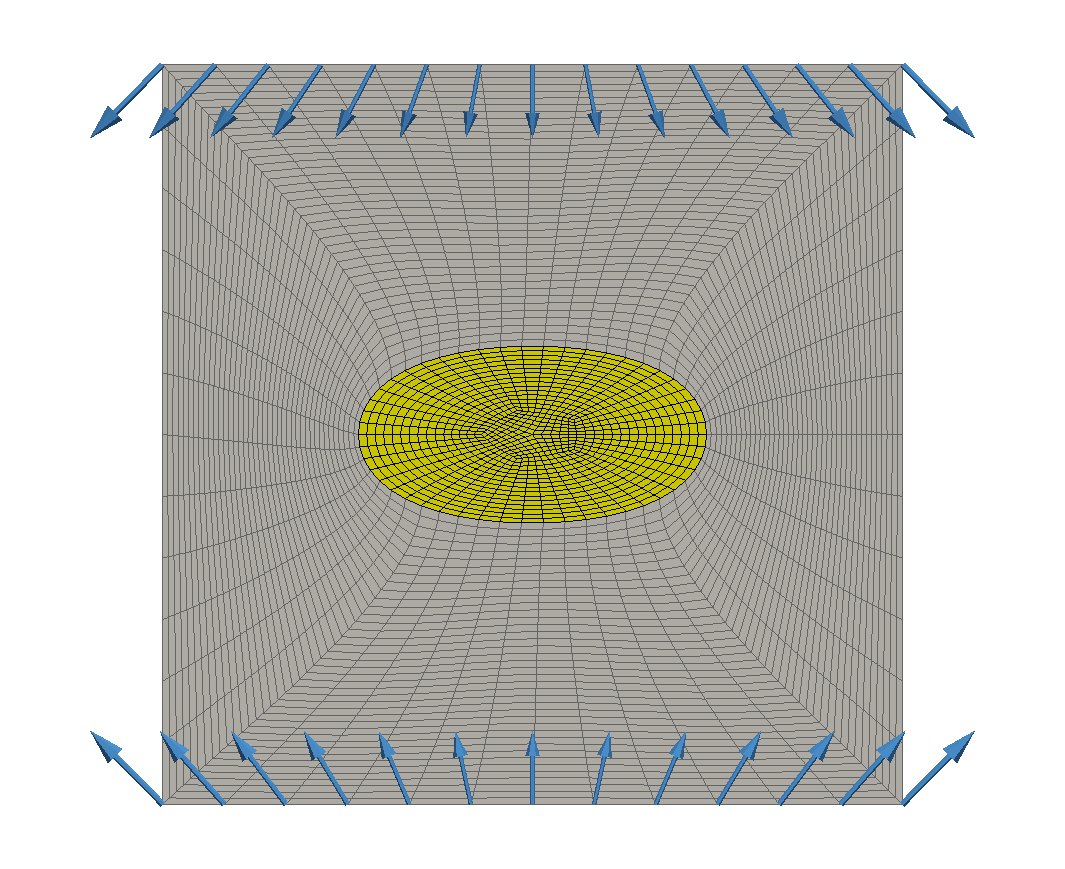}
\includegraphics[width=0.40\textwidth, trim=0.8cm 0cm 0.8cm 0cm, clip=true]{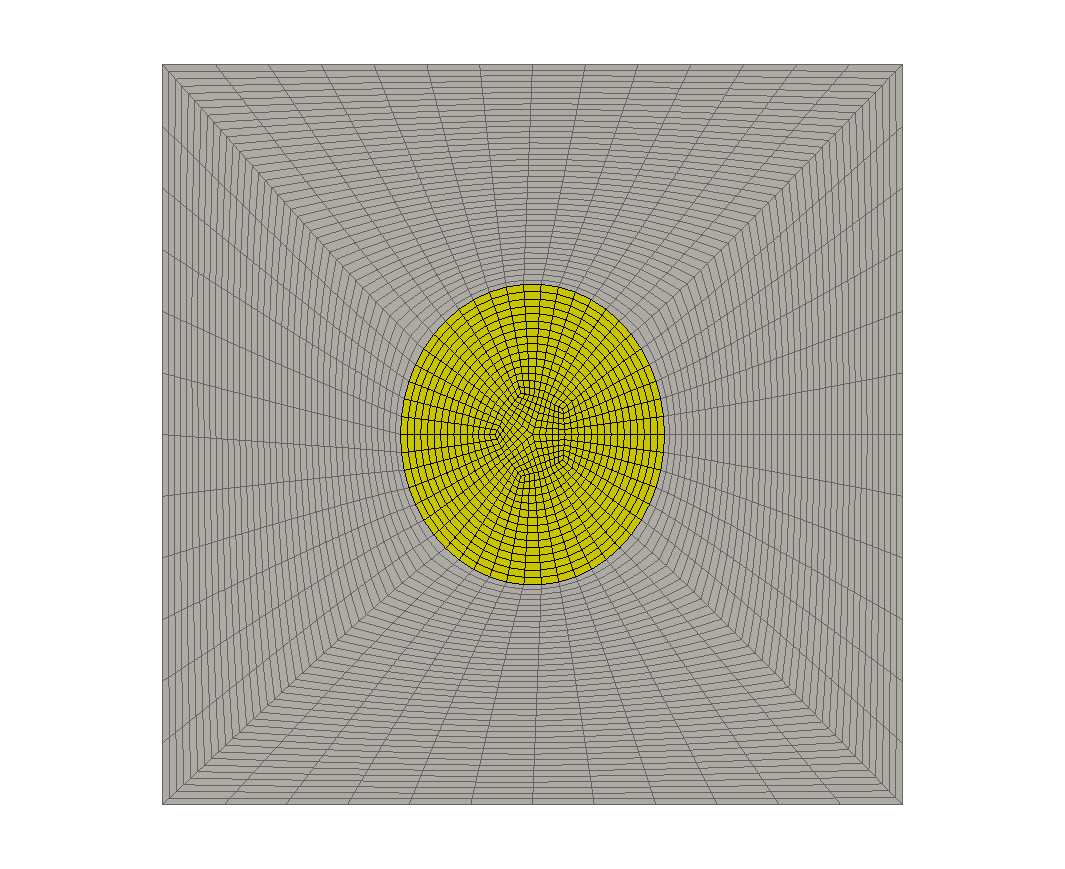}
\raisebox{1cm}[0pt][0pt]{\includegraphics[height=3cm, trim=19.0cm 18cm 12cm 5cm, clip=true]{pics/comp_struct/step400_mesh_alefsi.png}}
      \end{varwidth}
} \\
\subfloat[\name{CutFEM} based fixed-grid \name{FSI}.]{\label{fig:fsi:numex:compmeshxfsi}
      \begin{varwidth}{\linewidth}

\includegraphics[width=0.40\textwidth, trim=0.8cm 0cm 0.8cm 0cm, clip=true]{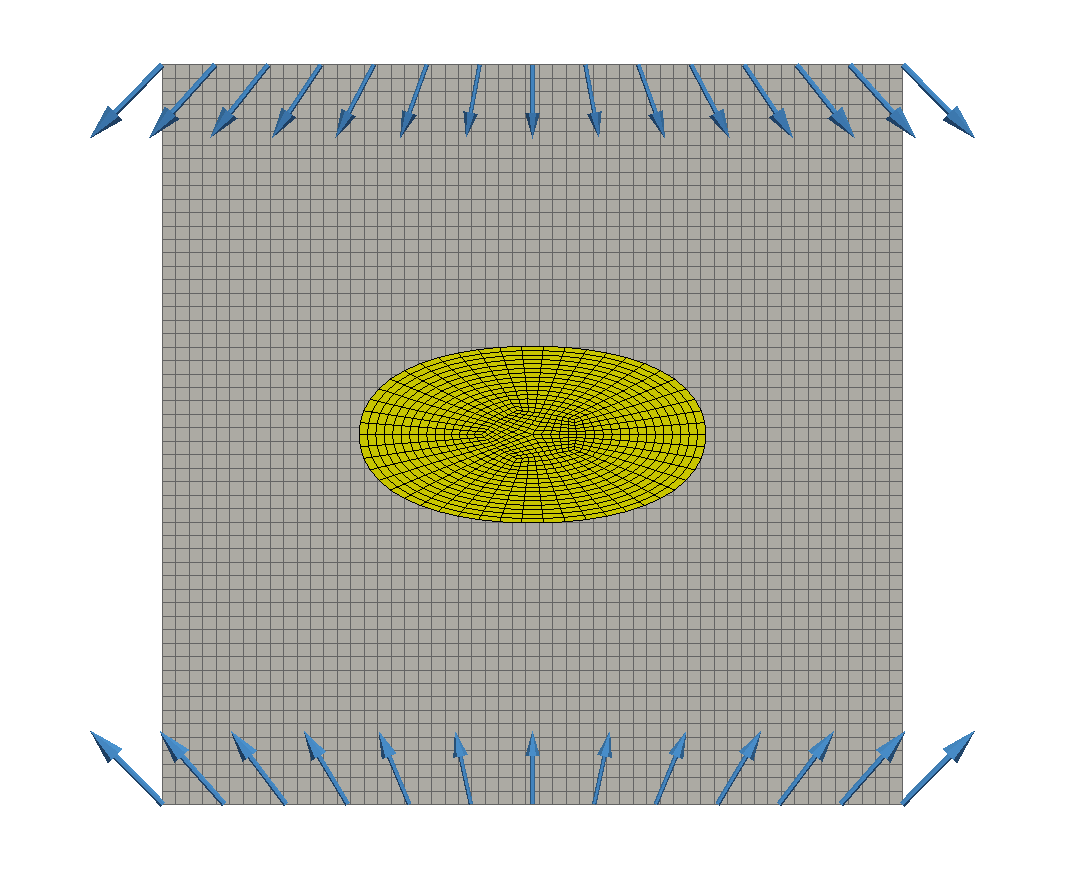}
\includegraphics[width=0.40\textwidth, trim=0.8cm 0cm 0.8cm 0cm, clip=true]{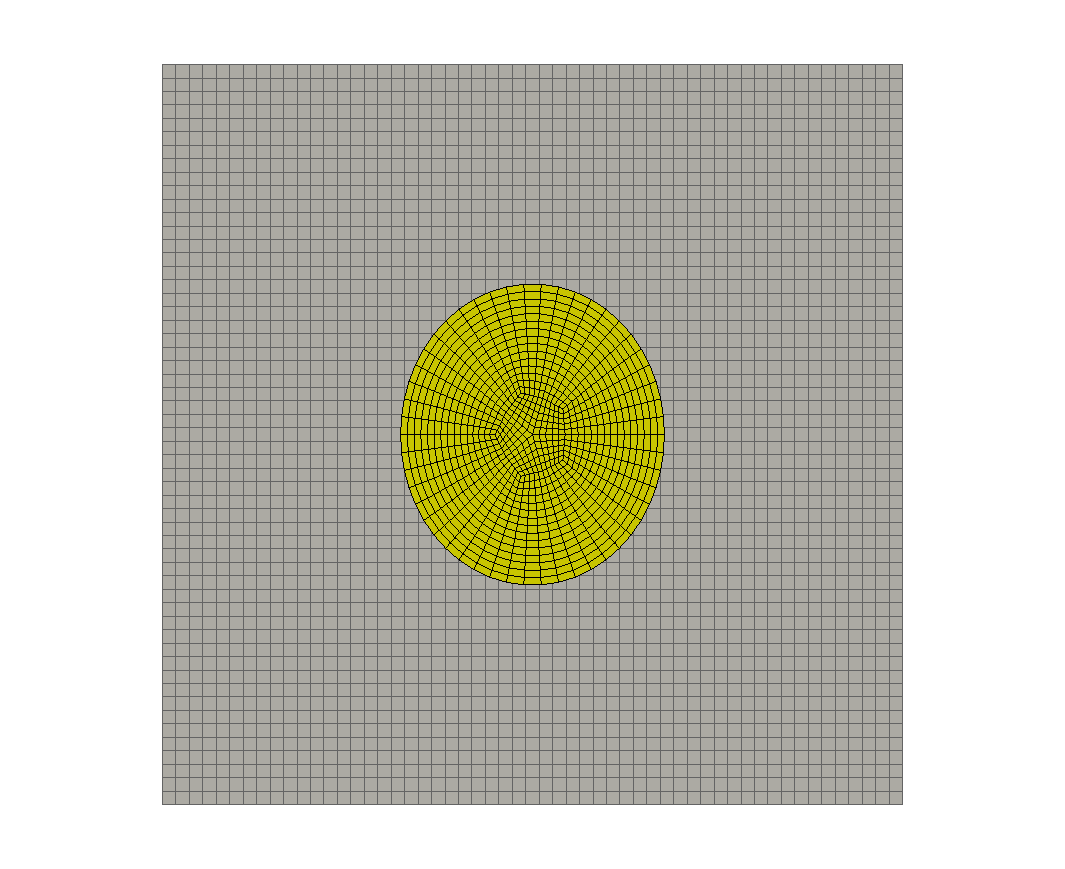}
\raisebox{1cm}[0pt][0pt]{\includegraphics[height=3cm, trim=19.0cm 18cm 12cm 5cm, clip=true]{pics/comp_struct/step400_mesh_xfsi.png}}
      \end{varwidth}
} \\
\subfloat[Hybrid Eulerian-\name{ALE} \name{FSI}.]{\label{fig:fsi:numex:compmeshxffsi}
      \begin{varwidth}{\linewidth}

\includegraphics[width=0.40\textwidth, trim=0.8cm 0cm 0.8cm 0cm, clip=true]{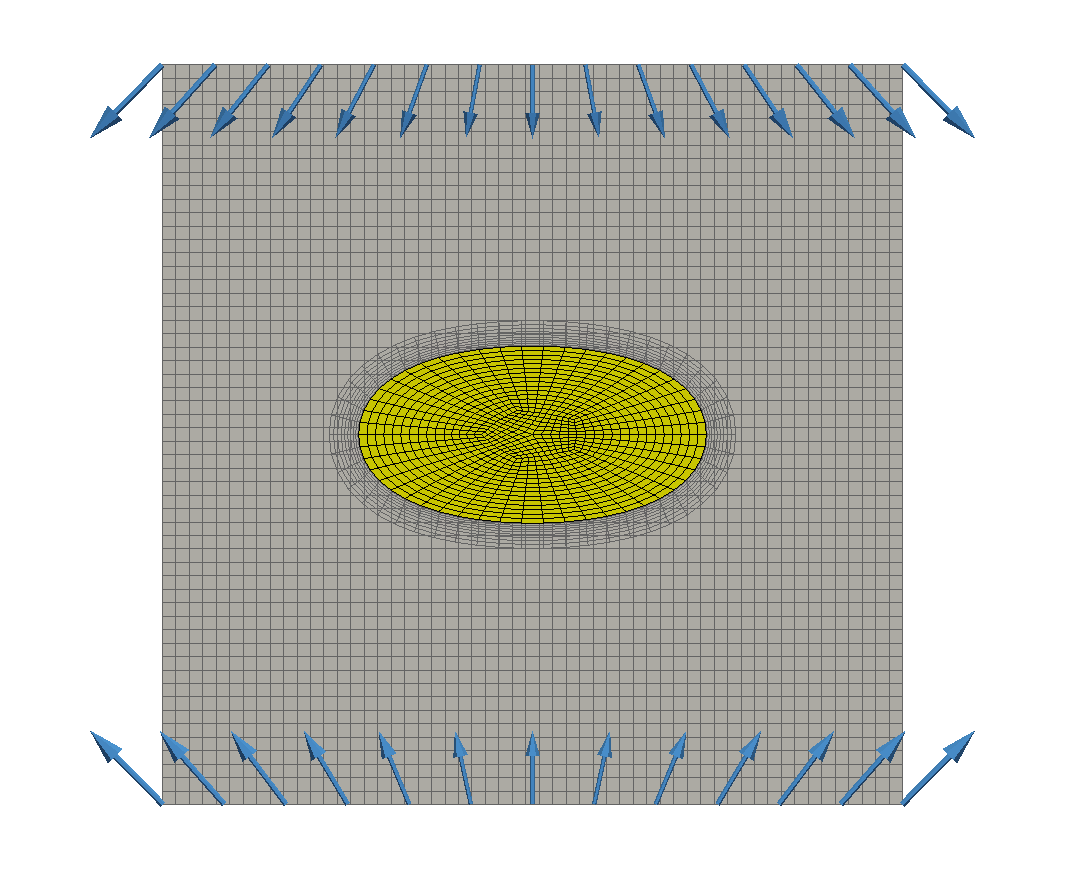}
\includegraphics[width=0.40\textwidth, trim=0.8cm 0cm 0.8cm 0cm, clip=true]{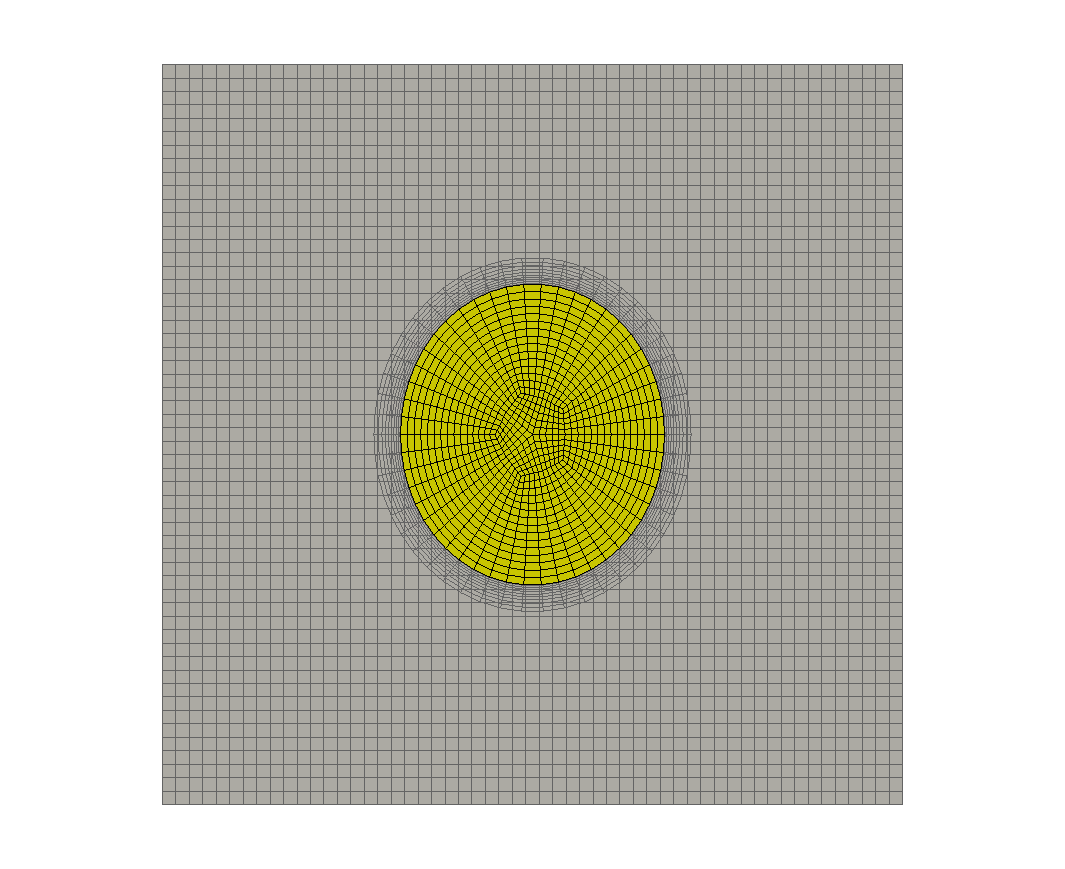}
\raisebox{1cm}[0pt][0pt]{\includegraphics[height=3cm, trim=19.0cm 18cm 12cm 5cm, clip=true]{pics/comp_struct/step400_mesh_xffsi.png}}
      \end{varwidth}
}
\caption{Compressing ball example: computational meshes for the approximation of solid and fluid domain at
different deformed states at times $t=3$ (left) and $t=4$ (middle) and snapshots at $t=4$ (right) for \protect\subref{fig:fsi:numex:compmeshalefsi}~classical \name{ALE} based moving mesh \name{FSI} setup,
\protect\subref{fig:fsi:numex:compmeshxfsi}~\name{CutFEM} based fixed-grid \name{FSI} setup and \protect\subref{fig:fsi:numex:compmeshxffsi}~hybrid Eulerian-\name{ALE} \name{FSI} setup.
Note that for visualization purposes, the depicted meshes are slightly coarser than those used for the simulations.
}
\end{figure}

A circular shaped structure with radius~$r=0.75$ and Neo-Hookean material
(Poisson's ratio $\nu^{\sd}=0.3$, Young's modulus $E^{\sd}=50$, density $\rho^{\sd}=1.0$)
is surrounded by a fluid (viscosity $\mu^{\fd}=1.0$ and density $\rho^{\fd}=1.0$)
within a square-shaped domain $[-2,2]^2$. Initially, the flow field is at rest.
The setup of this example for the classical \name{ALE} based \name{FSI} approach
and the \name{CutFEM} based fixed-grid approach is shown in \figref{fig:fsi:numex:compsettingfsi}.
The modified setup for the novel hybrid Eulerian-\name{ALE} \name{FSI} approach is depicted in \figref{fig:fsi:numex:compsettingxffsi}.
Periodic inflow
\begin{align}
u_2(x_1,t) = u_2^{\mathrm{max}}
\begin{cases}
  \tfrac{1}{2}(1+\sin(\pi t-\pi/2)) &\quad \foralls t\in[0,5], \\
  1  &\quad\foralls t\in[5,8]
\end{cases}
\end{align}
in opposite $x_2$-directions is prescribed at the top and bottom inlets,
indicated by~$\IntD^{\fd}$ in \figref{fig:fsi:numex:2Dcompcylinder:setup},
where it is set $u_2^{\mathrm{max}}=\pm 4$, respectively.
On both left and right outlets~$\IntN^{\fd}$, a zero Neumann boundary condition ($\bfhN=\bfzero$) is applied.
For the temporal discretization it is chosen $\Delta t =0.01$, and it is set $\theta=1.0$ for the fluid.
The structure is fixed at its middle point to avoid unstable \name{FSI} configurations initiated by mesh imperfections.
Identical structural meshes are used for all considered \name{FSI} approaches. 
It consists of $875$~bilinear elements with $50$~line segments along the circular surface.
For the spatial discretization of the flow field, three different setups are considered.
The classical moving mesh \name{ALE} approach (see \cite{Schott2017b}[Approach 1]) utilizes only one fluid grid consisting of $4000$~bilinearly interpolated elements.
Starting from the four outer walls, the mesh is refined towards the structural surface within $80$ segments distributed in radial direction,
resulting in a finest boundary layer thickness of \mbox{$h_{bl}\approx 0.01$}.
For the fixed-grid setup (see \cite{Schott2017b}[Approach 2]), a regular fluid mesh consisting of \mbox{$ 101 \times 101$} regular quadrilateral fluid elements is used.
Its active part is time dependent and depends on the interface location.
For the hybrid Eulerian-\name{ALE} approach, an initially ring-shaped fluid patch~$\mcT_h^2$ consisting of $10$~layers of quadrilateral elements within an outer radius $r_o=0.9$ and an inner radius $r_i=0.75$,
where it matches the structural surface node-wise, is embedded into a fixed non-moving background grid~$\mcT_h^1$. The latter consists of $55\times 55$ bilinearly interpolated elements,
where a portion of elements is deactivated as they are completely covered by the embedded fluid patch or the solid domain.
The set of active background elements changes when the embedded fluid patch deforms and follows the compression and decompression of the structural ball.
The different used computational meshes are visualized
in Figures~\ref{fig:fsi:numex:compmeshalefsi}--\ref{fig:fsi:numex:compmeshxffsi} at different times and deformed configurations.

\begin{figure}
\centering
\subfloat[\name{ALE} based moving mesh \name{FSI}.]{\label{fig:fsi:numex:compsimalefsi}
      \begin{varwidth}{\linewidth} 
\includegraphics[width=0.48\textwidth, trim=0.8cm 0cm 0.8cm 0cm, clip=true]{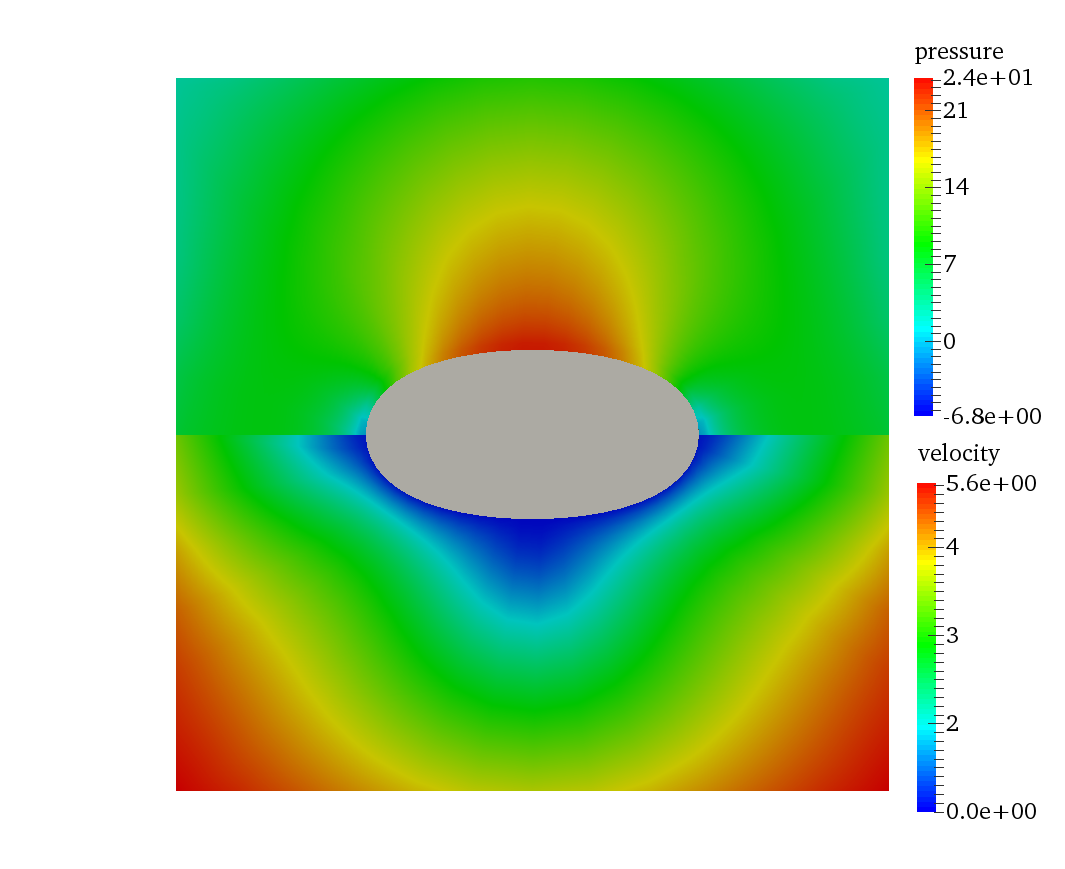}
\includegraphics[width=0.48\textwidth, trim=0.8cm 0cm 0.8cm 0cm, clip=true]{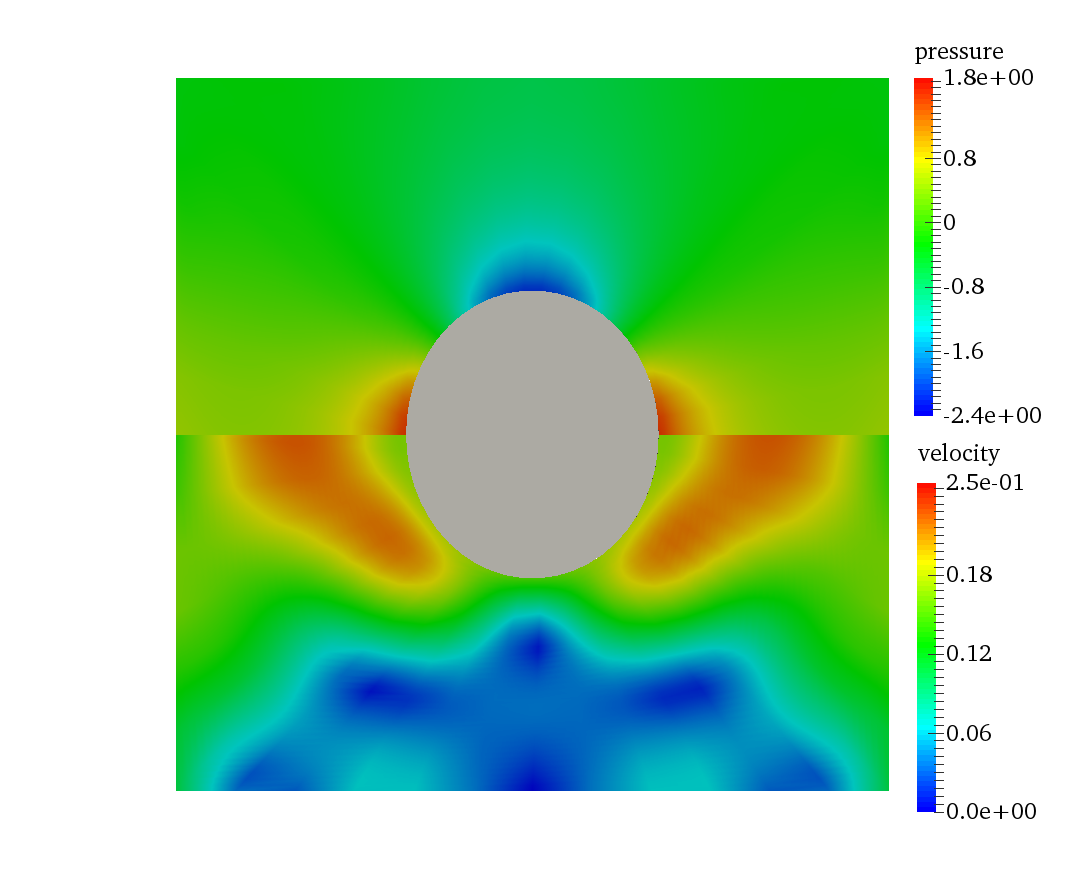}
      \end{varwidth}
}
\\
\subfloat[\name{CutFEM} based fixed-grid \name{FSI}.]{\label{fig:fsi:numex:compsimxfsi}
      \begin{varwidth}{\linewidth}
\includegraphics[width=0.48\textwidth, trim=0.8cm 0cm 0.8cm 0cm, clip=true]{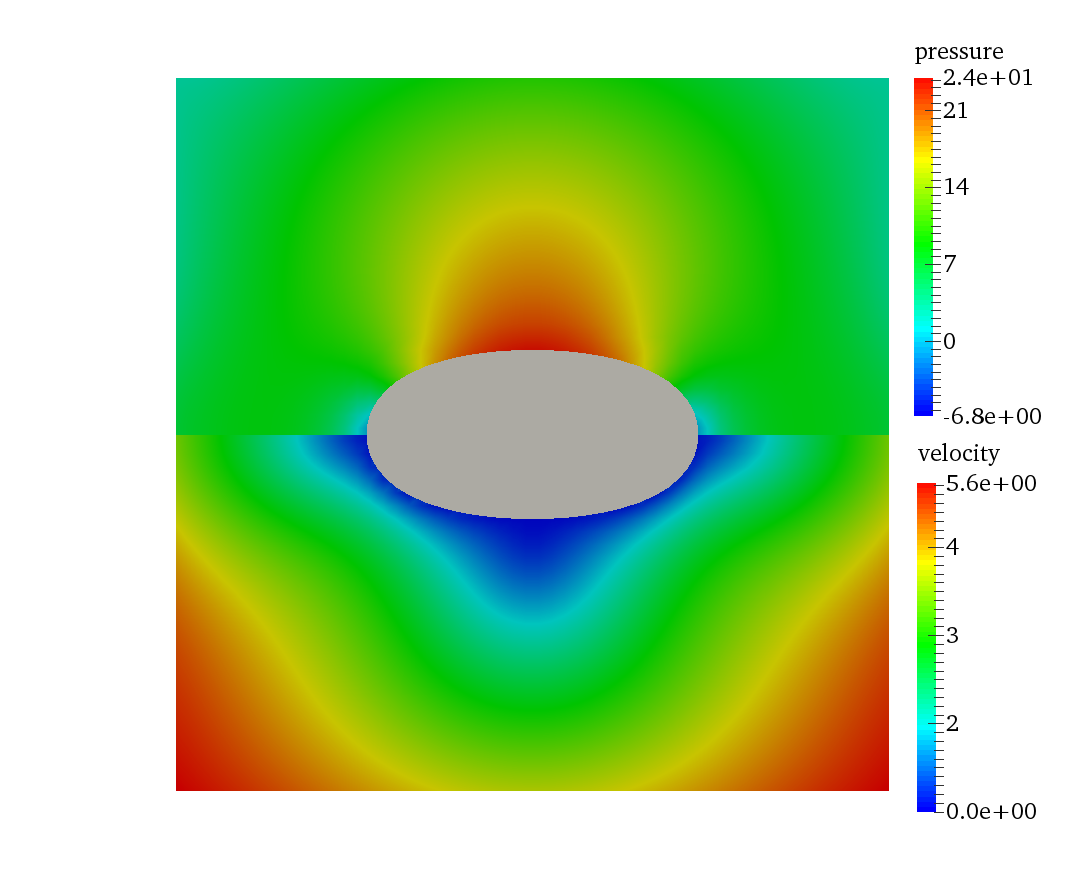}
\includegraphics[width=0.48\textwidth, trim=0.8cm 0cm 0.8cm 0cm, clip=true]{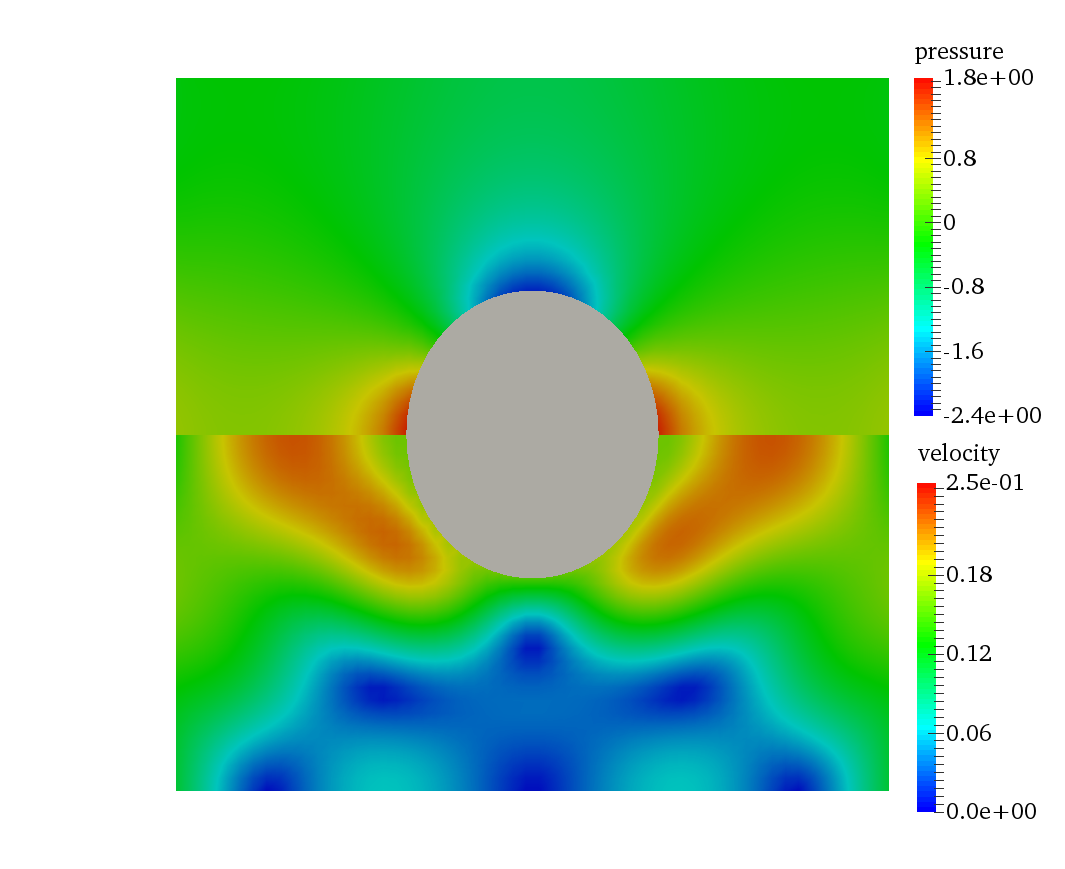}
      \end{varwidth}
}
\\
\subfloat[Hybrid Eulerian-\name{ALE} \name{FSI}.]{\label{fig:fsi:numex:compsimxffsi}
      \begin{varwidth}{\linewidth}
\includegraphics[width=0.48\textwidth, trim=0.8cm 0cm 0.8cm 0cm, clip=true]{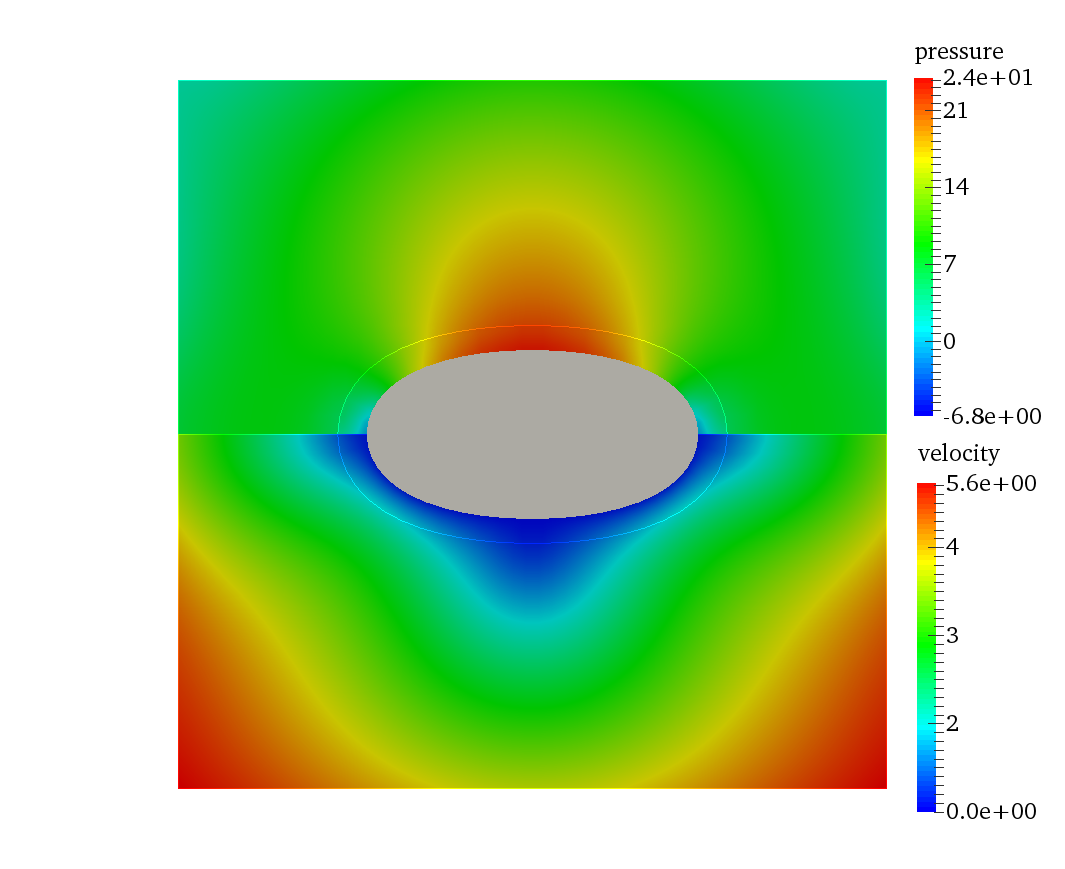}
\includegraphics[width=0.48\textwidth, trim=0.8cm 0cm 0.8cm 0cm, clip=true]{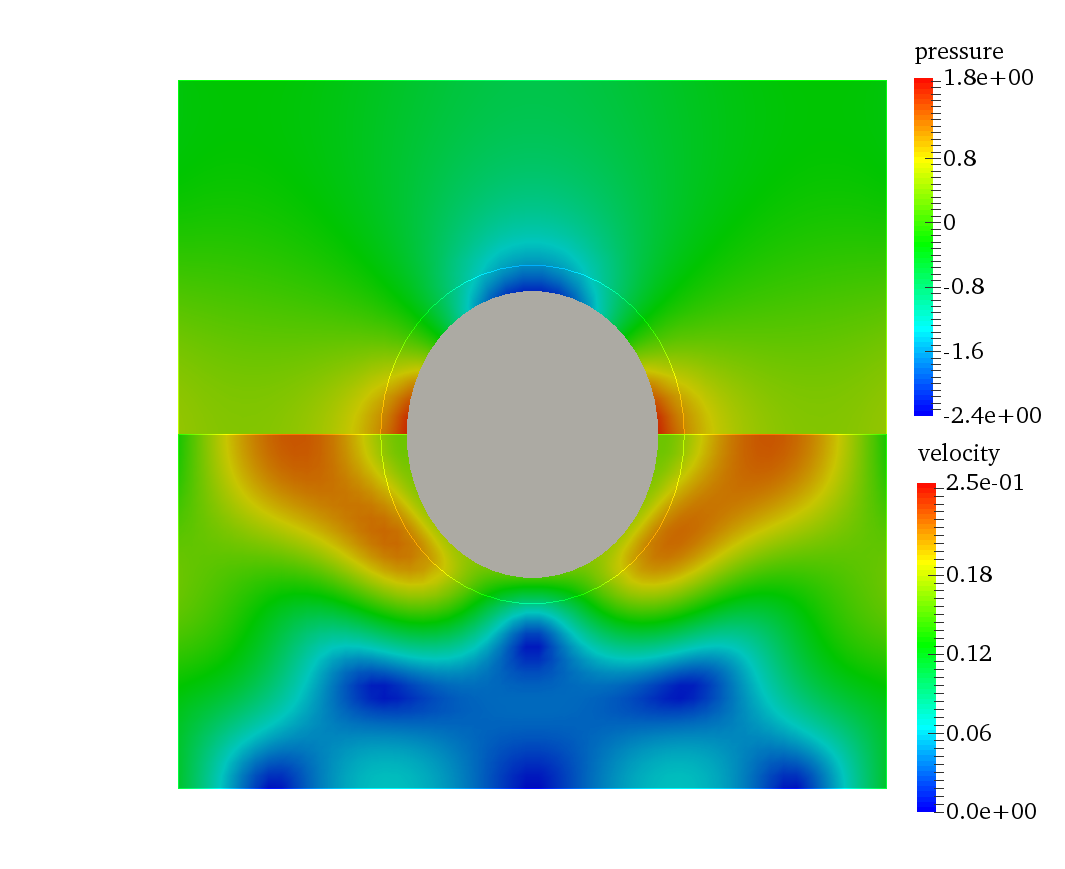}
      \end{varwidth}
}
\caption{Compressing ball example: velocity and pressure solution at
different deformed states at times $t=3$ (left) and $t=4$ (right)
for \protect\subref{fig:fsi:numex:compsimalefsi}~classical \name{ALE} based moving mesh \name{FSI} setup,
\protect\subref{fig:fsi:numex:compsimxfsi}~\name{CutFEM} based fixed-grid \name{FSI} setup
and \protect\subref{fig:fsi:numex:compsimxffsi}~hybrid Eulerian-\name{ALE} \name{FSI} setup.
}
\label{fig:fsi:numex:compsimfsi}
\end{figure}

\begin{figure}
\centering
\subfloat[Displacements over time $t$: $d_1$ for left-most point (left) and $d_2$ for top-most point (right).]{
\label{fig:comp_struct:displacements}
      \begin{varwidth}{\linewidth}
\includegraphics[width=6.0cm, trim=0cm 0cm 0cm 0cm, clip=true]{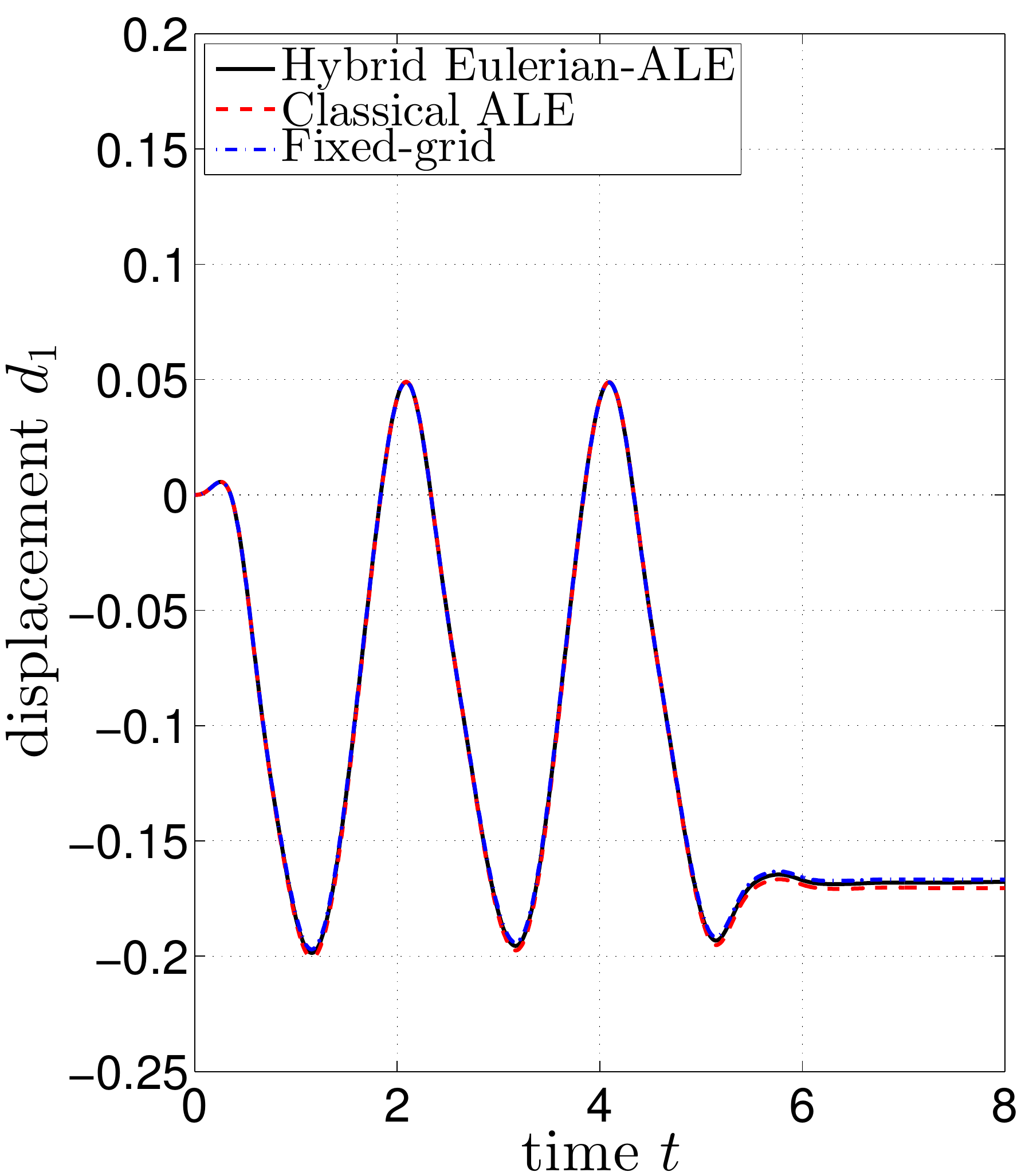}\qquad
\includegraphics[width=6.0cm, trim=0cm 0cm 0cm 0cm, clip=true]{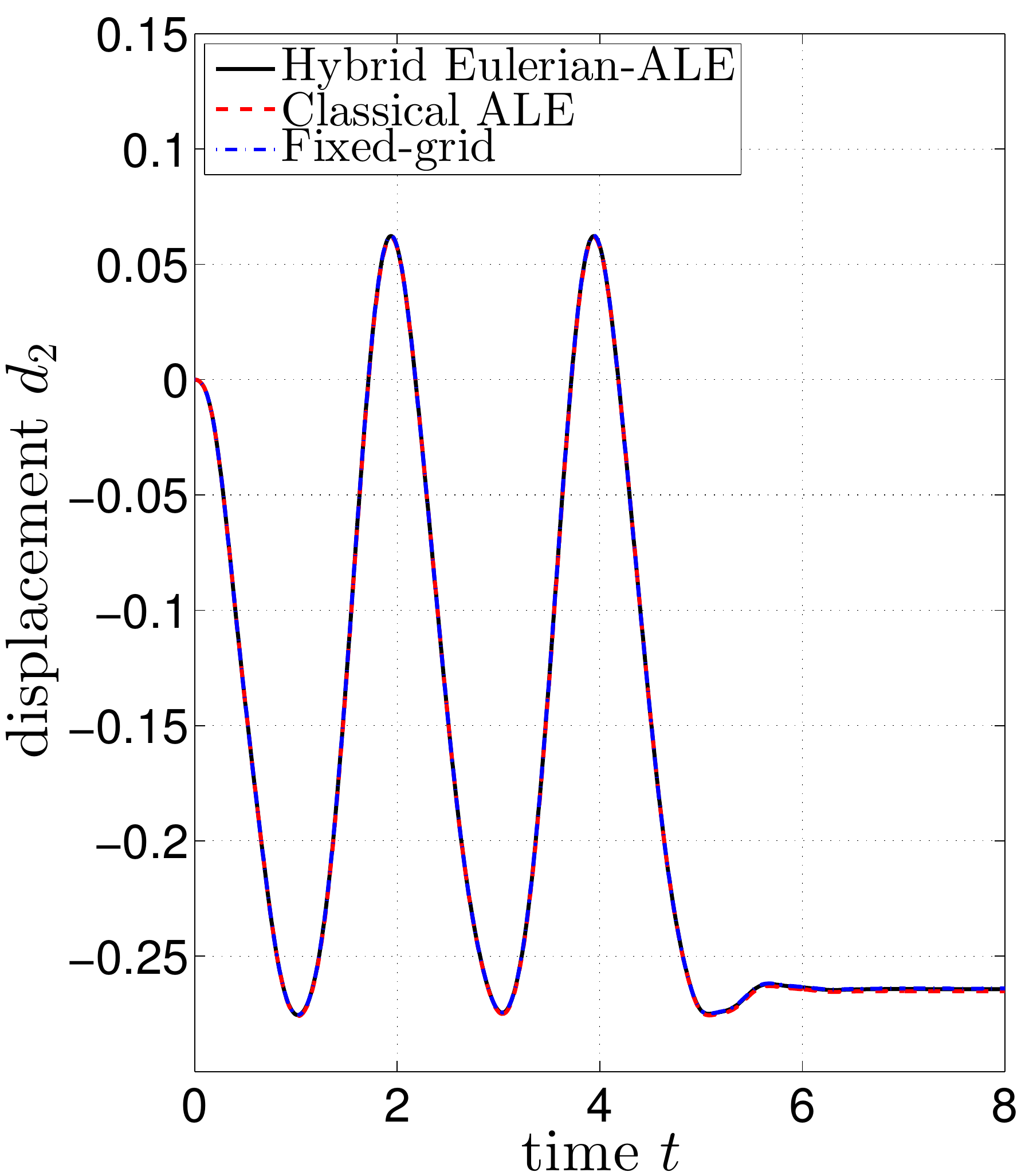}
      \end{varwidth}
}\\
\subfloat[Forces over time $t$: $f_1$ for left-most point (left) and $f_2$ for top-most point (right).]{
\label{fig:comp_struct:forces}
      \begin{varwidth}{\linewidth}
\includegraphics[width=6.0cm, trim=0cm 0cm 0cm 0cm, clip=true]{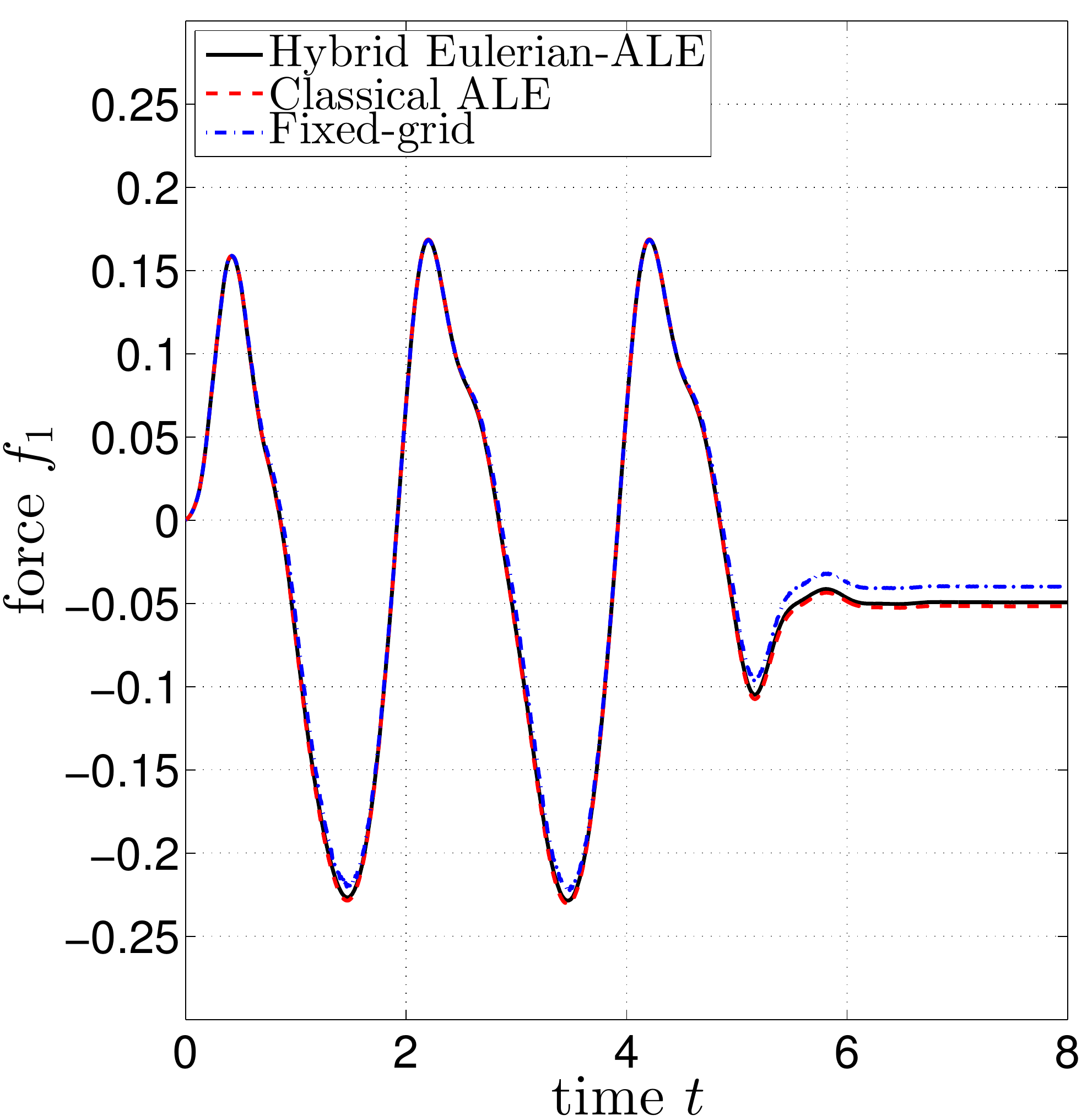}\qquad
\includegraphics[width=6.0cm, trim=0cm 0cm 0cm 0cm, clip=true]{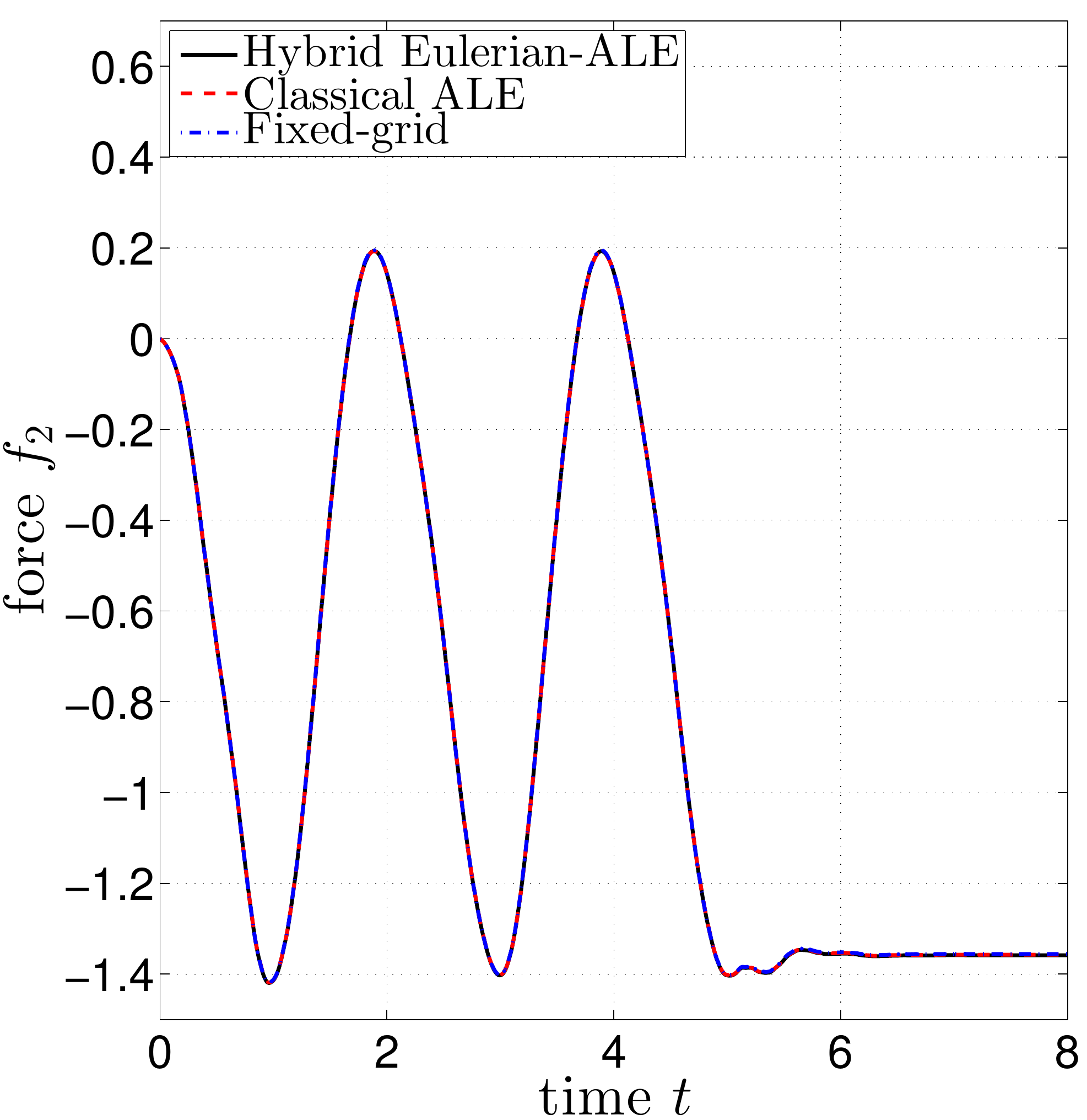}
      \end{varwidth}
}
  \caption{Compressing ball example: history of displacements~\protect\subref{fig:comp_struct:displacements}
and forces~\protect\subref{fig:comp_struct:forces} for
two characteristic control points at the \name{FSI} interface.
}
     \label{fig:comp_struct:displacements_forces}
\end{figure}

In Figures~\ref{fig:fsi:numex:compsimalefsi}--\ref{fig:fsi:numex:compsimxffsi}, equivalent velocity and pressure fields are shown for the three approaches, respectively.
To validate our hybrid approach, we compare the three simulations in more detail.
Results for displacements and interfacial forces at the left-most and the top-most structural point are plotted in \Figref{fig:comp_struct:displacements_forces}.
Due to the periodic excitation and the elastic material characteristics,
an oscillatory highly dynamic behavior can be observed for the computed values, before an equilibrium state is reached.
As it can be seen from the comparison, despite the different fluid meshes and computational approaches,
a good agreement of our novel hybrid Eulerian-\name{ALE} approach with the classical \name{ALE} based \name{FSI} approach and
the pure fixed-grid approach from \cite{Schott2017b} is obtained over the entire simulation time.

Comparing the fluid meshes of the different \name{FSI} approaches already indicates the advantage of the hybrid \name{FSI} approach. One can efficiently
generate a fine boundary layer mesh with high aspect ratio elements around the structure to suitably resolve potential high gradients in surface normal direction
as a prerequisite for an accurate approximation of viscous forces.
Please note that already in this very simple and low Reynolds  number example the pure fixed grid approach has problems in providing the correct forces (\Figref{fig:fsi:numex:compsimxfsi}, left) despite the fact that the fixed mesh is much finer.
Thereby, computational costs can be minimized by using fine-resolved grids just in the region of major interest.
These capabilities will become still more obvious for the subsequent test cases.

\subsection{Flow over a largely moving cylinder}
\label{sec:fsi:numex:2Dmovingcylinder}

A translationally moved rigid cylinder in a rectangular fluid domain serves as test case to demonstrate the
great capabilities of the proposed hybrid Eulerian-\name{ALE} \name{FSI} approach for highly dynamic higher-Reynolds-number flows.
Focus is put on the investigation of accuracy, cost efficiency, the ability to deal with large structural motions
and to sufficiently resolve boundary layer effects.
Even though the rigid body motion does not show all effects of a full fluid-structure interaction, it perfectly suites
to test the algorithmic setup of the hybrid Eulerian-\name{ALE} fluid solver and to analyze the accuracy of its flow approximation
in largely changing fluid domains.

A cylindrical rigid body~$\Dom^\sd$ with diameter~\mbox{$d=0.2$} and initial center point position \mbox{$(0.3,0.23)$} at~\mbox{$t=T_0$}
is placed in a box \mbox{$(0,2.2)\times (0,0.44)$}, surrounded by a fluid.
The location of the cylinder is slightly shifted from the vertical center in positive $x_2$-direction,
yielding a non-symmetric problem setup.
Since a much stronger vortex shedding will arise when moving the body, this makes the test case much more demanding.
The structure is pulled in positive $x_1$-direction, characterized by a prescribed displacement field \mbox{$d_1(t)=1.1+0.8\sin{(\tfrac{2}{3}\pi(t-0.75))}$}
imposed as Dirichlet constraint on the entire structural body.
The body will have returned at its starting point after a simulation time of \mbox{$T=3$}.
The flow is initially at rest and no-slip wall boundary conditions~\mbox{$\bfu=\bfzero$} are set strongly at top, bottom and left side of the fluid domain.
At~\mbox{$x_1=2.2$} the square fluid domain is opened and a zero-traction boundary condition is imposed, \ie~\mbox{$\bfhN=\bfzero$}.
For the temporal discretization, it is chosen \mbox{$\theta=1.0$} with \mbox{$\Delta t = 0.001$}.

\begin{figure}
\centering
\includegraphics[width=4.3cm]{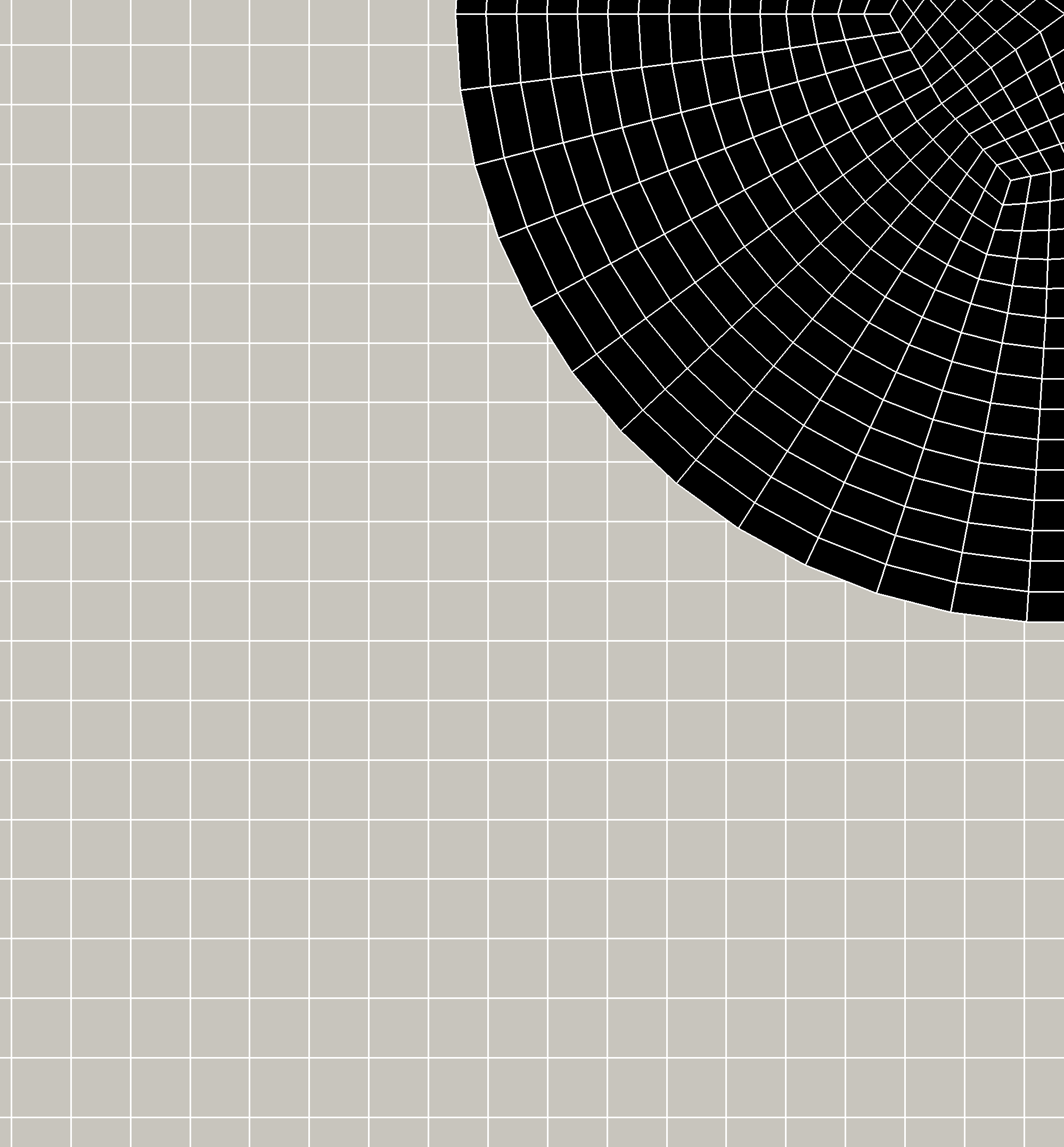}
\quad
\includegraphics[width=4.3cm]{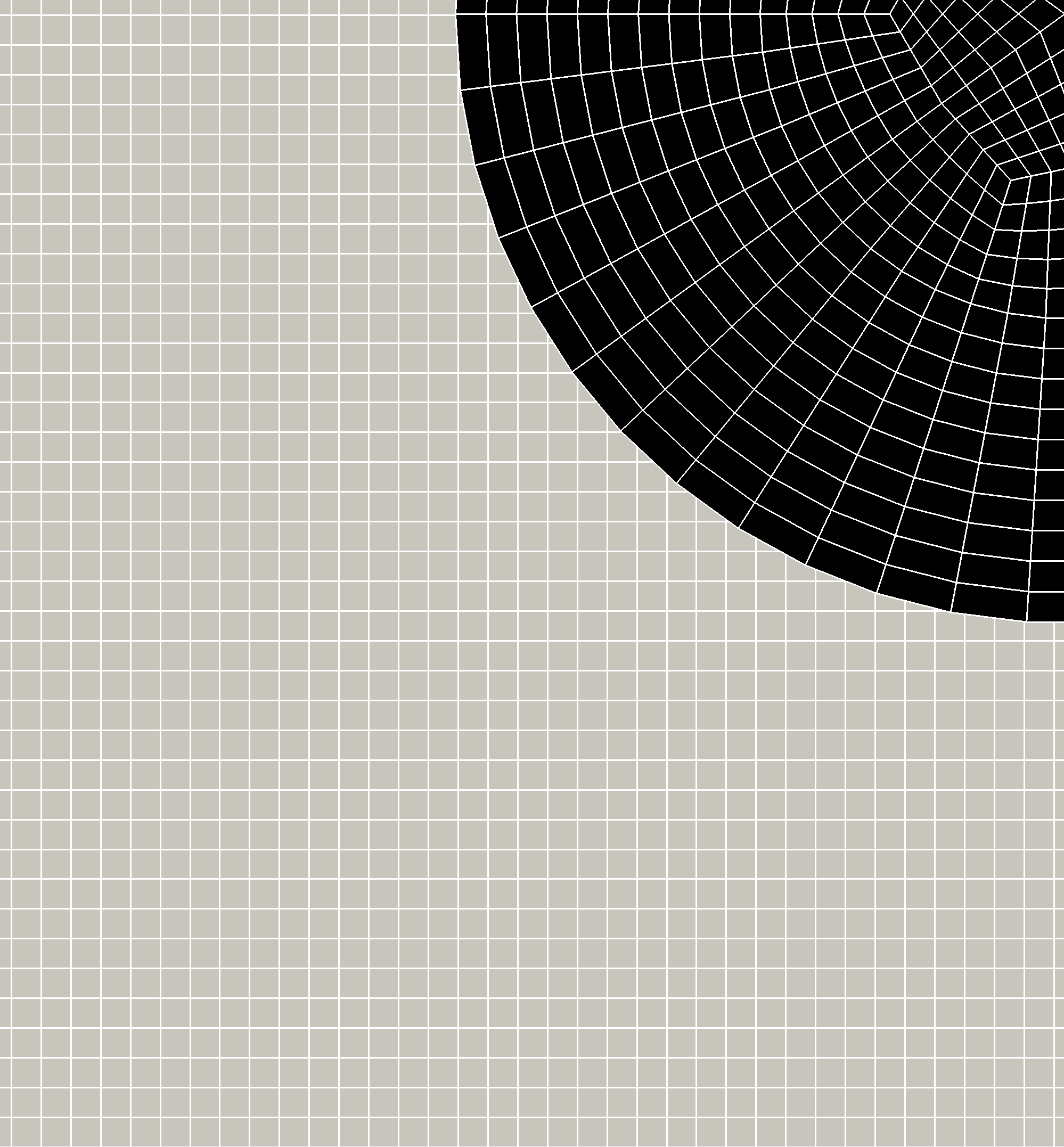}
\quad
\includegraphics[width=4.3cm]{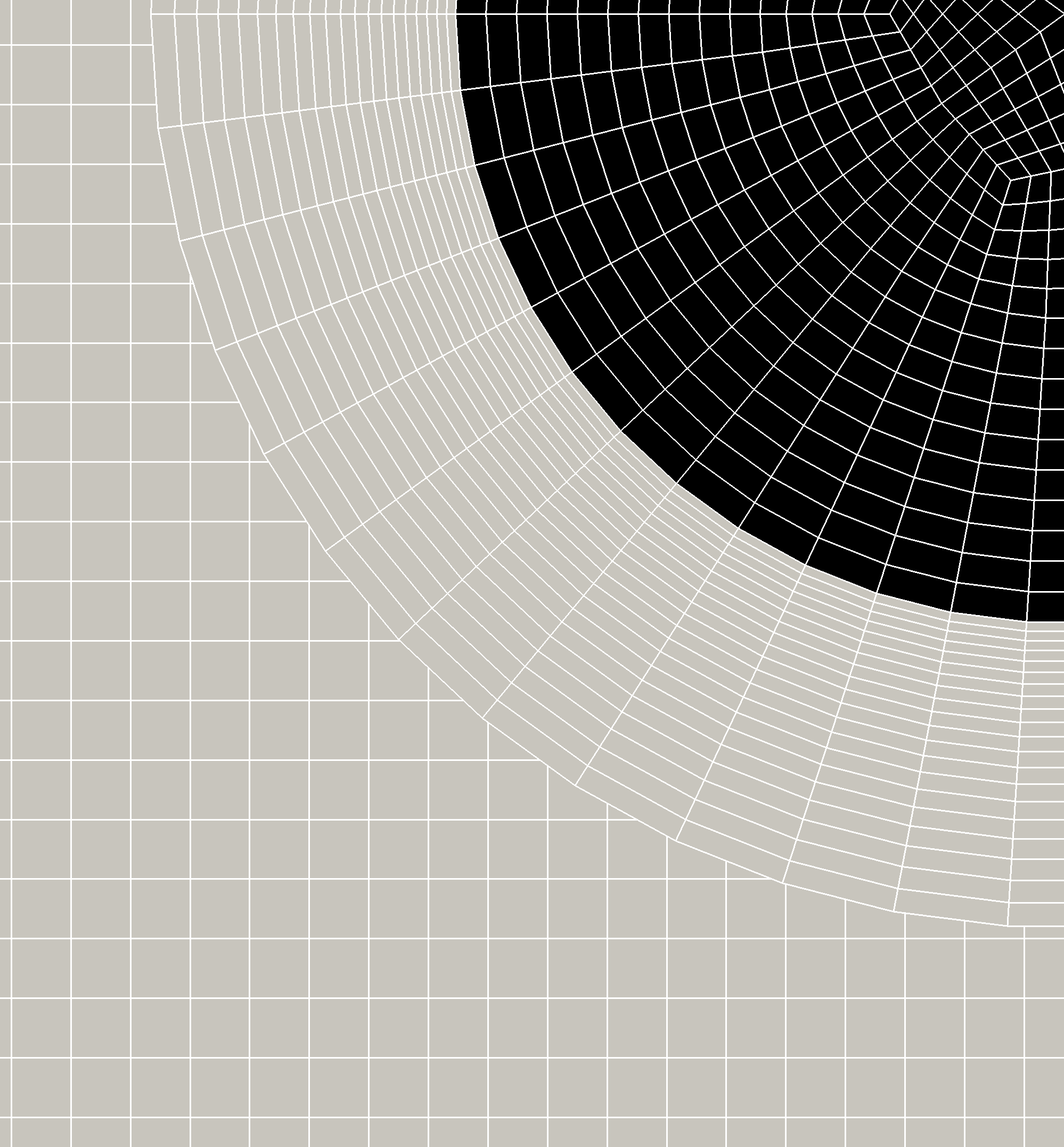}
  \caption{Flow over a largely moving cylinder: computational setup for \name{CutFEM} based fixed-grid approach from~\cite{Schott2017b} with $225\times 45$ elements (left),
  with $450\times 90$ elements (middle), and setup of the hybrid Eulerian-\name{ALE} approach with $225\times 45$ background elements and embedded boundary layer fluid patch (right).}
  \label{fig:moving_cylinder:computational_setup}
\end{figure}

Since the body is subjected to very large motions, a classical \name{ALE} based moving mesh approach cannot be used for reliable comparisons.
For the purpose of validation and to demonstrate the benefits of our novel scheme,
our hybrid Eulerian-\name{ALE} approach is compared to a pure fixed-grid \name{CutFEM} based method (see \cite{Schott2017b}[Approach~2]),
as already considered in \Secref{sec:fsi:numex:2Dcompcylinder}.
For the latter fixed-grid scheme, two different background mesh resolutions are considered, a fine mesh consisting of
\mbox{$450\times 90$} linearly-interpolated quadrilateral finite elements
and a coarser variant consisting of \mbox{$225\times 45$} elements.
The number of active elements depends on the positioning of the cylinder.
The coarser background mesh is also utilized for our hybrid approach,
into which a ring-shaped fluid patch with an outer diameter $d_o=0.3$ and
$20$~layers of thin boundary layer elements is embedded.
A refinement towards the fluid-solid interface $\Int^{\fd\sd}$ is chosen to accurately capture the boundary layer effects around the moving cylinder.
For all variants, identical structural meshes are used, where the interface is approximated with 
$50$~linear segments in structural circumferential direction.
The computational setups for the fixed-grid approach and the hybrid approach are visualized in \Figref{fig:moving_cylinder:computational_setup}.

\begin{figure}
\centering
\subfloat[Velocity norm~$\euclidian{\vel_h}$ (top) and pressure~$p_h$ (bottom) at time \mbox{$t=1.18$}.]{
\label{fig:moving_cylinder:vel_pres_1_a}
\begin{varwidth}{\linewidth}
\includegraphics[width=11.0cm]{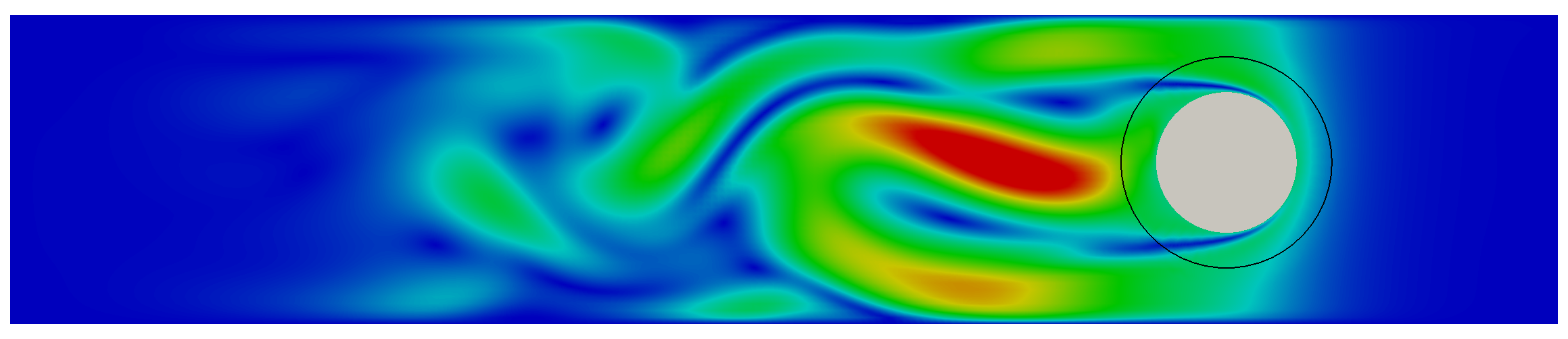}\\
\includegraphics[width=11.0cm]{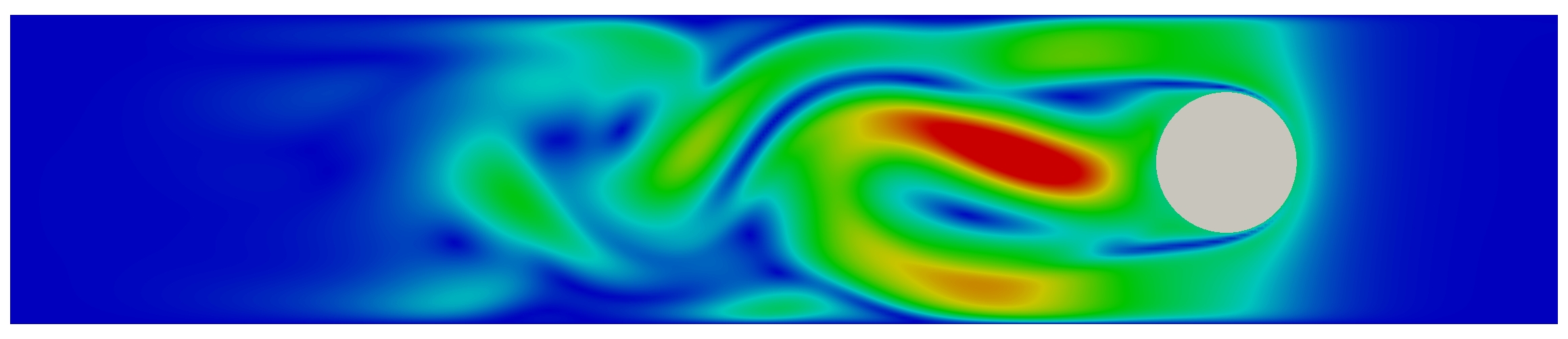}\\
~\\
\includegraphics[width=11.0cm]{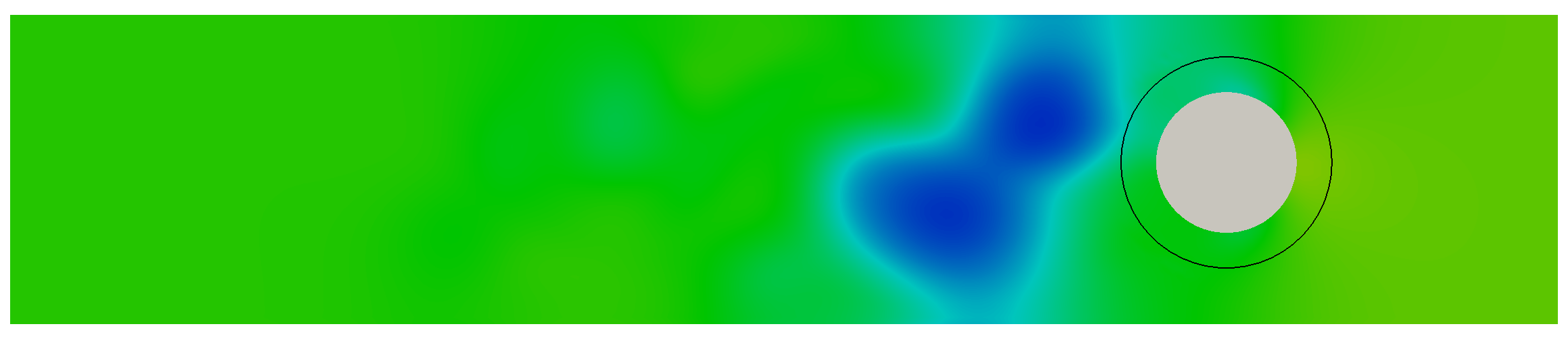}\\
\includegraphics[width=11.0cm]{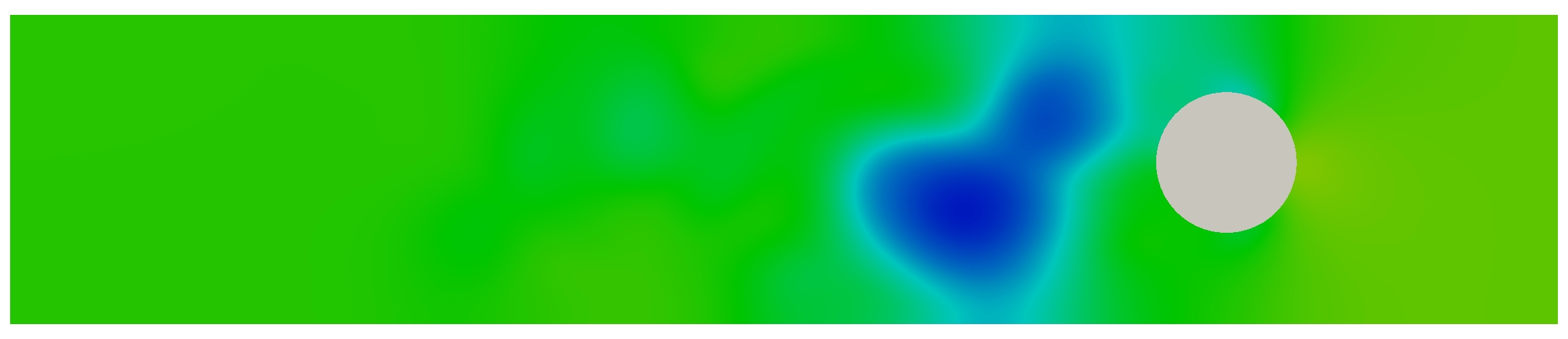}
\end{varwidth}
}
\\
\subfloat[Velocity norm~$\euclidian{\vel_h}$ (top) and pressure~$p_h$ (bottom) at time \mbox{$t=1.97$}.]{
\label{fig:moving_cylinder:vel_pres_1_b}
\begin{varwidth}{\linewidth}
\includegraphics[width=11.0cm]{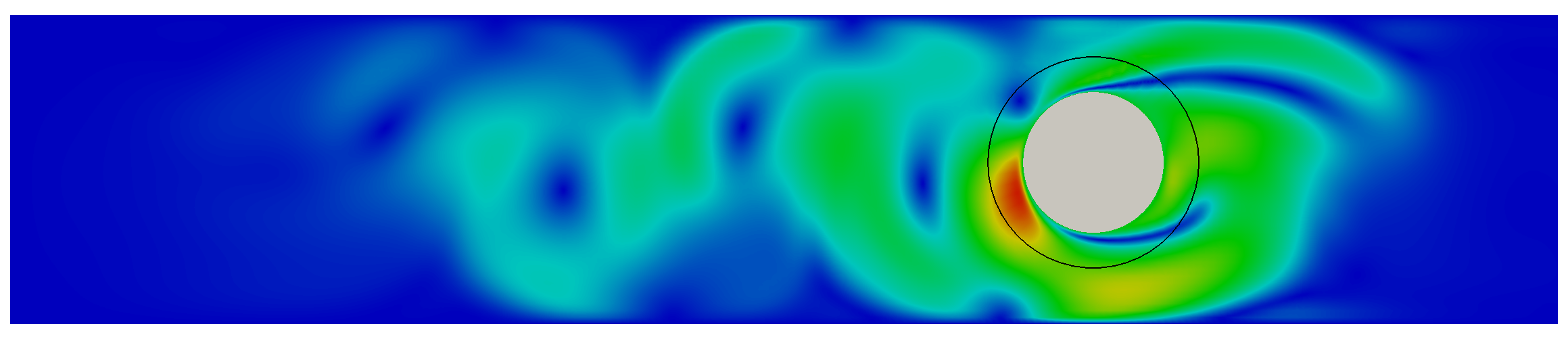}\\
\includegraphics[width=11.0cm]{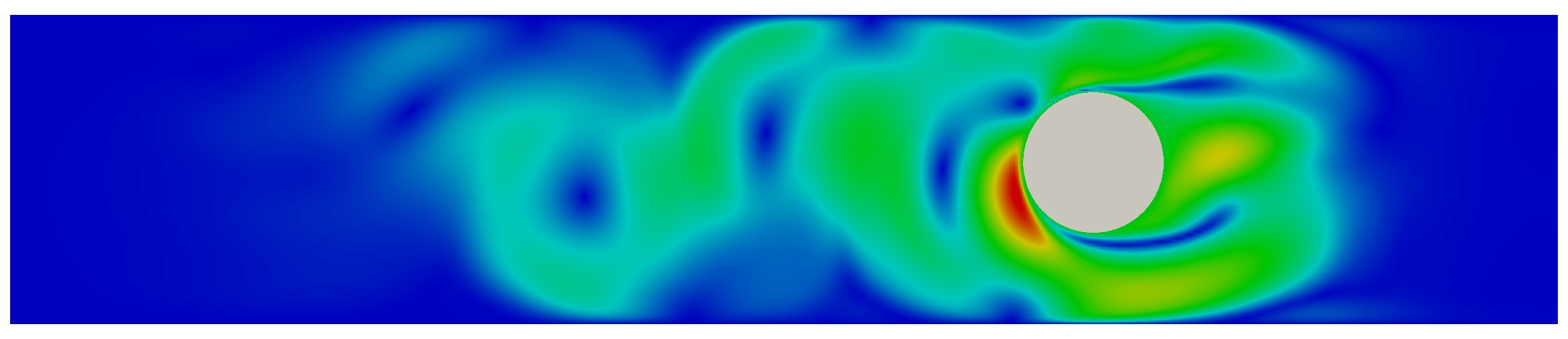}\\
~\\
\includegraphics[width=11.0cm]{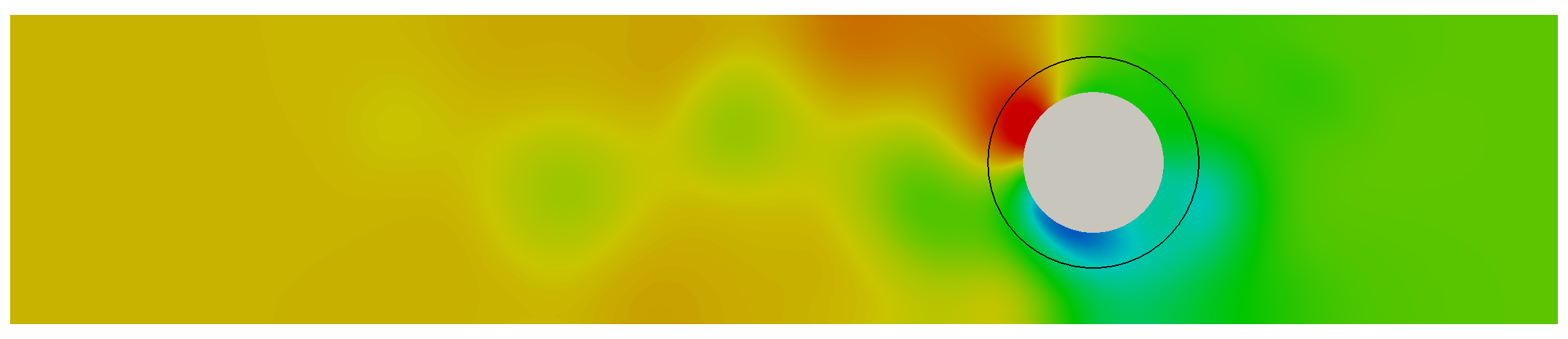}\\
\includegraphics[width=11.0cm]{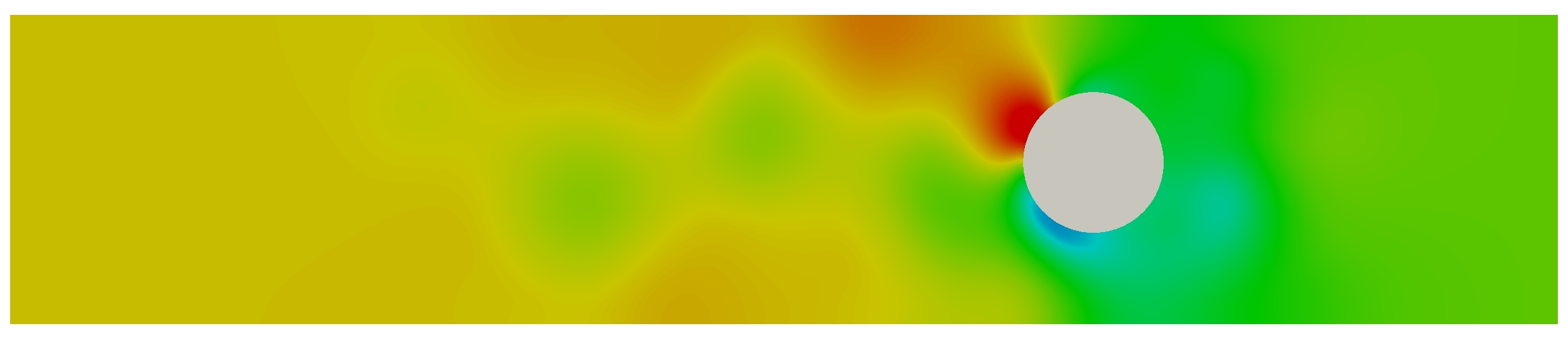}
\end{varwidth}
}
  \caption{Flow over a largely moving cylinder:
comparison of the hybrid Eulerian-\name{ALE} approach (with \mbox{$225\times 45$} background fluid elements
and embedded fluid patch) with the \name{CutFEM} based fixed-grid approach \cite{Schott2017b}[Approach~2] (with \mbox{$450\times 90$} fluid elements) at different times (\protect\subref{fig:moving_cylinder:vel_pres_1_a} and \protect\subref{fig:moving_cylinder:vel_pres_1_b}).
Velocity color scale \mbox{$[0,3]$}, pressure color scale \mbox{$[-5.6,3.5]$}. The top of all paired pictures shows the hybrid Eulerian-\name{ALE} approach.
}
\label{fig:moving_cylinder:vel_pres_1}
\end{figure}

\begin{figure}
\centering
\subfloat[Velocity norm~$\euclidian{\vel_h}$ (top) and pressure~$p_h$ (bottom) at time \mbox{$t=2.29$}.]{
\label{fig:moving_cylinder:vel_pres_2_a}
\begin{varwidth}{\linewidth}
\includegraphics[width=11.0cm]{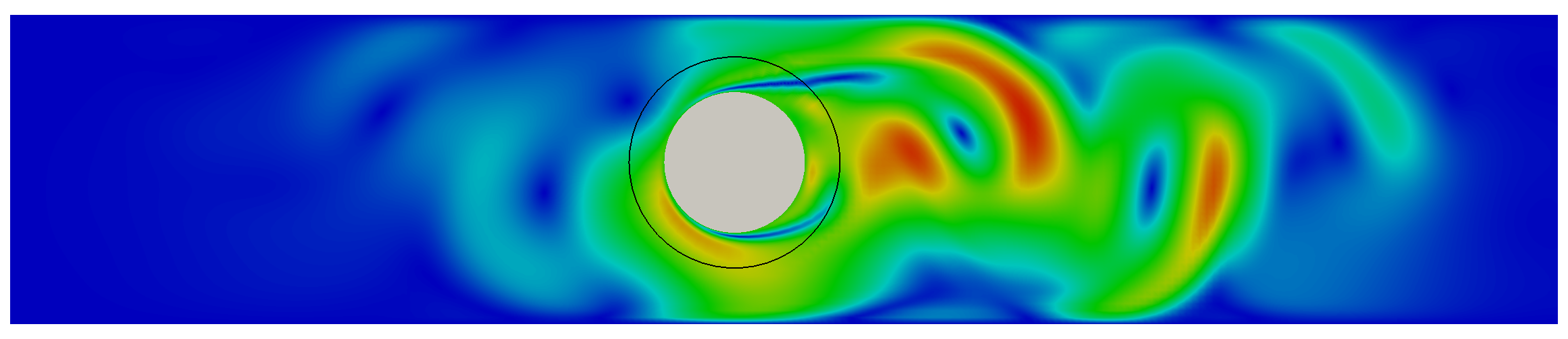}\\
\includegraphics[width=11.0cm]{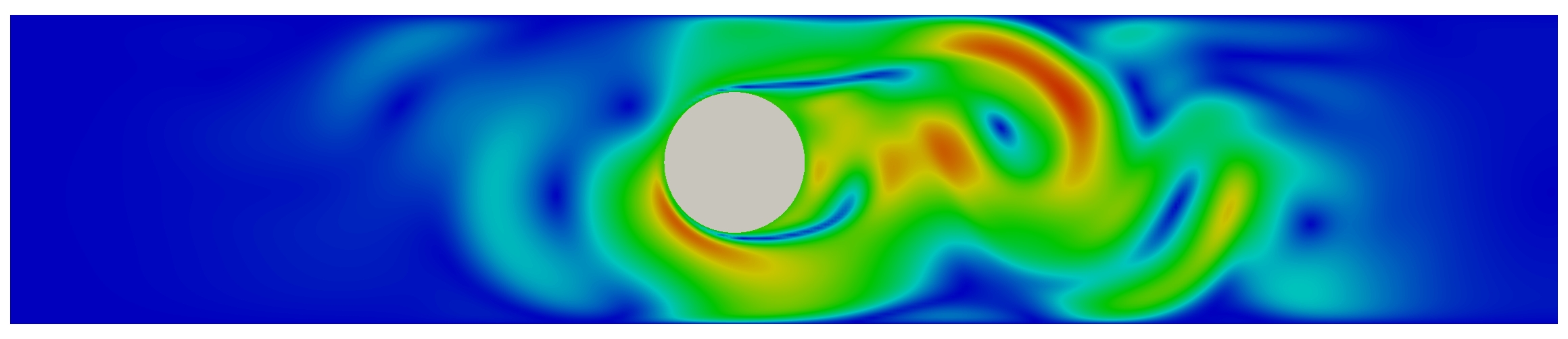}\\
~\\
\includegraphics[width=11.0cm]{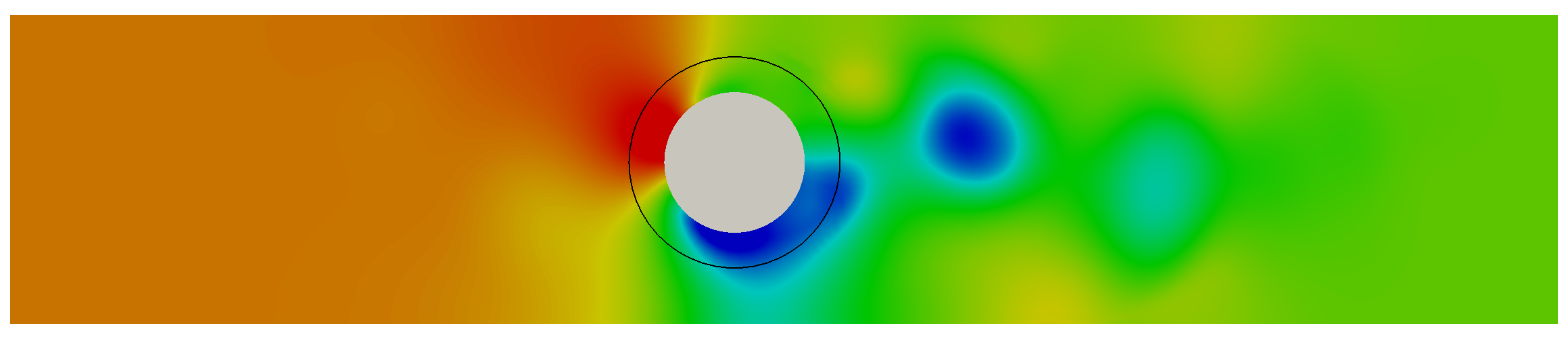}\\
\includegraphics[width=11.0cm]{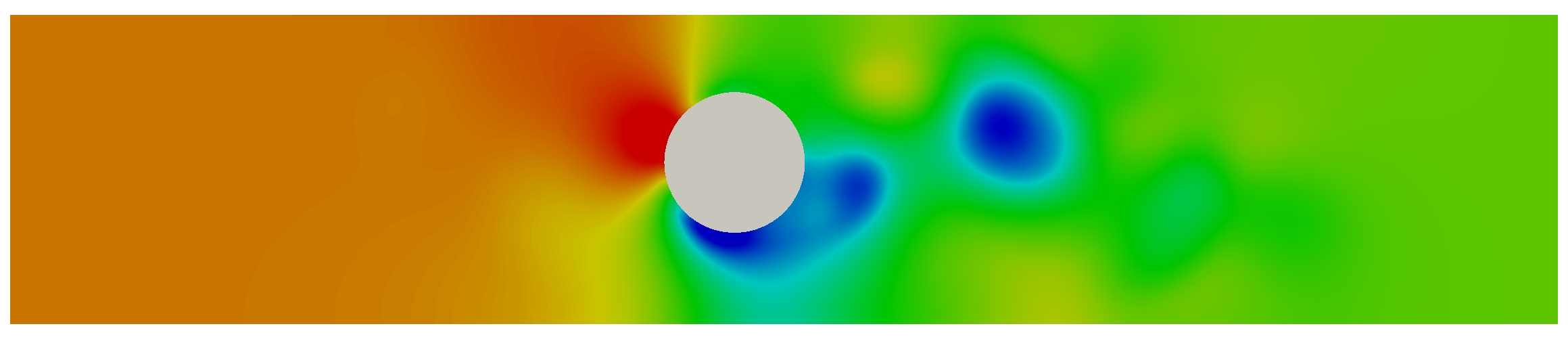}
\end{varwidth}
}
\\
\subfloat[Velocity norm~$\euclidian{\vel_h}$ (top) and pressure~$p_h$ (bottom) at time \mbox{$t=2.46$}.]{
\label{fig:moving_cylinder:vel_pres_2_b}
\begin{varwidth}{\linewidth}
\includegraphics[width=11.0cm]{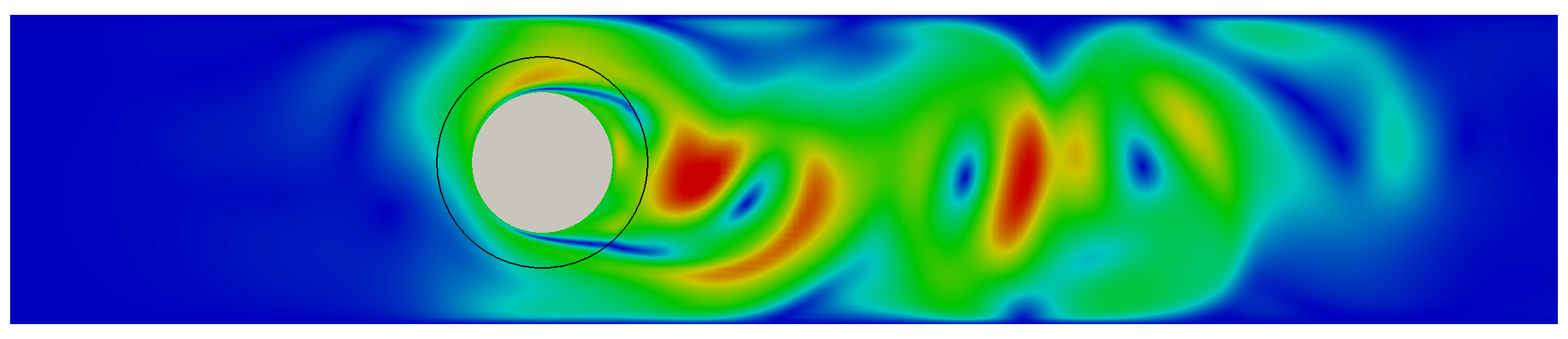}\\
\includegraphics[width=11.0cm]{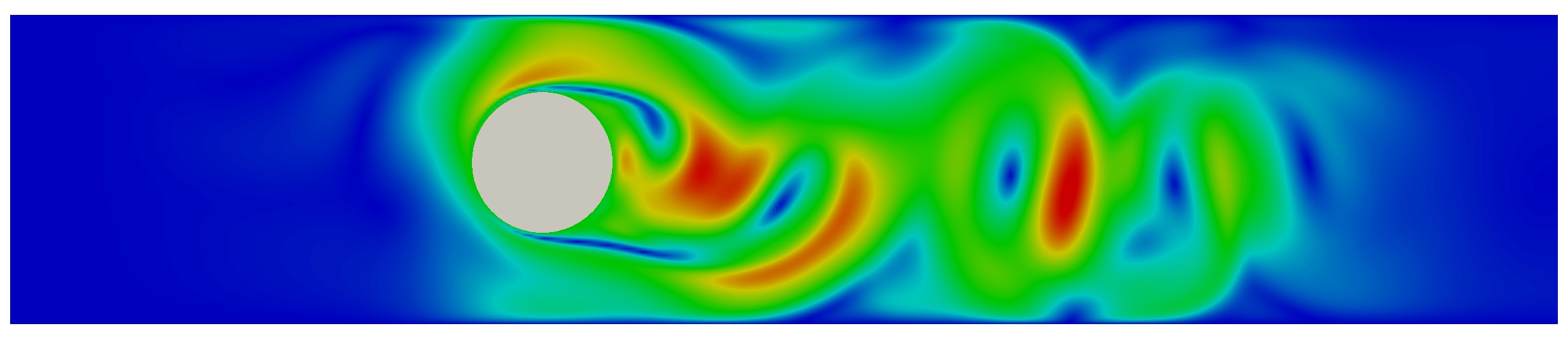}\\
~\\
\includegraphics[width=11.0cm]{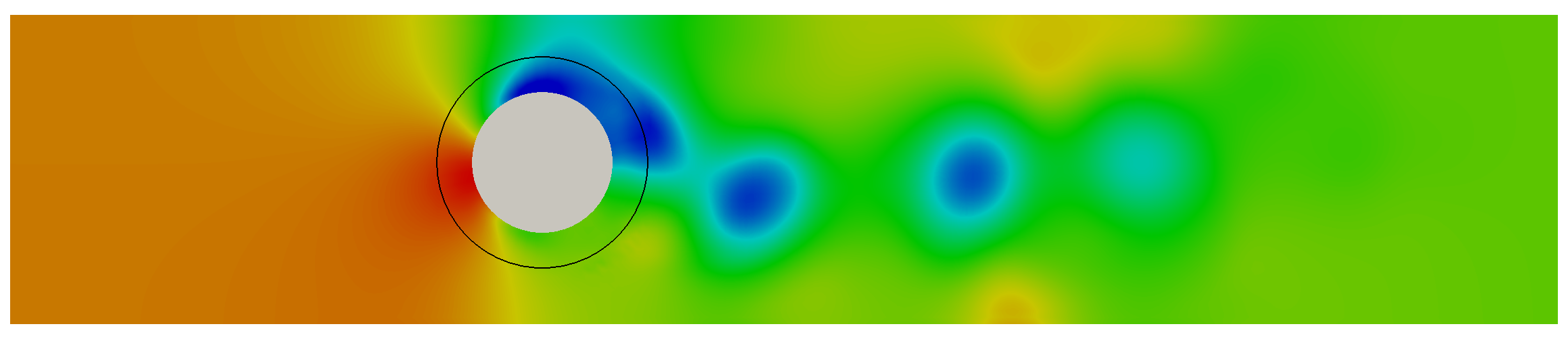}\\
\includegraphics[width=11.0cm]{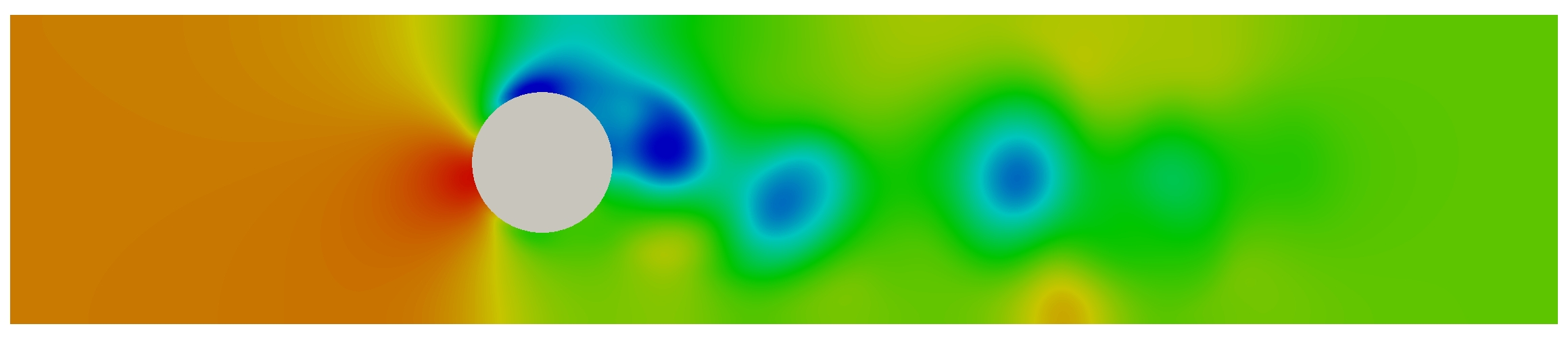}
\end{varwidth}
}
  \caption{Flow over a largely moving cylinder:
comparison of the hybrid Eulerian-\name{ALE} approach (with \mbox{$225\times 45$} background fluid elements
and embedded fluid patch) with the \name{CutFEM} based fixed-grid approach \cite{Schott2017b}[Approach~2] (with \mbox{$450\times 90$} fluid elements) at different times (\protect\subref{fig:moving_cylinder:vel_pres_2_a} and \protect\subref{fig:moving_cylinder:vel_pres_2_b}).
Velocity color scale \mbox{$[0,3]$}, pressure color scale \mbox{$[-5.6,3.5]$}. The top of all paired pictures shows the hybrid Eulerian-\name{ALE} approach.
}
\label{fig:moving_cylinder:vel_pres_2}
\end{figure}

The resulting velocity and pressure solutions are visualized in Figures~\ref{fig:moving_cylinder:vel_pres_1}~and~\ref{fig:moving_cylinder:vel_pres_2}
for different times showing characteristics of the evolution of this highly dynamic flow.
The comparison of the fine resolved fixed-grid approach with our novel hybrid Eulerian-\name{ALE} method
shows good agreement over the entire simulation time and thus validates the algorithmic steps of the fluid domain decomposition method
for largely moving fluid domains.
Within $t\in[0,1.5]$, the resulting profiles are characteristic for the flow around cylinders. 
Due to the low fluid viscosity of \mbox{$\nu=0.001$} and a density of $\rho=1$, a strong non-symmetric flow pattern develops
in the backflow of the cylinder when moving the body in positive $x_1$-direction.
The Reynolds number emerges to a maximum of approximately \mbox{$\RE\approx 300$}
based on a definition using the cylinder diameter~$d$ and the maximum cylinder velocity.
While moving the structure in opposite direction during \mbox{$t\in(1.5,3)$}, highly dynamic time-dependent
forces act on the structural surface. These are the result of the rigid body motion through a swirling flow.

\begin{figure}
\subfloat[$t=0.5$]{
\label{fig:moving_cylinder:boundary_layer_snapshot:t05}
      \begin{varwidth}{\linewidth}
\includegraphics[width=7.0cm]{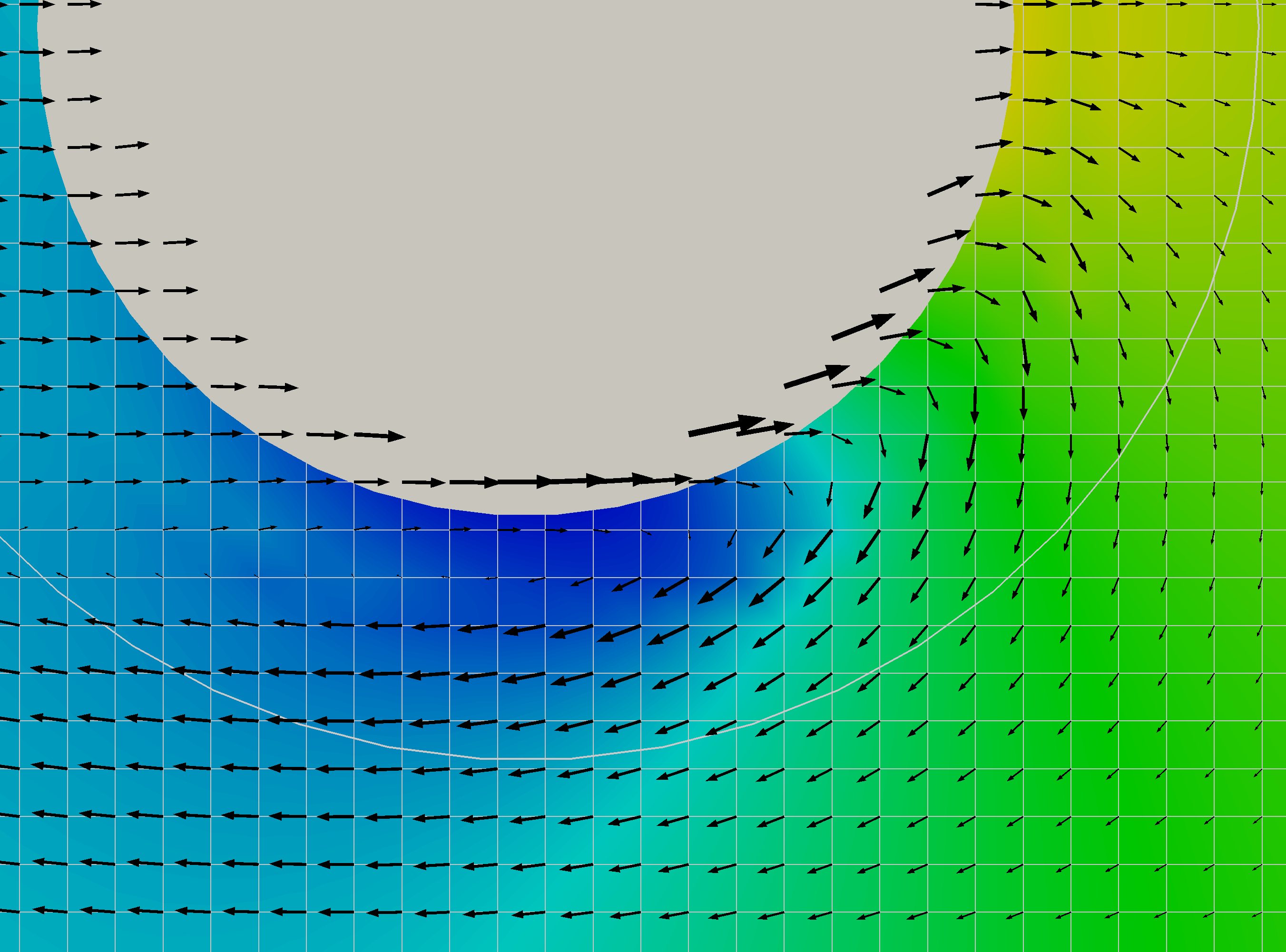} \\
\includegraphics[width=7.0cm]{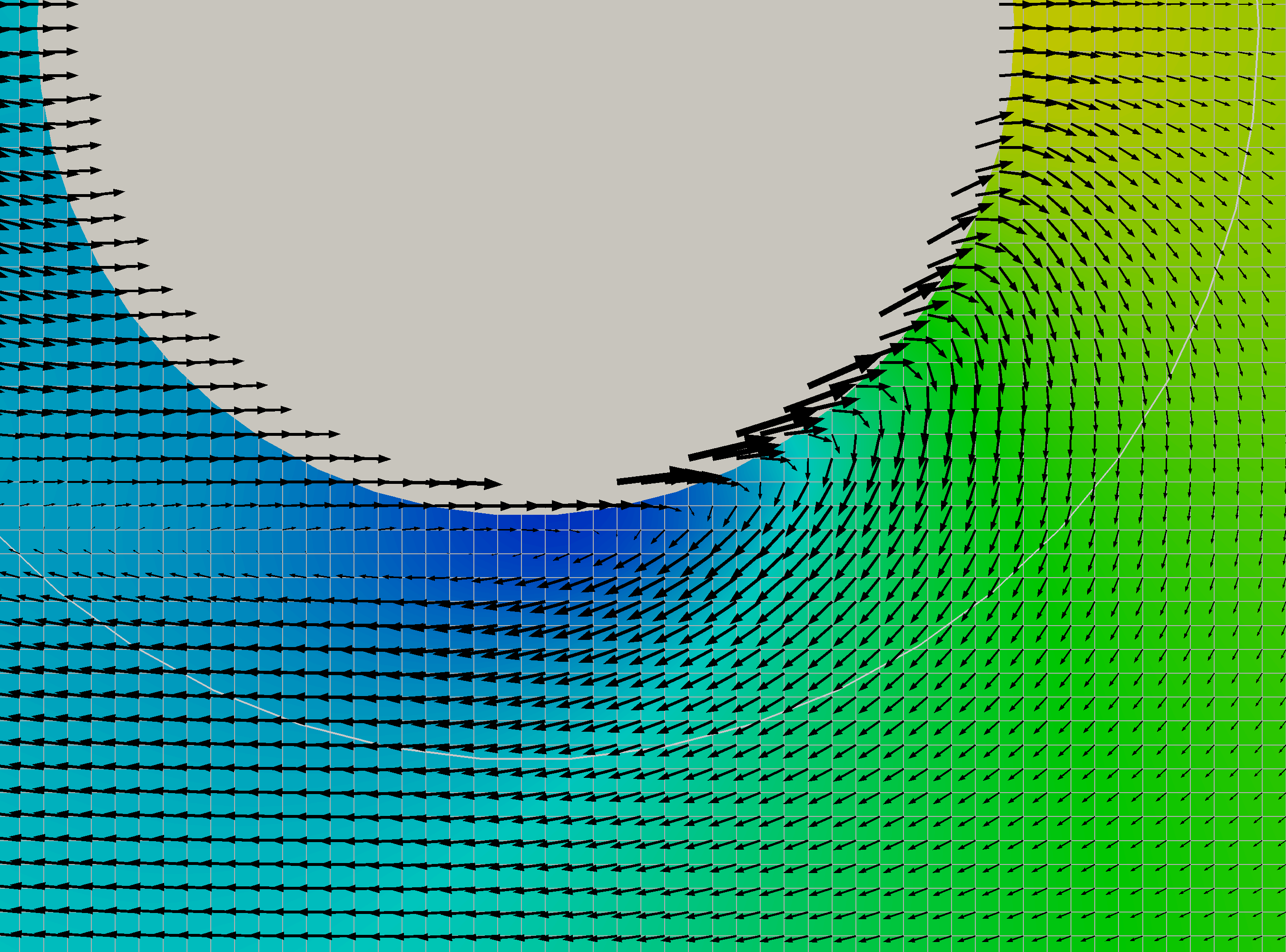}\\
\includegraphics[width=7.0cm]{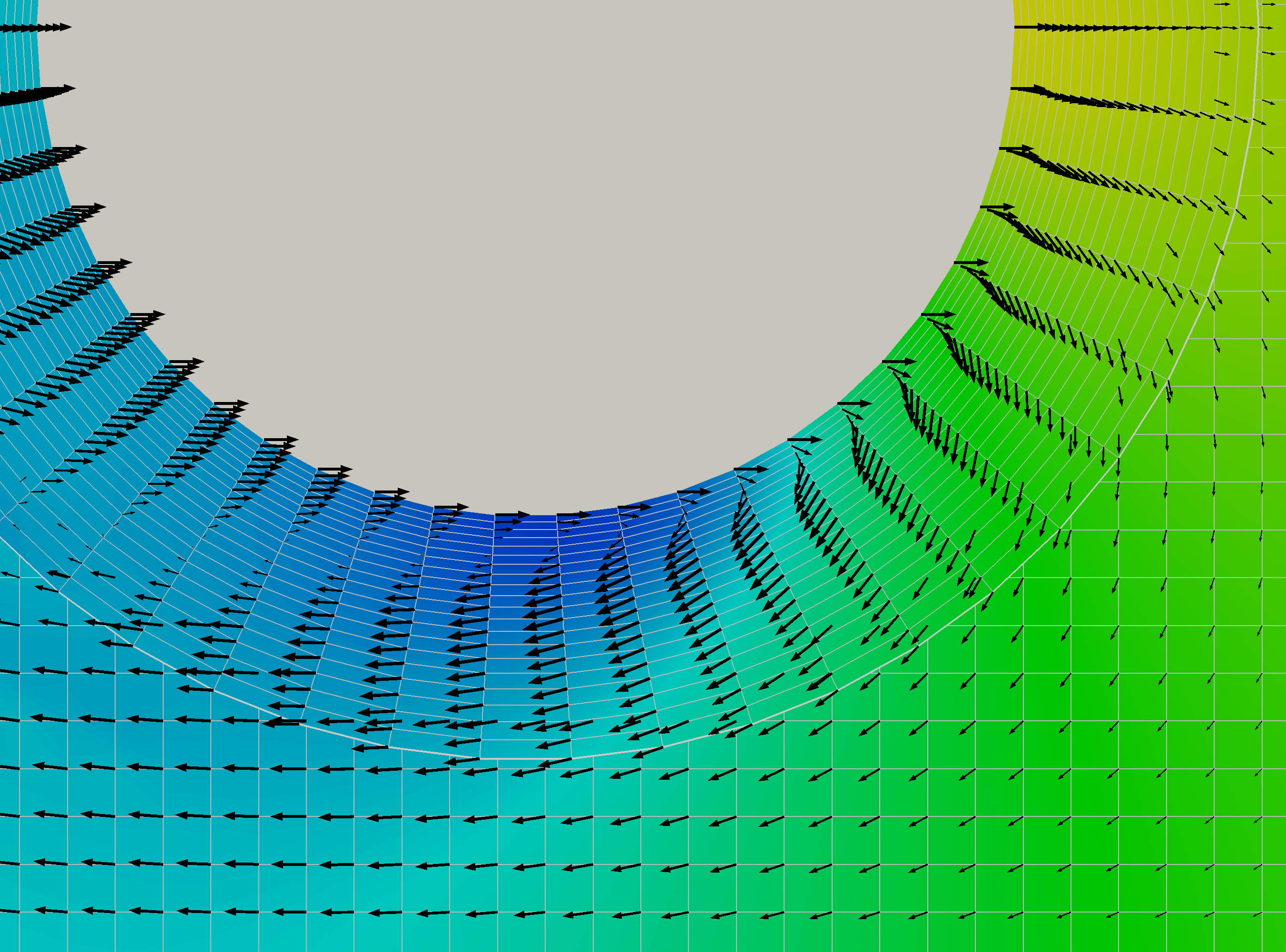}
      \end{varwidth}
}
\subfloat[$t=2.03$]{
\label{fig:moving_cylinder:boundary_layer_snapshot:t203}
      \begin{varwidth}{\linewidth}
\includegraphics[width=7.0cm]{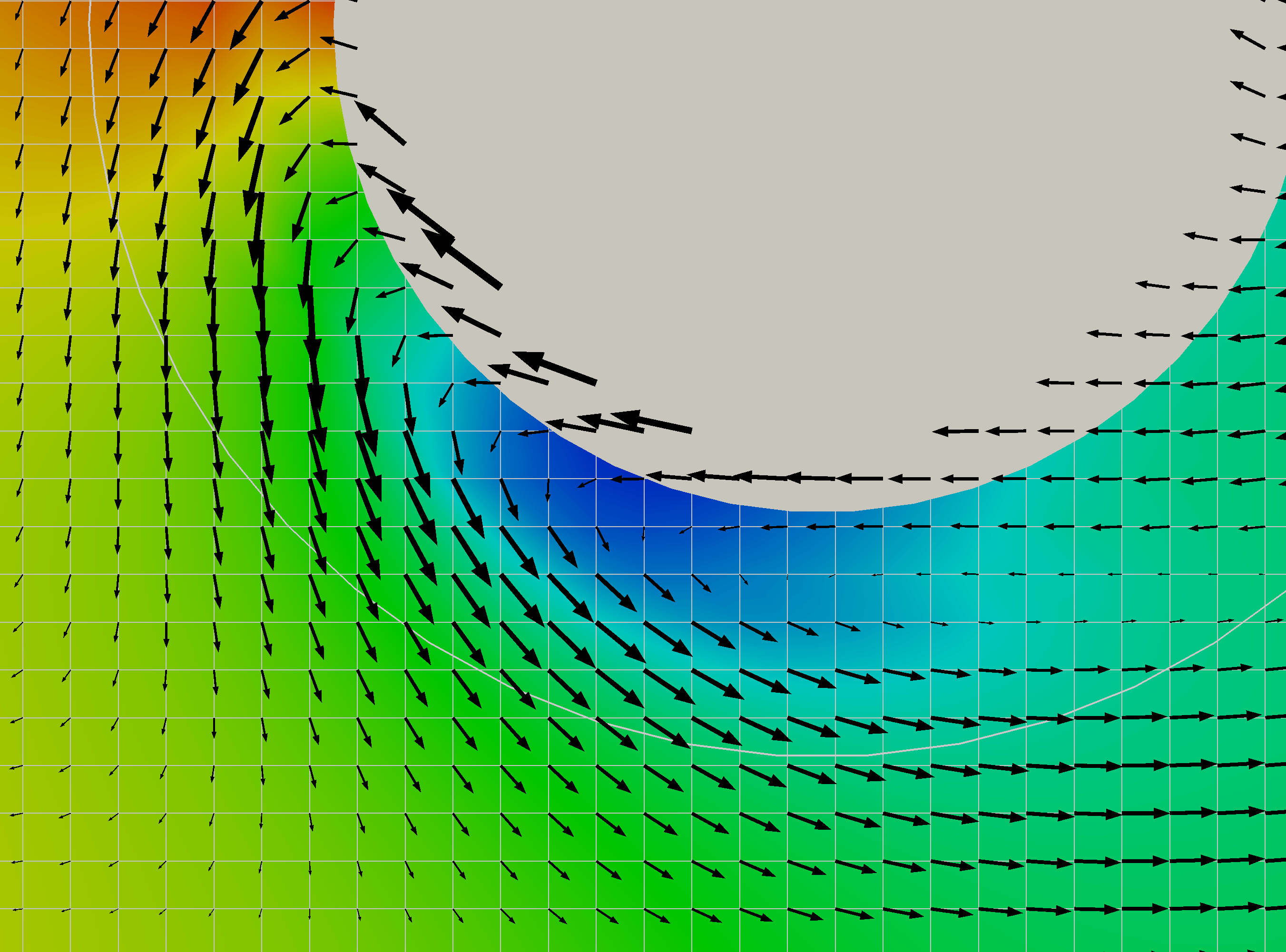}\\
\includegraphics[width=7.0cm]{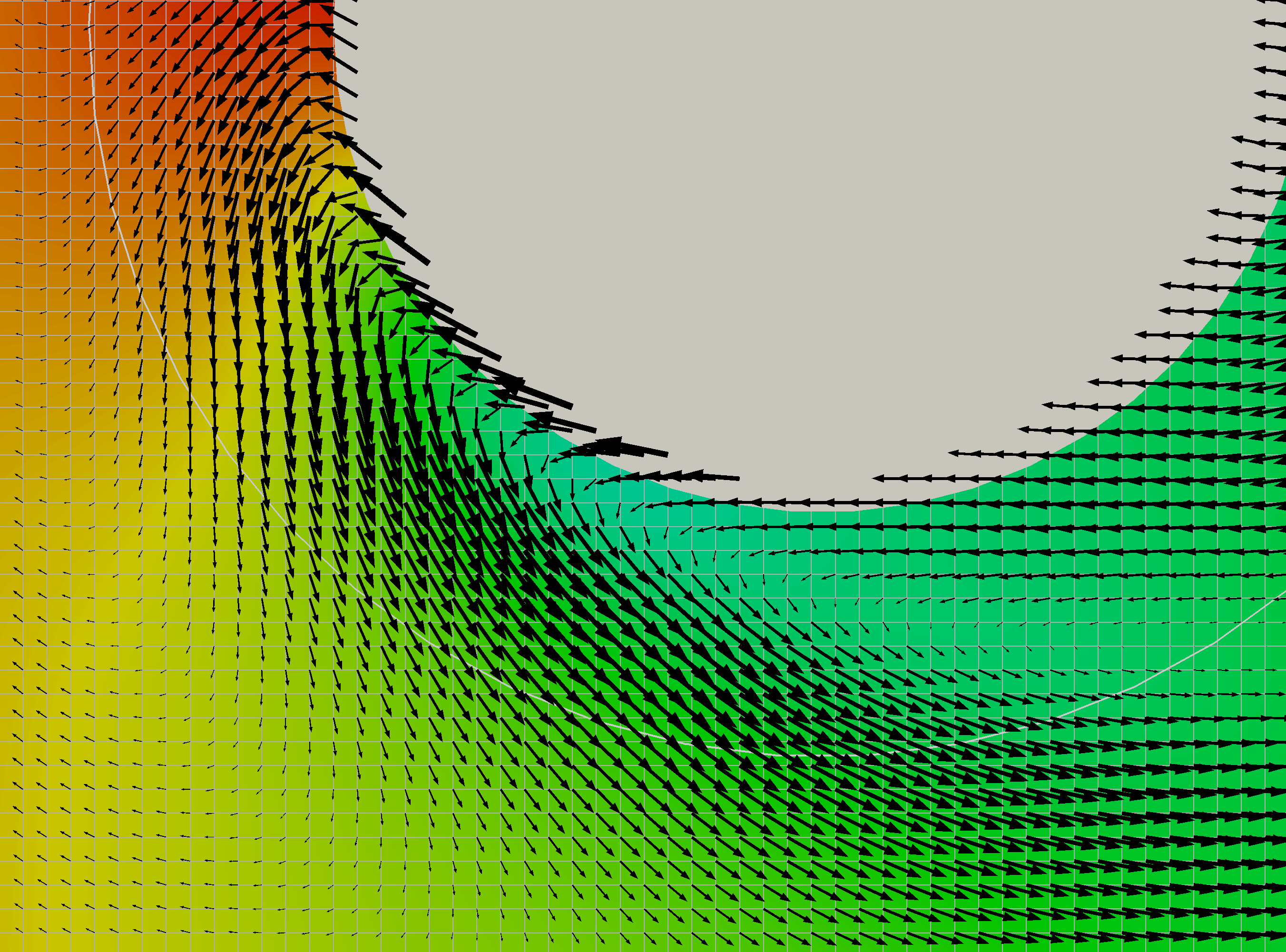}\\
\includegraphics[width=7.0cm]{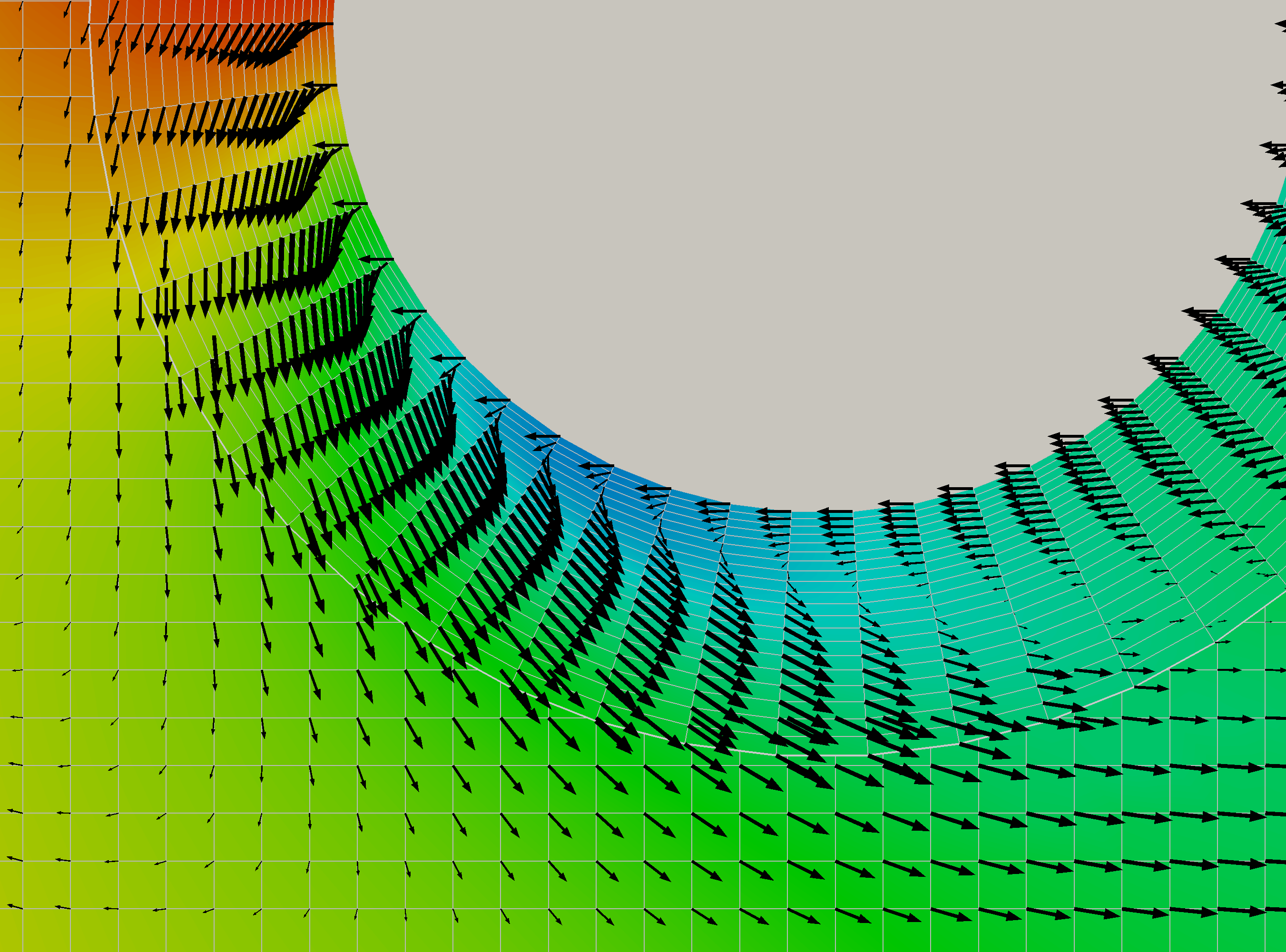}
      \end{varwidth}
}
  \caption{Flow over a largely moving cylinder:
close-up views of the boundary layer region at different times \protect\subref{fig:moving_cylinder:boundary_layer_snapshot:t05}~$t=0.5$ and \protect\subref{fig:moving_cylinder:boundary_layer_snapshot:t203}~$t=2.03$ for different approaches and mesh resolutions (from top to bottom);
coarse fixed-grid approach \mbox{$(45\times225)$} (top), fine fixed-grid approach \mbox{$(90\times450)$} (middle) and
hybrid Eulerian-\name{ALE} approach with coarse fixed background grid \mbox{$(45\times225)$} and a fine resolved moving embedded boundary layer fluid patch (bottom).
Arrows indicate velocity profile and colored fluid domain shows the pressure solution (pressure color scale \mbox{$[-5.6,3.5]$}).
}
  \label{fig:moving_cylinder:boundary_layer_snapshot}
\end{figure}

Close-up views of the interface region in \Figref{fig:moving_cylinder:boundary_layer_snapshot}
visualize how boundary layer effects are resolved by the three different considered approaches.
Steep wall-normal gradients~$\deriv{u_1}{x_2}$ arise during the pull forward phase, as shown at time~$t=0.5$.
At the fluid-solid interface, the kinematic coupling of the structural velocity to the fluid is weakly imposed by the Nitsche technique.
Due to the strong displacement of fluid during the motion,
high $u_1$-velocities in opposite $x_1$-direction arise near the interface.
Due to the low viscosity, these cause strong wall-normal gradients within a thin boundary layer.
This comes also along with a significant pressure drop.
Comparing the coarse fixed-grid method with the equivalently meshed hybrid method,
clearly shows the much more accurately captured flow field in the vicinity of the interface.
The ability of fixed-grid approaches to represent such effects strongly depends on the chosen mesh resolution,
as it is clearly visible from the results of the two fixed-grid variants.
This impressively demonstrates the benefit of the hybrid approach.
The latter method is able to deal with large structural mesh motions similar to the pure fixed-grid approach,
and thereby ensures a suitable mesh resolution over the entire solution time.
Moreover, it allows for a much coarser approximation of the far-field and thus allows to considerably save computational costs.
To demonstrate the capturing of a still more complex flow pattern during the pull back phase,
additional visualizations are provided for time~$t=2.03$.

\begin{figure}
\centering
\includegraphics[height=6.0cm, trim=0cm 0cm 0cm 0cm, clip=true]{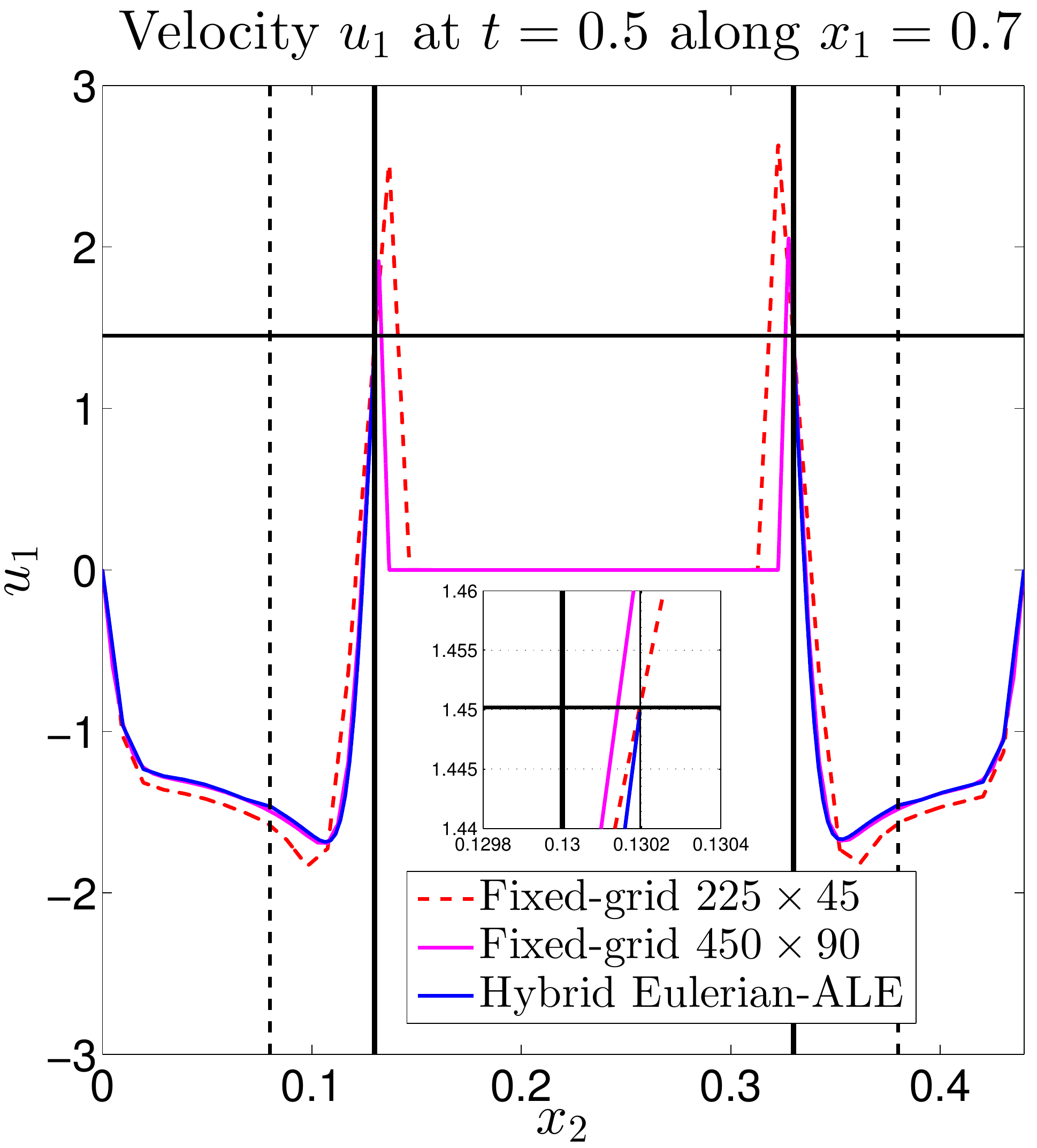}
\qquad
\includegraphics[height=6.0cm, trim=0cm 0cm 0cm 0cm, clip=true]{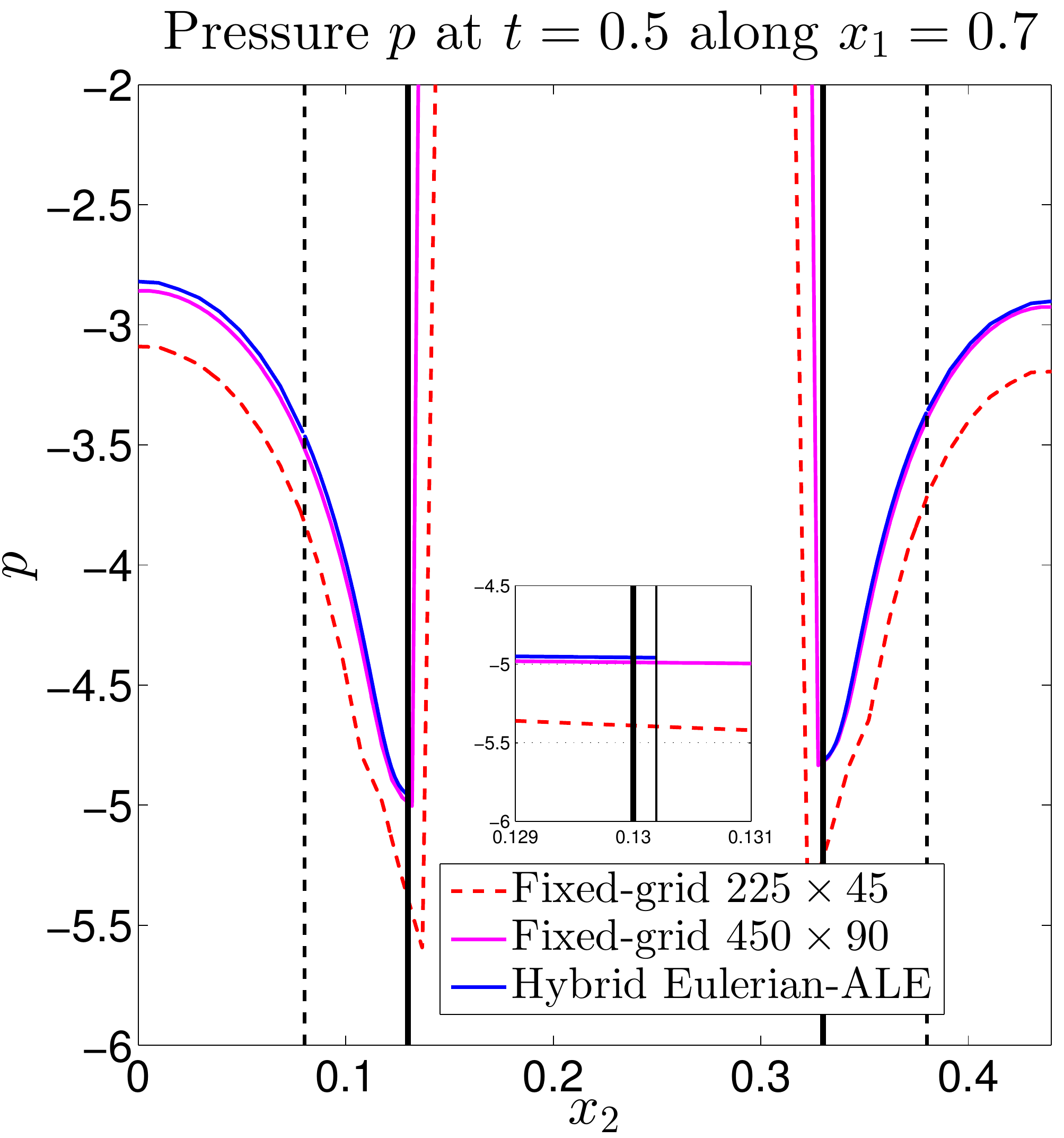}
  \caption{Flow over a largely moving cylinder:
velocity solution~$u_1$ (left) and pressure solution~$p$ (right) along structural vertical centerline \mbox{$x_1=0.7$} at time \mbox{$t=0.5$}.
Horizontal line indicates discrete interfacial velocity \mbox{$u_1 = 1.45016$} to be enforced weakly at the discrete \name{FSI} interface $\Int^{\fd\sd}$ (thick vertical lines).
Dashed vertical lines indicate the artificial fluid-fluid interface~$\Int^{\fd_1\fd_2}$ for the hybrid Eulerian-\name{ALE} method.
Zoom views show the velocity and pressure solution near the bottom-most discrete structural point with coordinates~\mbox{$(x_1,x_2)=(0.7,0.130197)$} (thin vertical line).
}
  \label{fig:moving_cylinder:t05_ux_at_line_x07}
\end{figure}

\begin{figure}
\centering
\includegraphics[height=6.0cm, trim=0cm 0cm 0cm 0cm, clip=true]{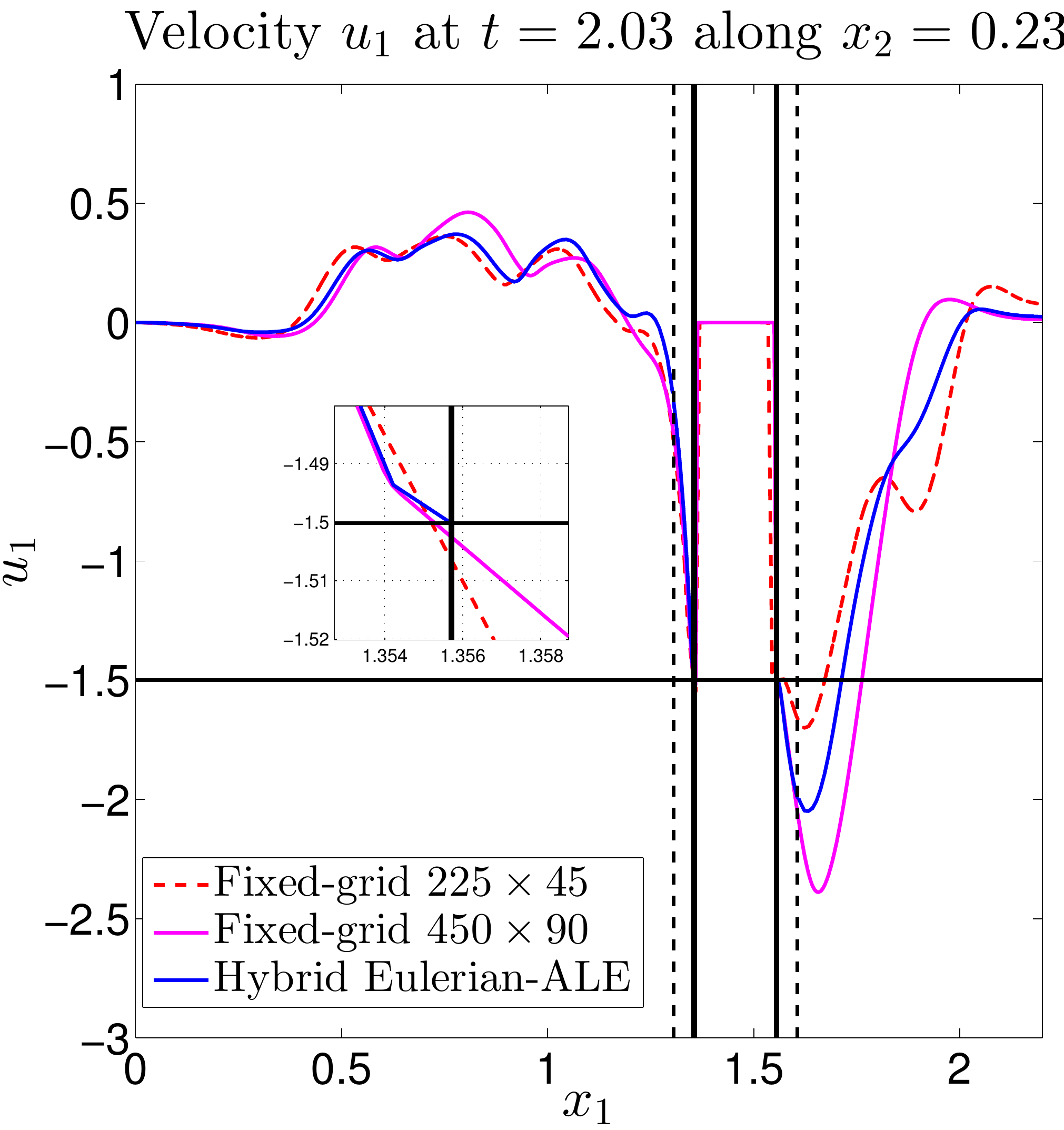}
\qquad
\includegraphics[height=6.0cm, trim=0cm 0cm 0cm 0cm, clip=true]{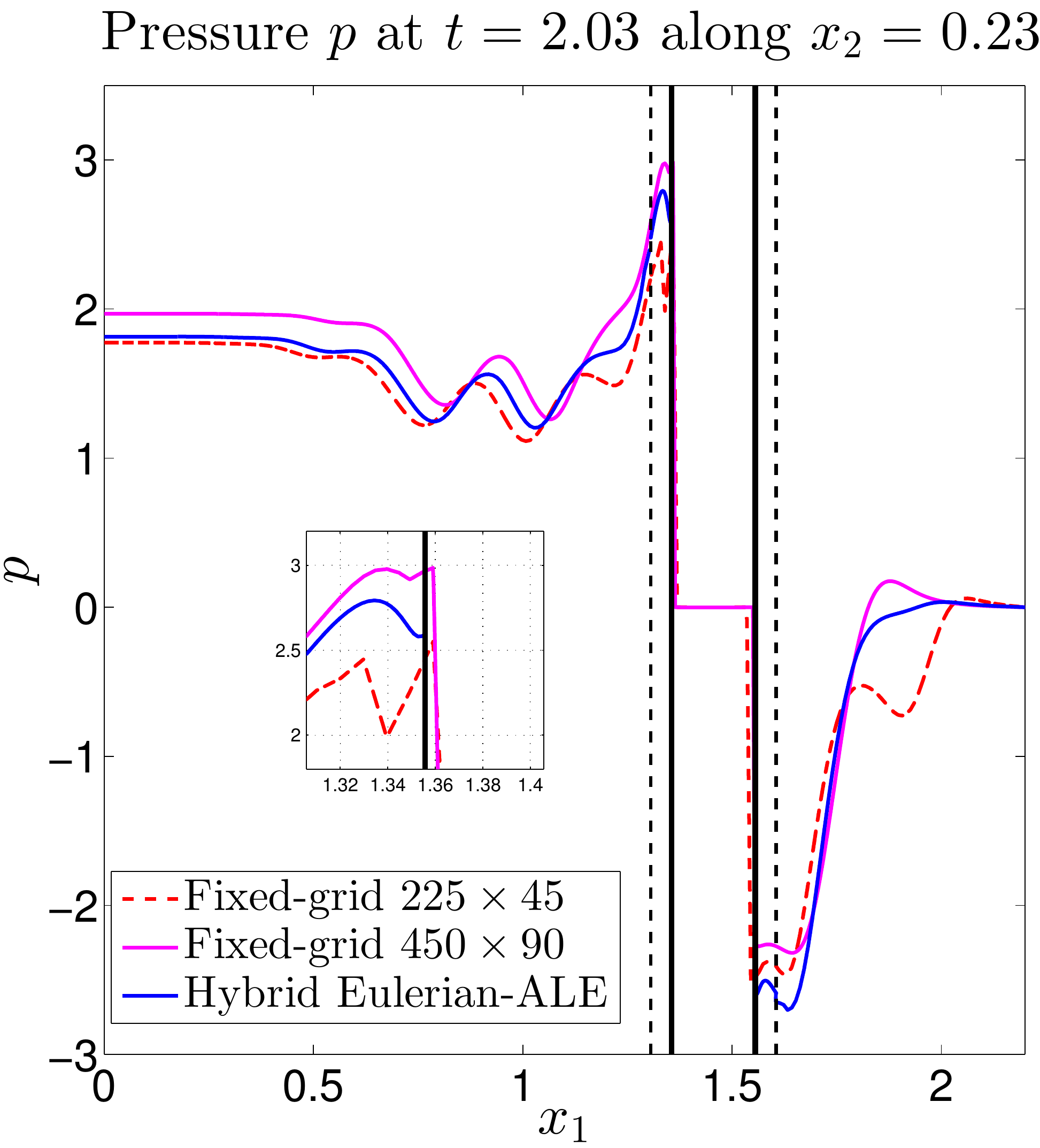}
  \caption{Flow over a largely moving cylinder:
velocity solution~$u_1$ (left) and pressure solution~$p$ (right) along structural horizontal centerline \mbox{$x_2=0.23$} at time \mbox{$t=2.03$}.
Horizontal line indicates discrete interfacial velocity \mbox{$u_1 = -1.50015$} to be enforced weakly at the discrete \name{FSI} interface $\Int^{\fd\sd}$ (thick vertical lines).
Dashed vertical lines indicate the artificial fluid-fluid interface~$\Int^{\fd_1\fd_2}$ for the hybrid Eulerian-\name{ALE} method.
Zoom views show the velocity and pressure solution near the left-most structural point with coordinates~\mbox{$(x_1,x_2)=(1.35571,0.23)$} (thick vertical line).
}
  \label{fig:moving_cylinder:t203_ux_at_line_y023}
\end{figure}

A detailed investigation of the enforcement of the interfacial fluid-structure constraint~\Eqref{eq:fsi_interface_condition_u_jump} and the representation of wall-normal gradients
of the velocity and the pressure solution is presented in \Figref{fig:moving_cylinder:t05_ux_at_line_x07}.
It shows the velocity profile~$u_1$ and the pressure solution along the line \mbox{$x_1=0.7$}, the positioning of the structural vertical centerline at time \mbox{$t=0.5$}.
The discrete solid velocity computed with a backward Euler scheme from the structural displacements,
emerges to \mbox{$u_1 = 1.45016$}.
Since for all approaches identical structural meshes are utilized, the discrete coordinates of the bottom-most structural point is \mbox{$x_2=0.130197$}.
At this coordinate, the velocity is imposed weakly via the Nitsche interface coupling.
The coarse fixed-grid approach clearly lacks accuracy within the bulk and at the interface,
which is obvious from the pressure solution.
Moreover, it is not able to represent the interface condition accurately.
In contrast, a quite good match between the finer resolved fixed-grid method and the hybrid Eulerian-\name{ALE} scheme can be observed for both, velocity and pressure.
Moreover, the continuity of velocity and pressure solution at the artificial fluid-fluid interface
demonstrates the accuracy of the weakly imposed coupling constraints \Eqref{eq:gov_eq_Fluid_Fluid_strong_form_5}--\eqref{eq:gov_eq_Fluid_Fluid_strong_form_6}
for the fluid domain decomposition.
Another comparison along a horizontal line $x_2=0.23$ is provided for a later time step at \mbox{$t=2.03$} for the velocity component $u_1$ and the pressure~$p$,
see \Figref{fig:moving_cylinder:t203_ux_at_line_y023}.

The reduced computational costs at a higher accuracy in the vicinity of the boundary layer makes the hybrid Eulerian-\name{ALE} clearly superior.
This aspect will play a still more decisive role, when increasing the Reynolds number, which comes along with a thinner boundary layer region
and much steeper wall-normal gradients, and also in more complex three-dimensional problems. For the hybrid approach, a finer resolution for fluid patch can be realized easily at a moderate increase of the computational costs.
In contrast, pure fixed-grid approaches would require a mesh refinement in the overall fluid domain, which in comparison yields an enormous loss of
computational efficiency.

\subsection{Vibrating flexible structure}

A further glimpse of the great capabilities and potential advantages of our hybrid \name{FSI} approach for highly dynamic mutual fluid-structure interaction might be also obtained from a rather classical example for \name{ALE} based \name{FSI},
a flow interacting with a vibrating flag-shaped structure.
The setup introduced in \cite{Wall1998} has been extensively used for validating \name{FSI} approaches.

\begin{figure}
\centering
\includegraphics[width=9.0cm]{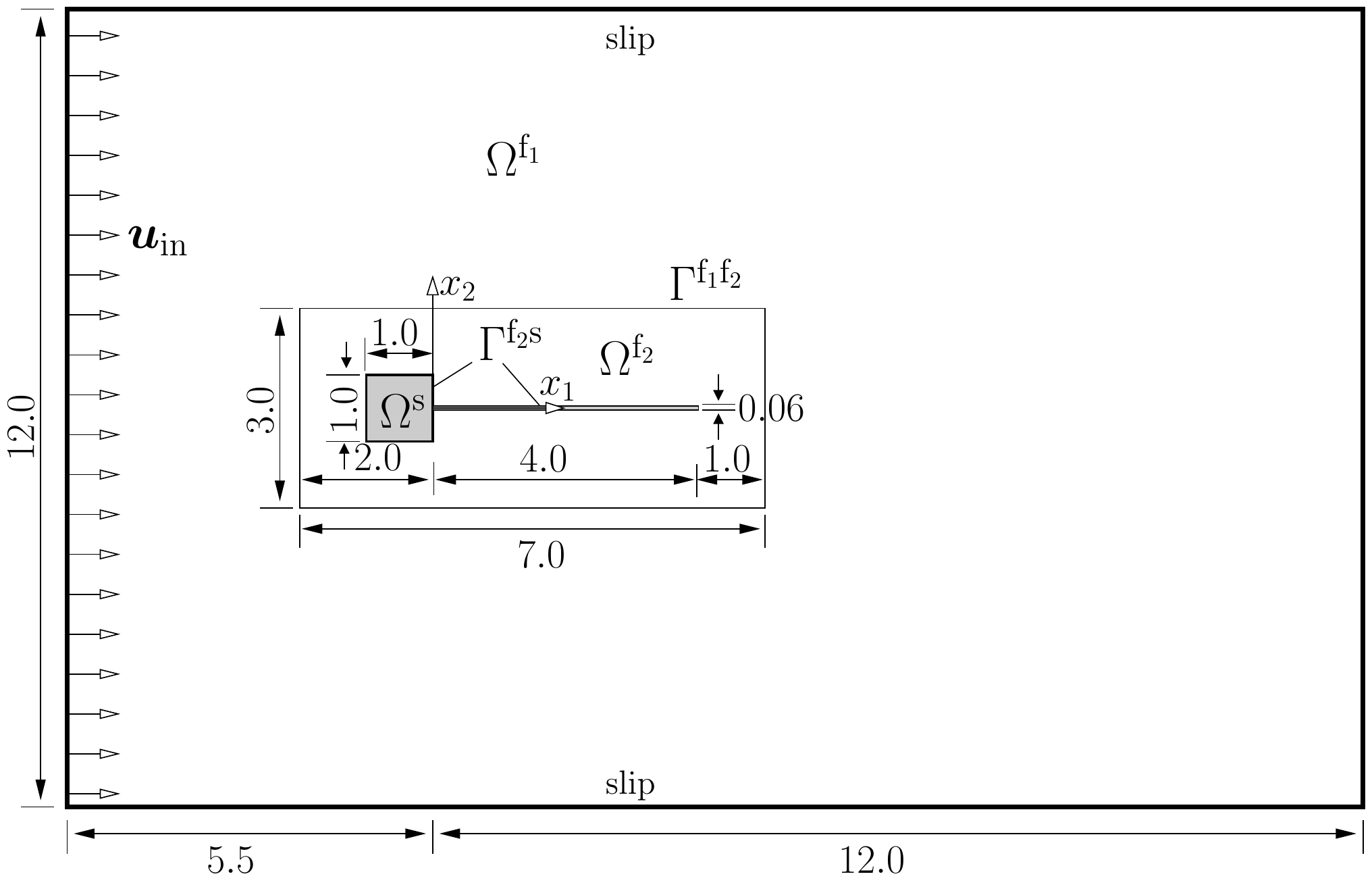}
  \caption{Vibrating flexible structure: geometric setup for hybrid Eulerian-\name{ALE} \name{FSI} approach.
A flexible tail is clamped by a fixed head and is embedded into a surrounding flow. The fluid domain is
artificially decomposed into an inner fluid domain $\Dom^{\fd_2}$ defined by a boundary layer fluid patch~$\mcT_h^{\fd_2}$, which fits to the structural
mesh~$\mcT_h^{\sd}$ at $\Int^{\fd_2 \sd}$.
The patch is embedded into a background mesh~$\mcT_h^{\fd_1}$ and defines the outer template-shaped domain~$\Dom^{\fd_1}$.}
  \label{fig:vibrating_struct:setup}
\end{figure}

\begin{figure}
\centering
\includegraphics[width=6.8cm, trim=0cm 0cm 0cm 0cm, clip=true]{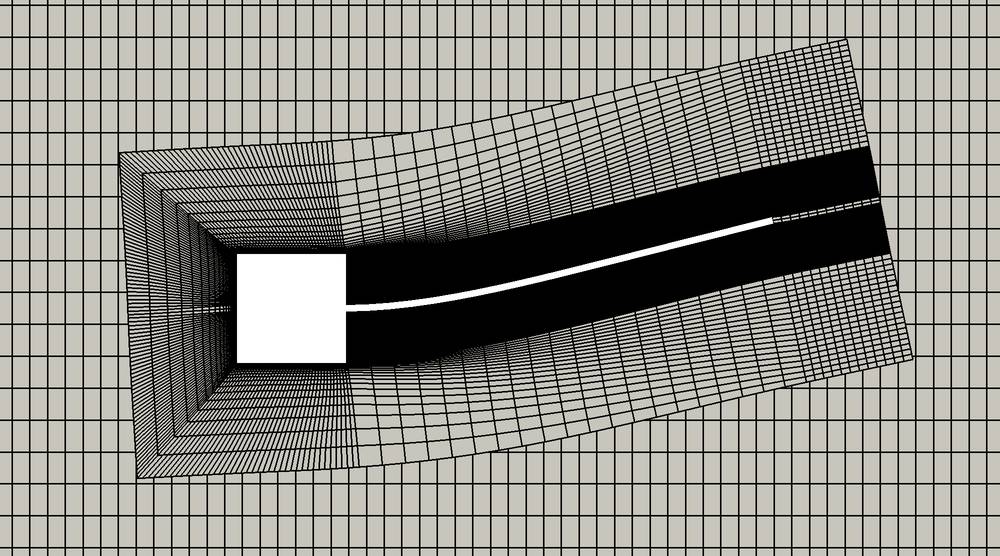}
\quad
\includegraphics[width=6.8cm, trim=0cm 0cm 0cm 0cm, clip=true]{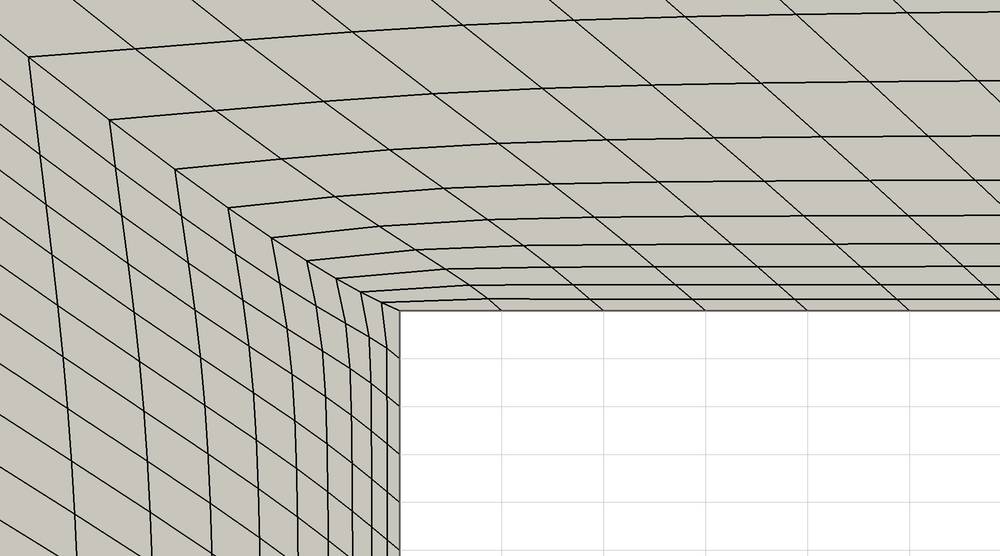}\\
~\\
\includegraphics[width=6.8cm, trim=0cm 0cm 0cm 0cm, clip=true]{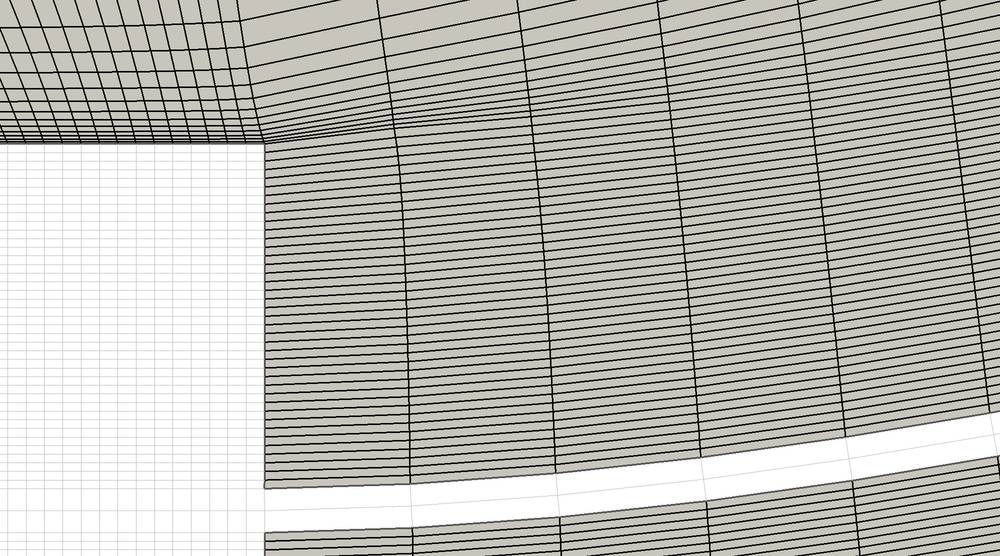}
\quad
\includegraphics[width=6.8cm, trim=0cm 0cm 0cm 0cm, clip=true]{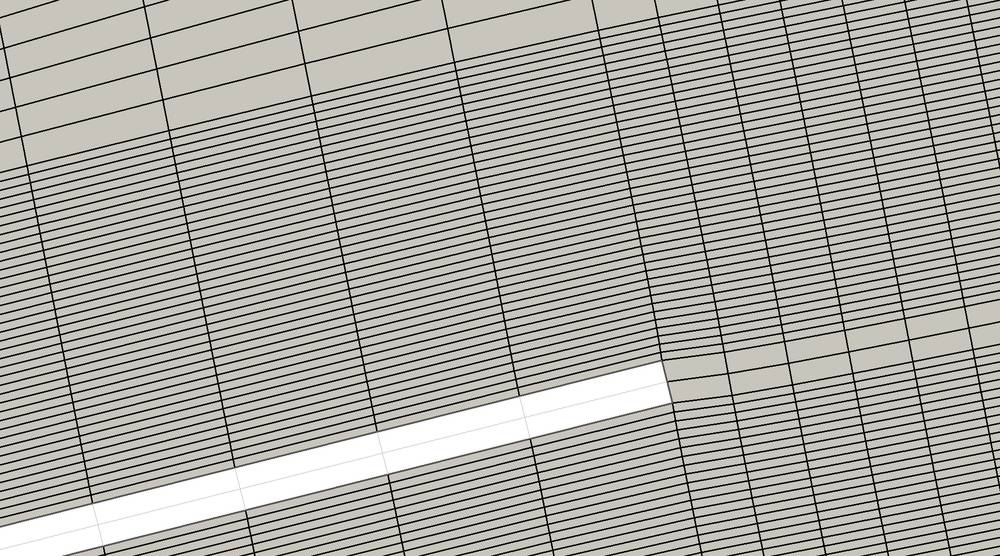}
  \caption{Vibrating flexible structure: close-up views of deformed computational meshes for the hybrid Eulerian-\name{ALE} approach at~\mbox{$t=5.7$}. From left top to right bottom:
solid-matching fluid patch~$\mcT_h^2$ embedded into background fluid grid~$\mcT_h^1$, left upper part around structural head corner, middle part around structural head and tail,
and right end of the flexible tail.}
  \label{fig:vibrating_struct:mesh_snapshot}
\end{figure}

The computational setup is taken from \cite{Wall1998} and is sketched in \Figref{fig:vibrating_struct:setup}.
A flexible structure of dimensions \mbox{$4.0\times 0.06$} is clamped at its front end by 
a square-shaped head of edge-length~$1.0$.
The origin is set to the midpoint of the left end of the tail.
The latter is approximated with \mbox{$20 \times 2$} linearly interpolated finite elements.
While the tail can interact arbitrarily with the surrounding flow, the head is kept fixed over the entire simulation time.
The composed flag is surrounded by a fluid~$\Domf$ whose outer boundaries
define a rectangle of dimensions \mbox{$[-5.5,12]\times[-6,6]$}.
For the hybrid Eulerian-\name{ALE} approach, the fluid domain is decomposed as \mbox{$\Domf=\Dom^{\fd_1}\dot{\cup}\Dom^{\fd_2}$}.
The domain~$\Dom^{\fd_2}$ is defined by a fluid patch~$\mcT_h^{\fd_2}$, which fits to the fluid-solid interface~$\Int^{\fd_2\sd}$.
Its outer dimensions are specified as \mbox{$[-2,5]\times[-1.5,1.5]$} and
enclose kind of boundary layer elements, which are refined towards the interface~$\Int^{\fd_2\sd}$.
Close-up views of the embedded fluid patch surrounding the body are visualized in \Figref{fig:vibrating_struct:mesh_snapshot}.
The background mesh~$\mcT_h^{\fd_1}$ contains \mbox{$120 \times 41$} linearly interpolated quadrilateral elements
covering the entire rectangular domain~$\Dom$.
The patch~$\mcT_h^{\fd_2}$ is embedded into $\mcT_h^{\fd_1}$ in a geometrically unfitted way
and intersects its elements that are located next to the interface~$\Int^{\fd_1 \fd_2}$,
and so defines the time-dependent active part of~$\mcT_h^{\fd_1}$.

Along the fixed head, the fluid-structure interaction reduces to a no-slip Dirichlet boundary condition for the fluid,
which is enforced weakly via the proposed Nitsche method.
At the remaining fluid-solid interface of the flexible tail and at the fluid-fluid interface, coupling constraints
\Eqref{eq:fsi_interface_condition_u_jump}--\eqref{eq:fsi_interface_condition_traction_jump} and \Eqref{eq:gov_eq_Fluid_Fluid_strong_form_5}--\eqref{eq:gov_eq_Fluid_Fluid_strong_form_6} are enforced, respectively.
At the inlet \mbox{$x_1=-5.5$}, a velocity of \mbox{$u_1^{\mathrm{max}}=51.3$} is initially ramped up
by a time curve factor \mbox{$g(t)=\frac{1}{2}(1-\cos(\pi t/0.1))$} within \mbox{$t\in [0,0.1]$}
and kept constant afterwards as \mbox{$\bfu_{\mathrm{in}}=(u_1^{\mathrm{max}},0)$}.
At the boarders perpendicular to the inlet, \ie~\mbox{$x_2=\pm 6.0$}, slip-conditions prevent the flow to escape and a zero-traction Neumann boundary condition \mbox{$\bfhN=\bfzero$}
is enforced at \mbox{$x_1=12.0$}.
A slight imperfection of the problem setup, given in terms of a small shift of the structure in positive $x_2$-direction by $\epsilon=10^{-3}$,
causes vortices in the backflow of the structure to detach slightly non-symmetric. These excite the structure to periodically vibrate
which in return cause complex vortex shedding near the flexible tail.
To accurately capture this behavior, a fine mesh resolution near the fluid-solid interface is required
for which our hybrid Eulerian-\name{ALE} \name{FSI} approach is perfectly suited and comes to its full extent.
Attempts based on \name{ALE} based approaches often suffer from the small elements next to the tip of the tail collapsing
once the structural deformations are getting larger.
The materials are chosen as follows: for the fluid, it is set \mbox{$\mu^\fd=1.82\cdot10^{-4}$} and
\mbox{$\rho^\fd=1.18\cdot 10^{-3}$} resulting in an approximate Reynolds number of $\RE\approx 333$ based on the structural head dimension.
The structural material properties are \mbox{$\rho^\sd=2.0$}, \mbox{$\nu^\sd=0.35$} and \mbox{$E^{\sd}= 2.0 \cdot 10^{-6}$}.
For the temporal discretization of the fluid it is chosen \mbox{$\theta=0.55$} with a time-step length of \mbox{$\Delta t = 0.001$}.

\begin{figure}
\centering
\subfloat[Displacements $d_2$ over time $t$.]{
\label{fig:vibrating_struct:displacements2}
      \begin{varwidth}{\linewidth}
\includegraphics[height=6.0cm, trim=0cm 0cm 0cm 0cm, clip=true]{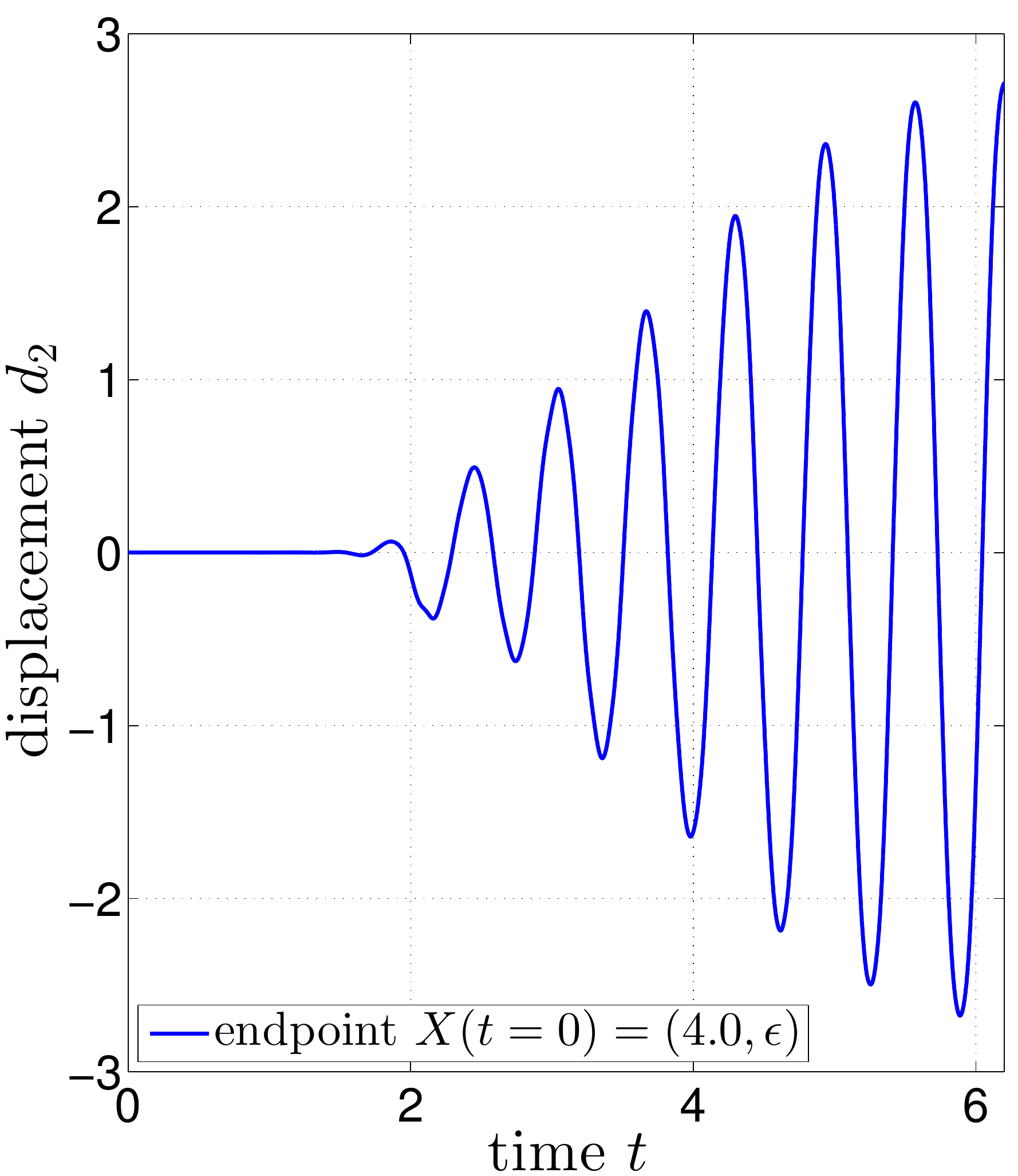}
      \end{varwidth}
}
\qquad
\subfloat[Displacements $d_1$ over time $t$.]{
\label{fig:vibrating_struct:displacements1}
      \begin{varwidth}{\linewidth}
\includegraphics[height=6.0cm, trim=0cm 0cm 0cm 0cm, clip=true]{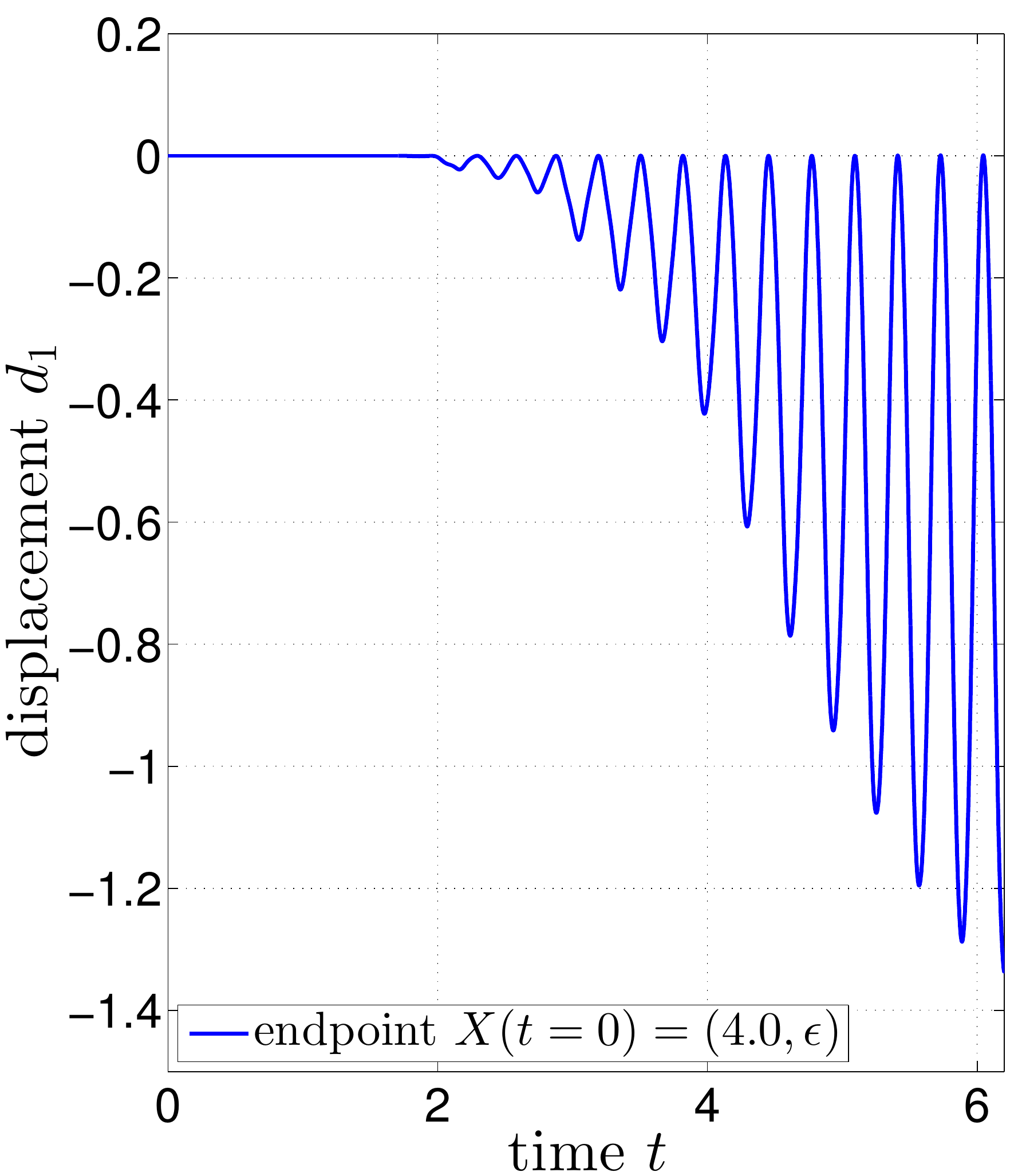}
      \end{varwidth}
}
  \caption{Vibrating flexible structure: history of displacements $d_2$ \protect\subref{fig:vibrating_struct:displacements2}
and $d_1$ \protect\subref{fig:vibrating_struct:displacements1} for the right tip of the tail.
}
  \label{fig:vibrating_struct:displacements}
\end{figure}

Flow entering the setup at the inlet drives the fluid-solid interaction.
Fluid streams around the structural head at which no-slip boundary conditions are imposed weakly.
At its sides, strong boundary layers arise which are accurately captured by the surrounding patch.
For this purpose, the fluid elements are refined strongly towards the walls.
Due to the introduced imperfection, a non-symmetric flow develops and cause vortices behind the structural steps to detach temporally shifted.
As a result, the tail is excited to slightly deform which further induces the creating of swirls that later will detach.
The oscillation amplitude of the flexible tail grows such that the flag starts to highly dynamically vibrate at a certain frequency.
The history of displacements at the right tip of the tail is shown in \Figref{fig:vibrating_struct:displacements}.

\begin{figure}
\centering
\subfloat[Velocity norm~$\euclidian{\vel_h}$ (left) and pressure~$p_h$ (right) at time \mbox{$t=5.7$}.]{
\label{fig:vibrating_struct:vel_pres:t5700}
      \begin{varwidth}{\linewidth}
\includegraphics[width=7.0cm]{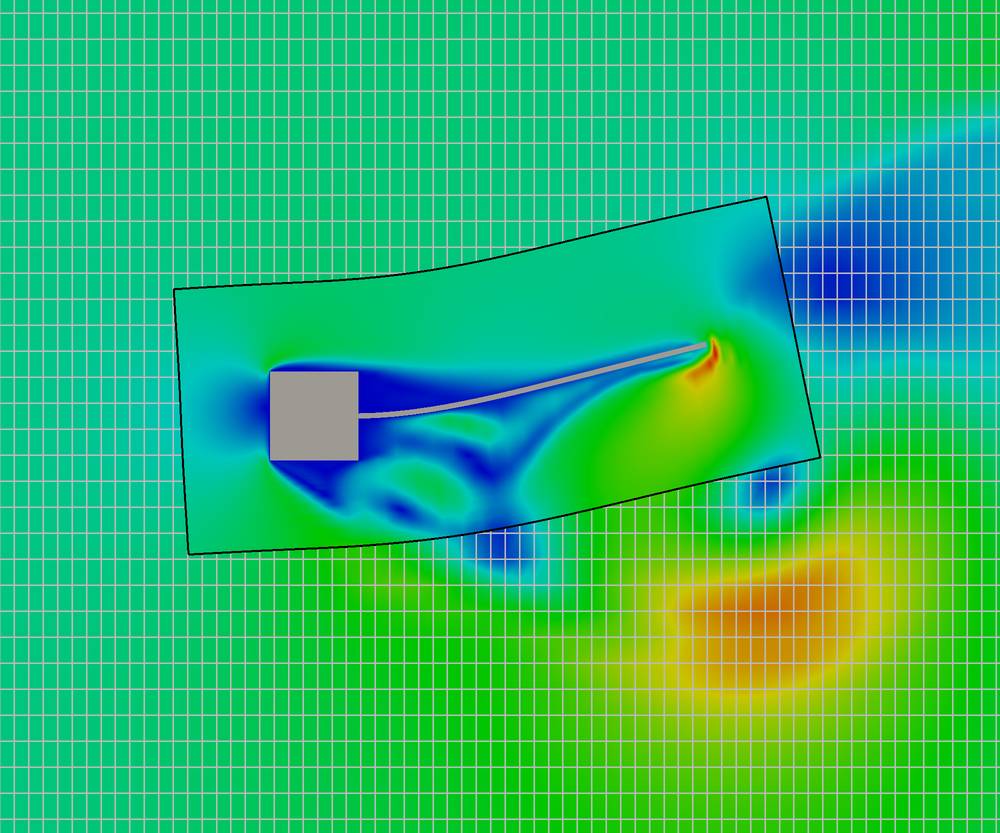}
\includegraphics[width=7.0cm]{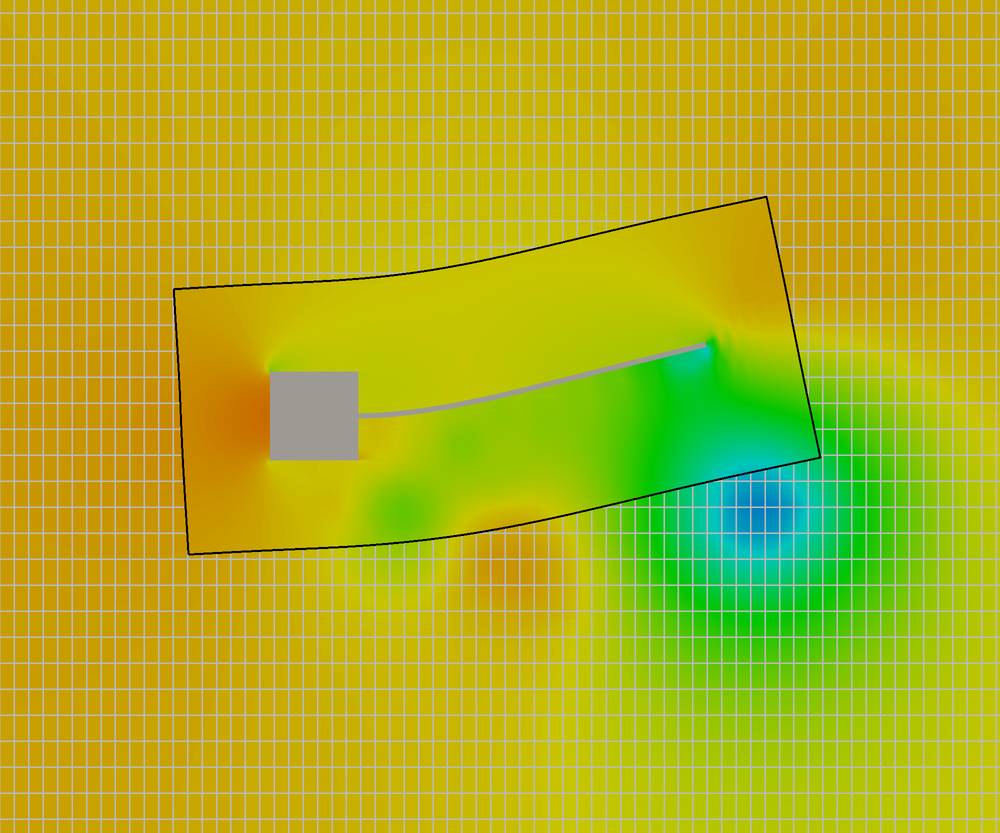}
      \end{varwidth}
}
\\
\subfloat[Velocity norm~$\euclidian{\vel_h}$ (left) and pressure~$p_h$ (right) at time \mbox{$t=5.88$}.]{
\label{fig:vibrating_struct:vel_pres:t5880}
      \begin{varwidth}{\linewidth}
\includegraphics[width=7.0cm]{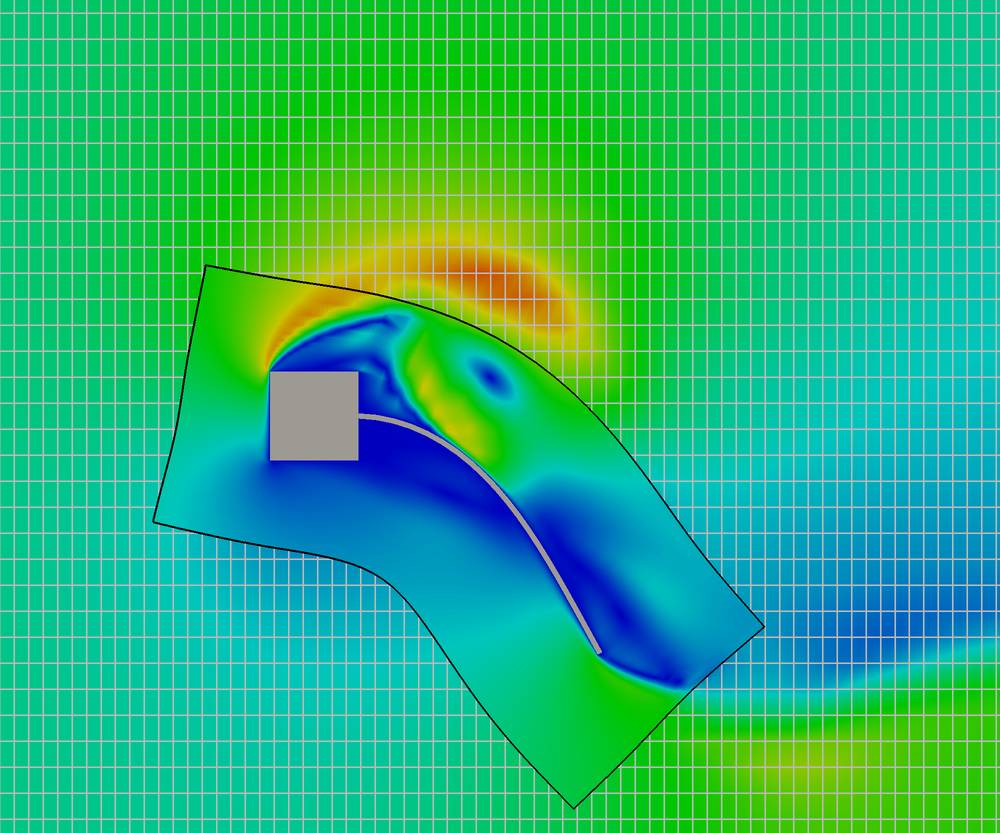}
\includegraphics[width=7.0cm]{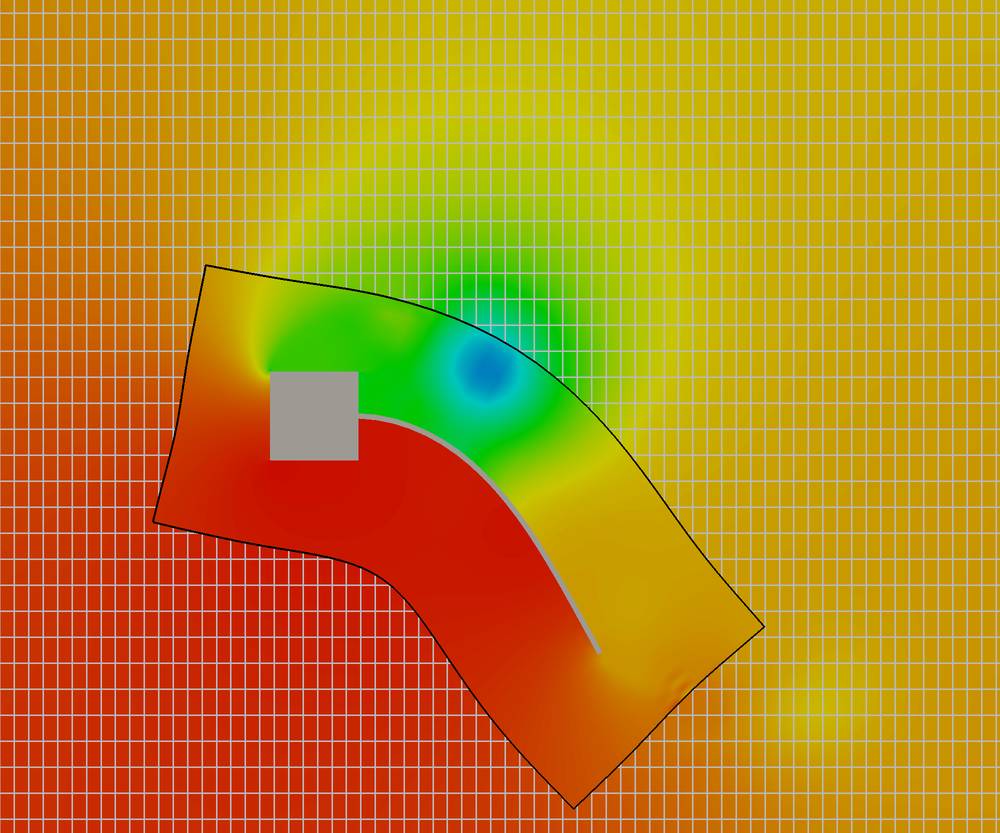}
      \end{varwidth}
}
\\
\subfloat[Velocity norm~$\euclidian{\vel_h}$ (left) and pressure~$p_h$ (right) at time \mbox{$t=6.2$}.]{
\label{fig:vibrating_struct:vel_pres:t6200}
      \begin{varwidth}{\linewidth}
\includegraphics[width=7.0cm]{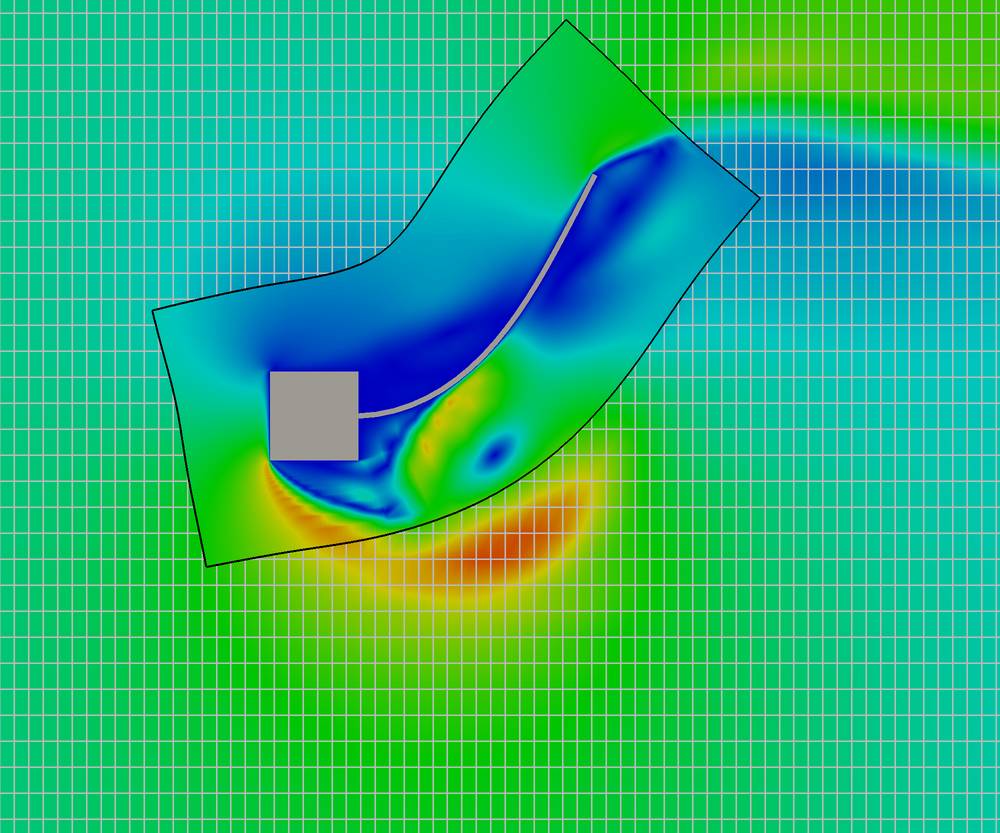}
\includegraphics[width=7.0cm]{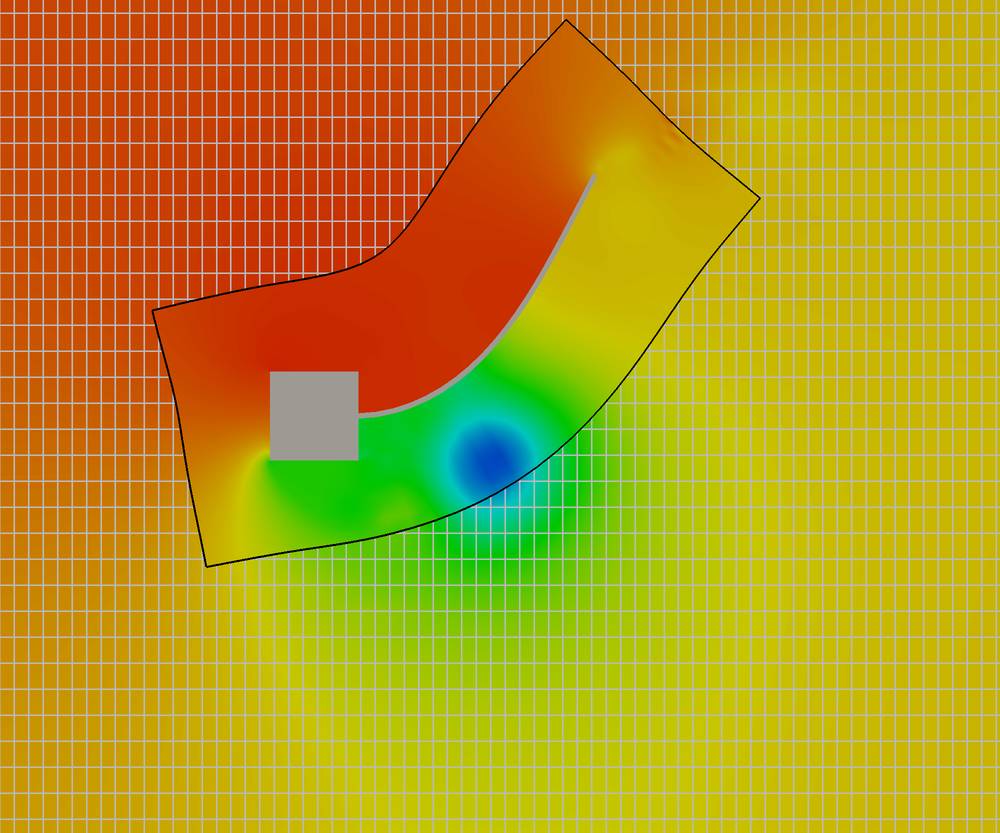}
      \end{varwidth}
}
  \caption{Vibrating flexible structure: hybrid Eulerian-\name{ALE} approach (with \mbox{$120\times 41$} background fluid elements
and a wall-refined embedded fluid patch) at different times (\protect\subref{fig:vibrating_struct:vel_pres:t5700},
\protect\subref{fig:vibrating_struct:vel_pres:t5880} and \protect\subref{fig:vibrating_struct:vel_pres:t6200}).
Velocity color scale \mbox{$[0,155]$}, pressure color scale \mbox{$[-18.0,7.5]$}.
}
  \label{fig:vibrating_struct:vel_pres}
\end{figure}

Snapshots of the simulation results at three times, which characterize different stages of the vibration evolution,
visualize the velocity magnitude and pressure solution and the deforming fluid patch
in \Figref{fig:vibrating_struct:vel_pres}.
As a great advantage of this hybrid Eulerian-\name{ALE} discretization concept, on the one hand,
large structural deformations and motions can be dealt with,
since arbitrary positions of the structural mesh~$\mcT_h^{\sd}$ and its surrounding fluid patch~$\mcT_h^{\fd_2}$
within the background grid $\mcT_h^{\fd_1}$ are allowed.
On the other hand, decomposing the fluid domain enables to utilize highly refined meshes in specific regions of interest,
without risking the collapse of these refined elements and keeping the computational costs at a minimum.
Moreover, providing appropriate boundary layer patches for the hybrid approach is a much easier task than
meshing the entire \name{FSI} setting in a classical \name{ALE} fashion.
The accuracy of the sharp \name{CutFEM} based interface-coupling of the two overlapping fluid meshes
is demonstrated by the continuity of the solution fields, even in the case when complex flow patterns develop in this region
and vortices are convected across~$\Int^{\fd_1 \fd_2}$.

\begin{figure}
\centering
\subfloat[$t=5.7$]{
\label{fig:vibrating_struct:boundary_layer_snapshot:t5700}
      \begin{varwidth}{\linewidth}
\includegraphics[width=7.0cm]{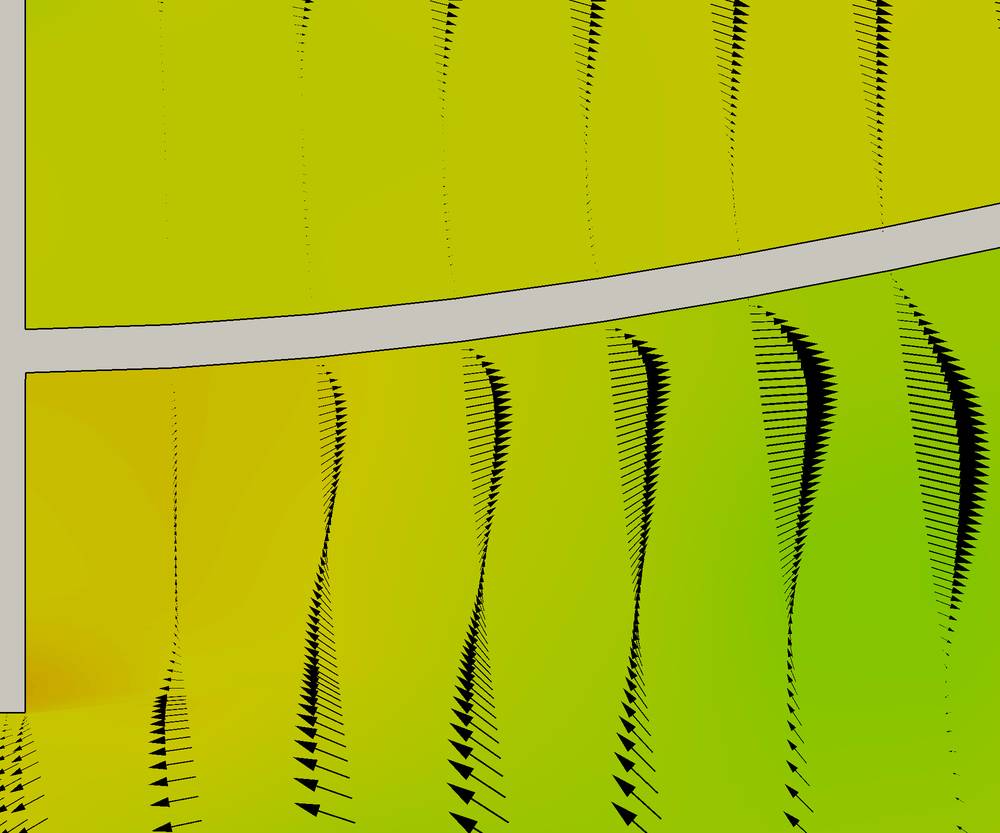}
\quad
\includegraphics[width=7.0cm]{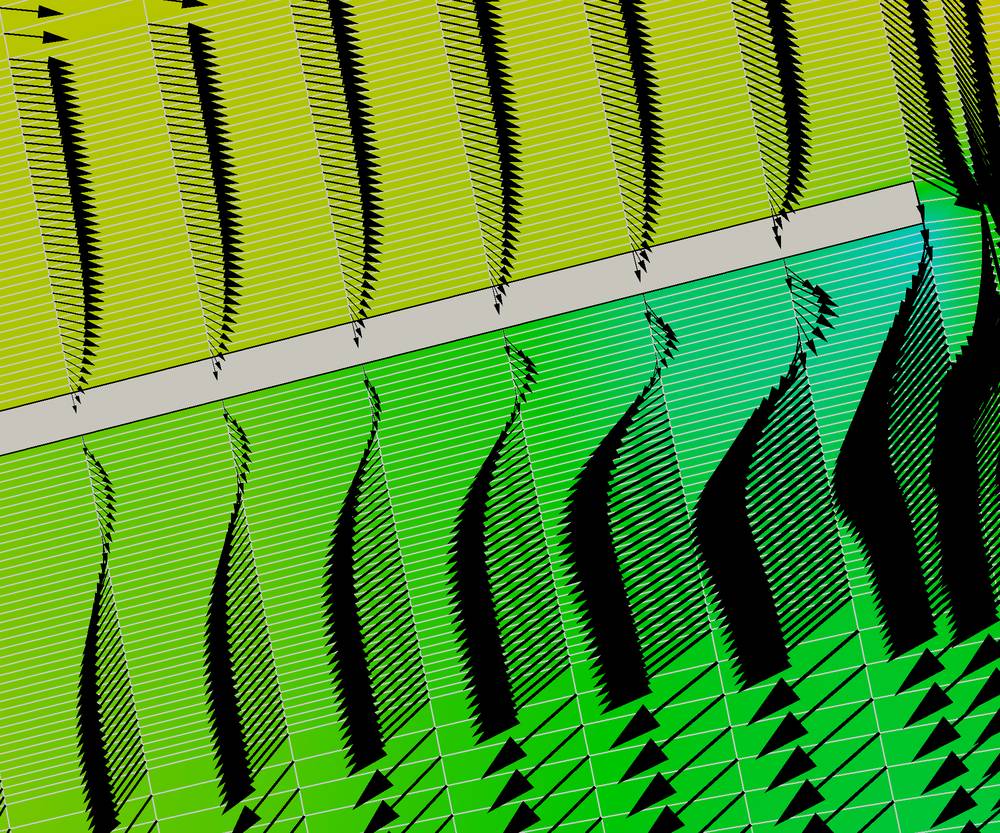}
      \end{varwidth}
}
\\
\subfloat[$t=5.88$]{
\label{fig:vibrating_struct:boundary_layer_snapshot:t5880}
      \begin{varwidth}{\linewidth}
\includegraphics[width=7.0cm]{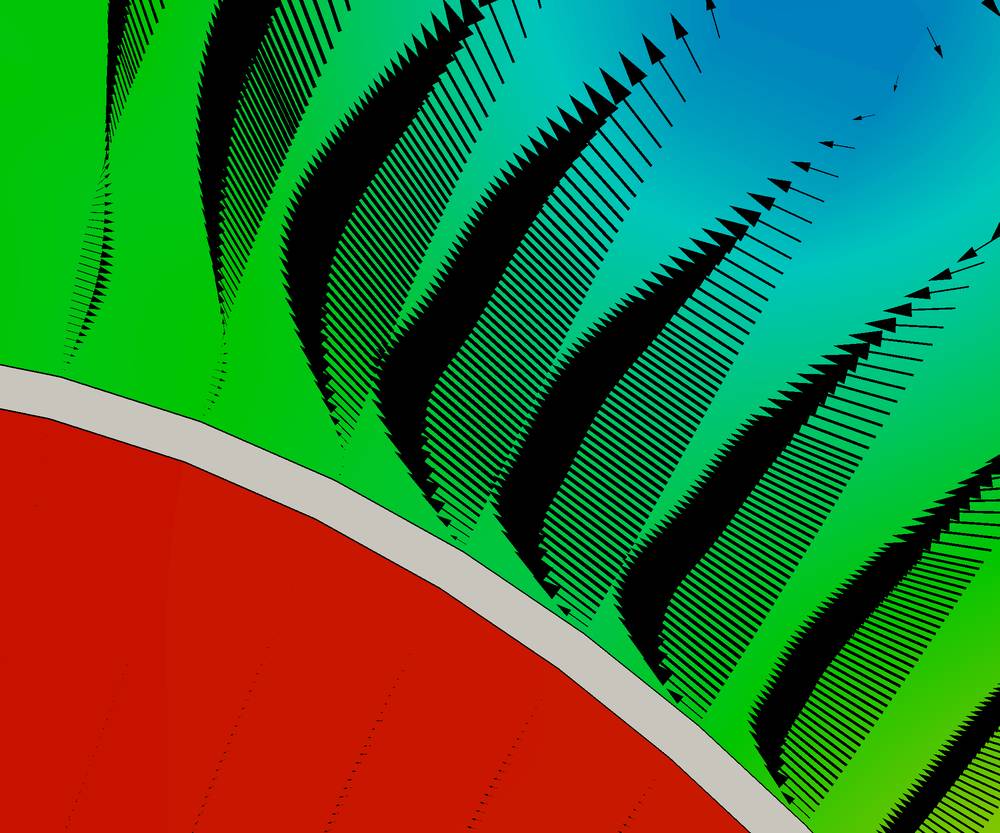}
\quad
\includegraphics[width=7.0cm]{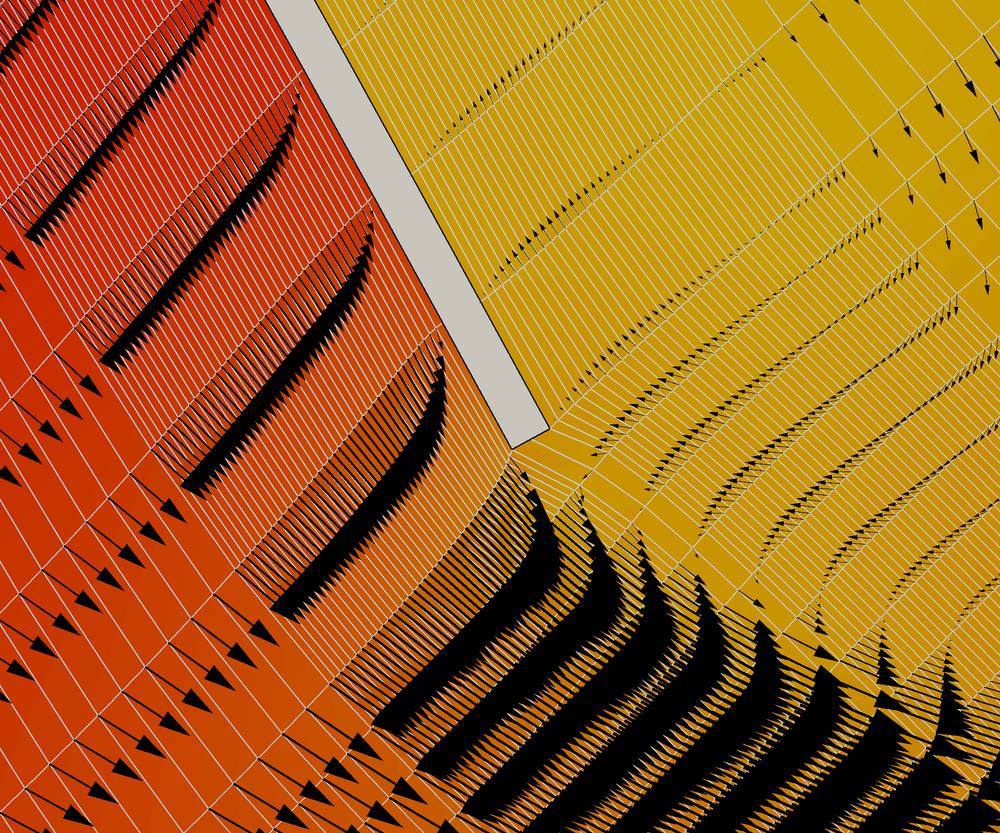}
      \end{varwidth}
}
\\
\subfloat[$t=6.2$]{
\label{fig:vibrating_struct:boundary_layer_snapshot:t6200}
      \begin{varwidth}{\linewidth}
\includegraphics[width=7.0cm]{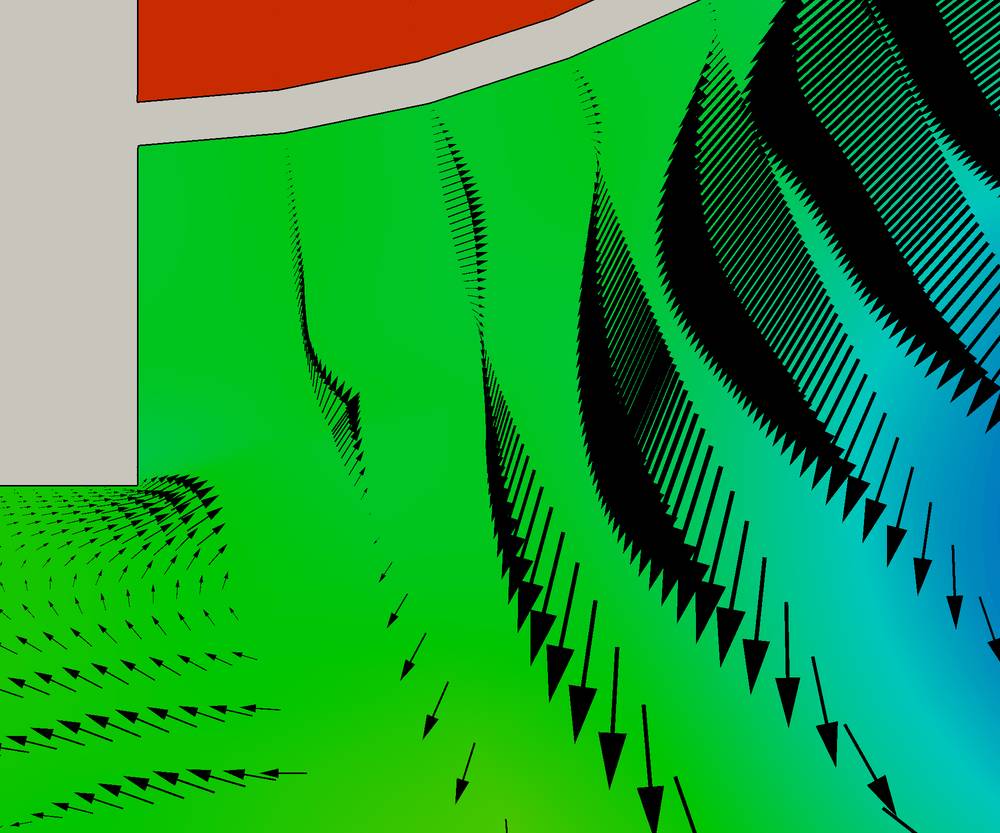}
\quad
\includegraphics[width=7.0cm]{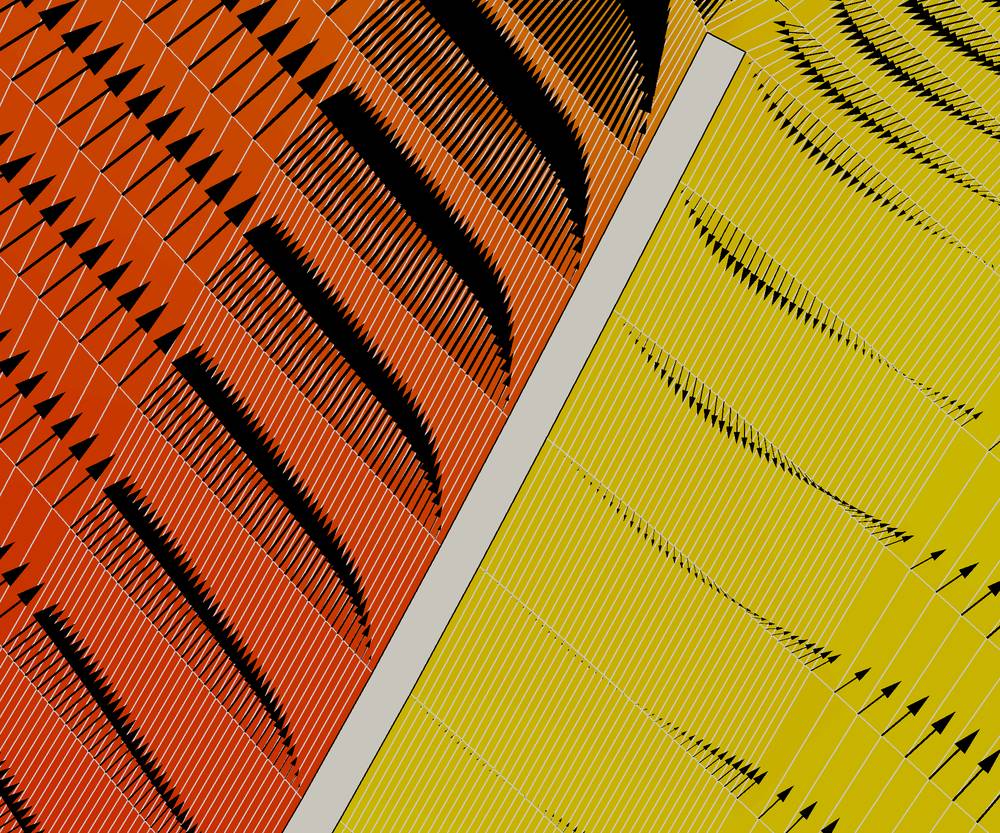}
      \end{varwidth}
}
  \caption{Vibrating flexible structure:
close-up views of boundary layer region at different locations next to the structural head (left) the structural tail (right)
at different times
\protect\subref{fig:vibrating_struct:boundary_layer_snapshot:t5700}
\protect\subref{fig:vibrating_struct:boundary_layer_snapshot:t5880}
and \protect\subref{fig:vibrating_struct:boundary_layer_snapshot:t6200}.
Arrows indicate velocity profile and colored fluid domain shows the pressure solution (pressure color scale \mbox{$[-18.0,7.5]$}).
}
  \label{fig:vibrating_struct:boundary_layer_snapshot}
\end{figure}

The complexity of the flow patterns developing in the vicinity of the structural head and the tail tip
is depicted in close-up views in \Figref{fig:vibrating_struct:boundary_layer_snapshot}.
Classical boundary layer velocity profiles, exhibiting steep near-wall gradients in the interface-tangential velocity solution,
are clearly visible. Thanks to the wall-refined fluid patch~$\mcT_h^{\fd_2}$, these characteristics are properly resolved,
independent of the deformation of the tail.
This is due to the benefit that the patch can track the structural deformation
and thus ensures a proper resolution over the entire simulation time.

\section{Conclusions}
\label{sec:conclusions}

A novel hybrid Eulerian-ALE discretization concept for large deformation and high Reynolds number fluid-structure interaction is proposed.
This approach combines the advantages of pure Eulerian unfitted fixed-grid approximations for the flow field with that of classical interface-fitted moving
mesh \name{ALE} methods.
A boundary layer patch of fluid elements, which fits to the structural mesh and follows its movement over time, ensures a
suitable resolution of wall-normal gradients of the solution fields in the vicinity of the structure. It thus allows to accurately capture boundary layer effects
as it is a prerequisite for challenging \name{FSI}.
By embedding this patch in a geometrically unfitted way into a second background fluid mesh enables to deal with large motions of the structure and its surrounding fluid patch.

In the present work, this approximation concept is realized with the Cut Finite Element Method applied to the composite fluid domain decomposition.
At the matching fluid-solid interface and the non-matching fluid-fluid interface, all interfacial constraints are imposed weakly using Nitsche-type techniques.
For efficiency and temporal stability reasons, the coupled \name{FSI} system is solved in a full-implicit monolithic way, for which new algorithmic aspects have been provided.
The method is validated by comparisons with established \name{ALE} based and \name{CutFEM} based fixed-grid schemes.
The characteristics and the high potentials and capabilities of this novel approximation scheme are indicated by means of different challenging \name{FSI} problem settings
subjected to moderate and large domain motions, which require an accurate capturing of boundary layer effects.

This novel method is not limited to finite element based schemes, but can be realized also within other frameworks, like finite volume or discontinuous Galerkin methods.
Moreover, this hybrid technique is not restricted to fluid-structure interaction, but offers vast new discretization concepts for challenging transport-dominated
multiphysics problems, whose domains are subjected to large changes or require special approximation schemes in certain regions of interest.

\section*{Acknowledgment}

The authors would like to thank S. Shahmiri and R. Kruse for their work on the fluid domain decomposition method.
\clearpage

\bibliography{bibliography}
\bibliographystyle{wileyj}

\end{document}